\documentclass[aps,preprintnumbers,floatfix,nofootinbib,preprint,superscriptaddress,secnumarabic]{revtex4}
\pdfoutput=1

\usepackage{mathrsfs}
\usepackage{amsmath, amsthm, amssymb}
\usepackage{color}
\usepackage{graphicx}
\usepackage{verbatim}

\usepackage{hyperref}	
\hypersetup{colorlinks,bookmarksopen,bookmarksnumbered,citecolor=blue,
linkcolor=blue,pdfstartview=FitH,urlcolor=blue}

\newcommand{\be}{\begin{equation}}
\newcommand{\ee}{\end{equation}}
\newcommand{\bea}{\begin{eqnarray}}
\newcommand{\eea}{\end{eqnarray}}

\newcommand{\fig}{Fig.}

\newcommand{\NO}{{\rm NO}}
\newcommand{\IO}{{\rm IO}}
\newcommand{\ie}{\emph{i.e.}}
\newcommand{\eg}{\emph{e.g.}}

\DeclareMathOperator{\erfc}{erfc}
\DeclareMathOperator{\erf}{erf}
\newcommand{\NOvA}{NO$\nu$A}
\newcommand{\NOvAt}{NO$\mathbf{\nu}$A}
\newcommand{\kt}{kt}
\newcommand{\LBNE}[1]{LBNE-#1~\kt}

\numberwithin{equation}{section}
\begin{document}

\title{\vspace*{2cm}
Quantifying the sensitivity of oscillation experiments to the neutrino mass 
ordering
\vspace*{1cm}
}

\author{Mattias Blennow}
\email{emb@kth.se}
\affiliation{Department of Theoretical Physics, School of Engineering Sciences, KTH Royal Institute of Technology, AlbaNova University Center, 106 91 Stockholm, Sweden}
\author{Pilar Coloma}
\email{pcoloma@vt.edu}
\author{Patrick Huber}
\email{pahuber@vt.edu}
\affiliation{Center for Neutrino Physics, Virginia Tech, Blacksburg, VA 24061, USA}
\author{Thomas Schwetz}
\email{schwetz@fysik.su.se}
\affiliation{Max-Planck-Institut f{\"u}r Kernphysik, Saupfercheckweg 1, 69117 Heidelberg, Germany}
\affiliation{Oskar Klein Centre for Cosmoparticle Physics, 
Department of Physics, Stockholm University, SE-10691 Stockholm, Sweden
\vspace*{1cm}}

\begin{abstract}
\vspace*{0.5cm} Determining the type of the neutrino mass ordering
(normal versus inverted) is one of the most important open questions
in neutrino physics. In this paper we clarify the statistical
interpretation of sensitivity calculations for this measurement.  We
employ standard frequentist methods of hypothesis testing in order to
precisely define terms like the median sensitivity of an
experiment. We consider a test statistic $T$ which in a certain limit
will be normal distributed. We show that the median sensitivity in
this limit is very close to standard sensitivities based on
$\Delta\chi^2$ values from a data set without statistical
fluctuations, such as widely used in the literature. Furthermore, we
perform an explicit Monte Carlo simulation of the INO, JUNO, LBNE,
\NOvA, and PINGU experiments in order to verify the validity of the
Gaussian limit, and provide a comparison of the expected sensitivities
for those experiments.
\end{abstract}

\maketitle

\tableofcontents

\section{Introduction}
\label{sec:intro}

The ordering of neutrinos masses constitutes one of the major open
issues in particle physics. The mass ordering is called ``normal''
(``inverted'') if $\Delta m^2_{31} \equiv m^2_3 - m^2_1$ is positive
(negative). Here and in the following we use the standard
parameterization for the neutrino mass states and PMNS lepton mixing
matrix \cite{PDG}. Finding out which of these two possibilities is realized in
Nature has profound implications for the flavor puzzle, as well as
phenomenological consequences for cosmology, searches for neutrino
mass, and for neutrinoless double-beta decay. Therefore, the
determination of the mass ordering is one of the experimental
priorities in the field.  In particular, with the discovery of a large
value of $\theta_{13}$~\cite{An:2012eh, Ahn:2012nd, Abe:2012tg,
Abe:2013xua} an answer within a decade or so is certainly possible and
first hints may be obtained even sooner in global fits to the world's
neutrino data. 

New information is expected to come from long-baseline experiments,
like T2K~\cite{Abe:2011ks} and
\NOvA~\cite{Ayres:2004js,Patterson:2012zs}, which look for the
appearance of $\nu_e (\bar\nu_e)$ in a beam of $\nu_\mu
(\bar\nu_\mu)$. Proposals for a more long-term time frame include
LBNE~\cite{Akiri:2011dv,CDR,Adams:2013qkq}, LBNO~\cite{Stahl:2012exa},
a superbeam based on the ESS~\cite{Baussan:2013zcy}, and eventually a
neutrino factory~\cite{Choubey:2011zzq}.  Matter effects
\cite{Wolfenstein:1977ue, Barger:1980tf, Mikheev:1986gs} will induce
characteristic differences between the neutrino and antineutrino
channels, which in turn will allow inference of the mass ordering, see
\eg, Refs.~\cite{Freund:1999gy, Barger:2000cp} for early
references. The fact that a comparison of neutrino and antineutrino
channels is performed also implies that the leptonic CP phase $\delta$
cannot be ignored and has to be included in the analysis as well.  A
selective set of recent sensitivity studies for present and future
proposed long baseline oscillation experiments can be found in
Refs.~\cite{Qian:2013nhp,Bass:2013hla,Barger:2013rha,Agarwalla:2013hma,
Messier:2013sfa,Agarwalla:2012bv,Blennow:2013swa, 
Agarwalla:2011hh,Coloma:2012ut,Coloma:2012ma,Dusini:2012vc,Ghosh:2013pfa}. 

Another possibility to determine the mass ordering arises from
observing the energy and zenith angle dependence of atmospheric
neutrinos in the GeV range, which will also have the mass ordering
information imprinted by matter effects \cite{Petcov:1998su,Akhmedov:1998ui,
Akhmedov:1998xq, Chizhov:1998ug, Chizhov:1999az, Banuls:2001zn}.  The
flux of atmospheric neutrinos follows a steep power law with energy
and thus the flux in the GeV range is quite small and requires very
large detectors. IceCube technology can be adapted to neutrino
energies in the GeV range by reducing the spacing of optical modules,
eventually leading to the PINGU extension~\cite{Aartsen:2013aaa} and a
similar low-energy modification can also be implemented for neutrino
telescopes in the open ocean, called ORCA~\cite{orca}.  Another way to
overcome the small neutrino flux is to separate neutrino and
antineutrino events using a magnetic field like in the ICal@INO
experiment~\cite{INO,Ghosh:2013mga} (INO for short in the following).
Mass ordering sensitivity calculations have been performed for
instance in Refs.~\cite{Mena:2008rh, FernandezMartinez:2010am,
Akhmedov:2012ah, Agarwalla:2012uj, Franco:2013in, Ribordy:2013xea,
Winter:2013ema, Blennow:2013vta, Ge:2013zua} for PINGU/ORCA and
in Refs.~\cite{TabarellideFatis:2002ni, PalomaresRuiz:2004tk,
Indumathi:2004kd, Petcov:2005rv, Samanta:2006sj, Gandhi:2007td,
Kopp:2007ai, Blennow:2012gj, Ghosh:2012px} for INO or similar setups.

Finally, the interference effects between the oscillations driven by $\Delta m_{21}^2$ and $\Delta m_{31}^2$ in the disappearance of $\bar\nu_e$
provide a third potential avenue for this measurement. In particular,
this approach has been put forward in the context of reactor
neutrinos~\cite{Petcov:2001sy}. JUNO~\cite{JUNO, Li:2013zyd} will
comprise a 20~\kt\ detector at a baseline of about 52\,km of several
nuclear reactors. A similar project is also discussed within the RENO
collaboration~\cite{reno50}.  The possibility to use a precision
measurement of the $\bar\nu_e$ survival probability at a nuclear
reactor to identify the neutrino mass ordering has been considered by
a number of authors, \eg, Refs.~\cite{Schonert:2002ep, Choubey:2003qx,
Learned:2006wy,
Zhan:2008id, Zhan:2009rs, Ghoshal:2010wt,
Qian:2012xh, 
Ge:2012wj, Ciuffoli:2013ep, Li:2013zyd, Blennow:2013vta, 
Kettell:2013eos, Capozzi:2013psa}. 

This impressive experimental (and phenomenological) effort has also
resulted in a renewed interest in potential issues arising from the
statistical interpretation of the resulting data~\cite{Qian:2012zn,
Ciuffoli:2013rza} (see also~\cite{Ge:2012wj}), which can be summarized as: Given that the
determination of the mass ordering is essentially a binary yes-or-no
type question, are the usual techniques relying on a Taylor expansion
around a single maximum of the likelihood applicable in this case? The
goal of this paper is to answer this question within a frequentist
framework for a wide range of experimental situations, including
disappearance as well as appearance measurements. The answer we find
in this paper can be stated succinctly as: The currently accepted
methods yield approximately the expected frequentist coverage for the
median experimental outcome; quantitative corrections typically lead
to a (slightly) increased sensitivity compared to the standard
approach prediction. The methods applied in the following are
analogous to the ones from Ref.~\cite{Schwetz:2006md}, where similar
questions have been addressed for the discovery of $\theta_{13}$ and
CP violation. In the present work we strictly adhere to frequentist
methods; Bayesian statistics is used to address the neutrino mass
ordering question in Ref.~\cite{Blennow:2013kga}, see also Refs.~\cite{Qian:2012zn,
Ciuffoli:2013rza} for Bayesian considerations.

The outline of our paper is as follows.  We first review the
principles of hypothesis testing in a frequentist framework in
Sec.~\ref{sec:hypothesis-testing}, apply them to the case of the mass
ordering, define the sensitivity of the median experiment and discuss
the relation to the standard sensitivity based on $\Delta\chi^2$
values from the Asimov data set. In Sec.~\ref{sec:gauss} we consider
the Gaussian limit, where all relevant quantities, such as
sensitivities can be expressed analytically. Details of the derivation
can be found in App.~\ref{app:Tgauss}, as well as a discussion of
conditions under which the Gaussian approximation is expected to
hold. In Sec.~\ref{sec:numericalresults} we present results from Monte
Carlo simulations of the INO, PINGU, JUNO, \NOvA, and LBNE
experiments. The technical details regarding the simulations are
summarized in App.~\ref{app:simulation}.  We show that for most cases
the Gaussian approximation is justified to good accuracy, with the
largest deviations observed for
\NOvA. In Sec.~\ref{sec:comparison} we present a 
comparison between the sensitivities expected for the different
proposals, illustrating how the sensitivities may evolve with date.
We summarize in Sec.~\ref{sec:summary}, where we also provide a table
which allows to translate the traditional performance indicator for
the mass ordering ($\Delta\chi^2$ without statistical fluctuations)
into well defined frequentist sensitivity measures under the Gaussian
approximation. We also comment briefly on how our results
compare to those in Refs.~\cite{Qian:2012zn, Ciuffoli:2013rza}.

\section{Terminology and statistical methods}
\label{sec:statanalysis}

\label{sec:hypothesis-testing}

\subsection{Frequentist hypothesis testing}

Let us start by reviewing the principles of frequentist hypothesis testing, see \eg, Ref.~\cite{PDG}. First we consider the case of so-called ``simple hypotheses'', where the hypothesis we want to test, $H$, as well as the alternative hypothesis $H'$ do not depend on any free parameters. $H$ is
conventionally called null hypothesis. In order to test whether data can reject the null hypothesis $H$ we have to choose a test statistic $T$. A test statistic is a stochastic variable depending on the data which is chosen in such a way that the more extreme the outcome is considered to be, the larger (or smaller) the value of the test statistic is. Once the distribution of $T$ is known under the assumption of $H$ being true, we decide to reject $H$ at confidence level (CL) $1-\alpha$ if the observation is within the $\alpha$ most extreme results, \ie, if $T > T_c^\alpha$, where the critical value $T_c^\alpha$ is defined by 
\begin{equation}
 \int_{T_c^\alpha}^\infty p(T|H) dT = \alpha \,,
\end{equation}
with $p(T|H)$ being the probability distribution function of $T$
given that $H$ is true. The probability $\alpha$ is the probability
of making an ``error of the first kind'' (or type-I error rate), \ie, rejecting $H$ although
it is true. It is custom to convert this probability into a number of
Gaussian standard deviations. In this work we will adopt the
convention to use a double-sided Gaussian test for this conversion, such that a
hypothesis is rejected if the data is more than $n\sigma$ away (on
either side) from the mean. This leads to the following conversion
between $n\sigma$ and the value of $\alpha$:\footnote{Note that we are
using the complementary error function $\erfc(x) \equiv 1 - \erf(x)$.}
\begin{equation}\label{eq:sigma-alpha}
\alpha(n) = \frac{2}{\sqrt{2\pi}} \int_{n}^{\infty} dx \, e^{-x^2/2} = 
\erfc\left(\frac n{\sqrt 2}\right) \quad \Leftrightarrow \quad n = \sqrt 2 \erfc^{-1}(\alpha).
\end{equation}
This definition implies that we identify, for instance, $1\sigma,2\sigma,3\sigma$
with a CL $(1-\alpha)$ of 68.27\%, 95.45\%, 99.73\%,
respectively, which is a common convention in neutrino physics. However,
note that $n\sigma$ is sometimes defined differently, as
a one-sided Gaussian limit, see e.g., Eq.~(1) of Ref.~\cite{Cowan:2010js}. This
leads to a different conversion between $n\sigma$ and $\alpha$, namely
\begin{equation}\label{eq:sigma-alpha-1sided}
 n_\text{1-sided} = \sqrt 2 \erfc^{-1}(2\alpha) \,,
\end{equation}
which would lead to a CL of 84.14\%, 97.73\%, 99.87\% for $1\sigma,2\sigma,3\sigma$.

In order to quantify how powerful a given test is for rejecting $H$ at a given CL we have to compute the so-called ``power'' of the test or, equivalently, the probability of making an ``error of the second kind'' (or type-II error rate). This is the probability $\beta$ to accept $H$ if it is not true:
\begin{equation}
 \beta = P(T < T_c^\alpha|H') = \int_{-\infty}^{T_c^\alpha} p(T|H') dT \,,
\end{equation}
where now $p(T|H')$ is the probability distribution function of $T$
assuming that the alternative hypothesis $H'$ is true. Obviously, $\beta$
depends on the CL $(1-\alpha)$ at which we want to reject $H$. A small
value of $\beta$ means that the rate for an error of the
second kind is small, \ie, the power of the test (which is defined as
$1-\beta$) is large.

The case we are interested in here (neutrino mass ordering) is slightly different, since both hypotheses (normal and inverted) may depend on additional parameters $\theta$, a situation which is called ``composite hypothesis testing''. This is for instance the case of long baseline oscillation experiments, where the value of $\delta$ has a large impact on the sensitivities to the neutrino mass ordering. In this case the same approach is valid while keeping a few things in mind:
\begin{itemize}
\item We can reject the hypothesis $H$ only if we can reject all $\theta \in H$. Thus, with
\begin{equation}
	 \int_{T_c^\alpha(\theta)}^\infty p(T|\theta\in H) dT = \alpha,
\end{equation}
we must chose 
\begin{equation}\label{eq:Tc-composite}
T_c^\alpha = \max_{\theta \in H} T_c^\alpha(\theta) \,.
\end{equation}
This ensures that all $\theta \in H$ are rejected at confidence level
$(1-\alpha)$ if $T > T_c^\alpha$.\footnote{Here we assume that for
given data the value of the observed test statistic $T$ is independent
of the true parameter values. This is the case for the statistic $T$ 
introduced in Eq.~\eqref{eq:T}, but it will not be true for instance
for the statistic $T'$ mentioned in footnote~\ref{foot:Tprime}. }
\item The rate of an error of the second kind will now depend on the true parameters in the alternative hypothesis:
\begin{equation}
 \beta(\theta) = P(T < T_c^\alpha|\theta \in H') = 
 \int_{-\infty}^{T_c^\alpha} p(T|\theta \in H') dT \,,
\end{equation}
with $T_c^\alpha$ defined in Eq.~\eqref{eq:Tc-composite}. It is important to
note that in a frequentist framework this cannot be averaged in any
way to give some sort of mean rejection power, as this would require an
assumption about the distribution of the parameters implemented in
Nature (which is only possible in a Bayesian analysis~\cite{Blennow:2013kga}). Sticking to
frequentist reasoning, we can either give $\beta$ as a function of the
parameters in the alternative hypothesis, or quote the highest and/or
lowest possible values of $\beta$ within the alternative hypothesis.
\end{itemize}

\subsection{Application to the neutrino mass ordering}
\label{sec:MO}

In the search for the neutrino mass ordering, we are faced with two
different mutually exclusive hypotheses, namely $H_\NO$ for normal ordering and
$H_\IO$ for inverted ordering. Both hypotheses will depend
on the values of the oscillation parameters (which we collectively denote
by $\theta$) within the corresponding ordering. In particular,
appearance experiments depend crucially on the CP-violating
phase $\delta$. Hence, we have to deal with the situation of composite
hypothesis testing as described above. Let us now select a specific
test statistic for addressing this problem.

A common test statistic is the $\chi^2$ with $n$ degrees of freedom, which describes the deviation from the expected values of the outcome of a series of measurements $x_i$ of the normal distributions $\mathcal{N}(\mu_i,\sigma_i)$:
\begin{equation}
 \chi^2 = \sum_{i=1}^n \frac{(x_i-\mu_i)^2}{\sigma_i^2}\,. \label{eq:chi2}
\end{equation}
The further the observations are from the expected values, \ie, the more extreme the outcome, the larger is the $\chi^2$. If the mean values $\mu_i$ depend on a set of $p$ parameters $\theta$ whose values have to be estimated from the data one usually considers the minimum of the $\chi^2$ with respect to the parameters:
\begin{equation}
  \chi^2_{\rm min} = \min_{\theta} \chi^2(\theta) \,.
\end{equation}
According to Wilk's theorem~\cite{Wilks} this quantity will follow a $\chi^2$ distribution with $n-p$ degrees of freedom, whereas $\Delta\chi^2(\theta) = \chi^2(\theta) - \chi^2_{\rm min}$ will have a $\chi^2$ distribution with $p$ degrees of freedom. The $\chi^2$ distributions have known properties, and in physics we often encounter situations where data can be well described by this method and the conditions for Wilk's theorem to hold are sufficiently fulfilled, even when individual data points are not strictly normal distributed. In general, however, it is not guaranteed and the actual distribution of those test statistics has to be verified by Monte Carlo simulations~\cite{Feldman:1997qc}.

Coming now to the problem of identifying the neutrino mass ordering,
one needs to select a test statistic which is well suited to
distinguish between the two hypotheses $H_\NO$ and $H_\IO$. Here we
will focus on the following test statistic, which is based on a
log-likelihood ratio and has been used in the past in the literature:
\begin{equation}
T = \min_{\theta \in \IO} \chi^2(\theta) - \min_{\theta \in \NO} \chi^2(\theta) 
\equiv \chi^2_\IO - \chi^2_\NO, \label{eq:T}
\end{equation}
where $\theta$ is the set of neutrino oscillation parameters which
are confined to a given mass ordering during the minimization. Let us
stress that the choice of $T$ is not unique. In principle one is free
to chose any test statistic, although some will provide more powerful
tests than others.\footnote{In the case of simple hypotheses the
  Neyman Pearson lemma~\cite{NeymanPearson1933} implies that the test
  based on the likelihood ratio is most powerful. For composite
  hypotheses in general no unique most powerful test is known. 
  An alternative choice for a test statistic could be for instance the statistic 
  $T'(\theta) = \chi^2(\theta) - \chi^2_{\rm min}$, where $\chi^2_{\rm min}$ 
  is the absolute minimum including minimization over the two mass 
  orderings, and $\theta$ generically denotes the (continuous) oscillation 
  parameters. This statistic is based on parameter estimation and amounts 
  to testing whether a parameter range for $\theta$ remains at a given CL 
  in a given mass ordering. We have checked by explicit Monte Carlo simulations 
  that typically the distribution of $T'$ is close to a $\chi^2$ distribution 
  with number of d.o.f.\ corresponding to the non-minimized parameters in 
  the first term (the approximation is excellent for JUNO but somewhat 
  worse for LBL experiments). Sensitivity results for the mass ordering 
  based on $T'$ will be reported elsewhere.
  \label{foot:Tprime}}

\begin{figure}
\begin{center}
\includegraphics[height=0.365\textwidth]{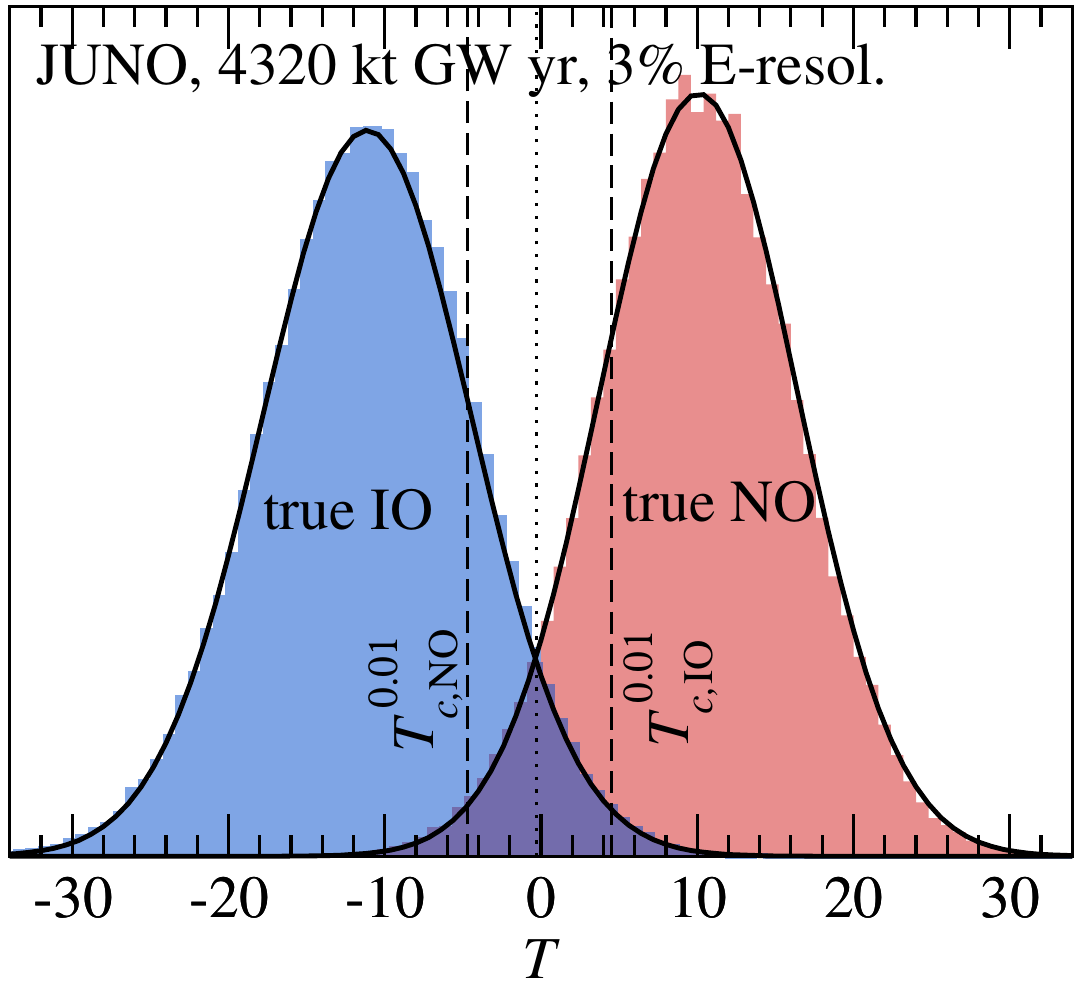} \quad
\raisebox{-2mm}{\includegraphics[height=0.39\textwidth]{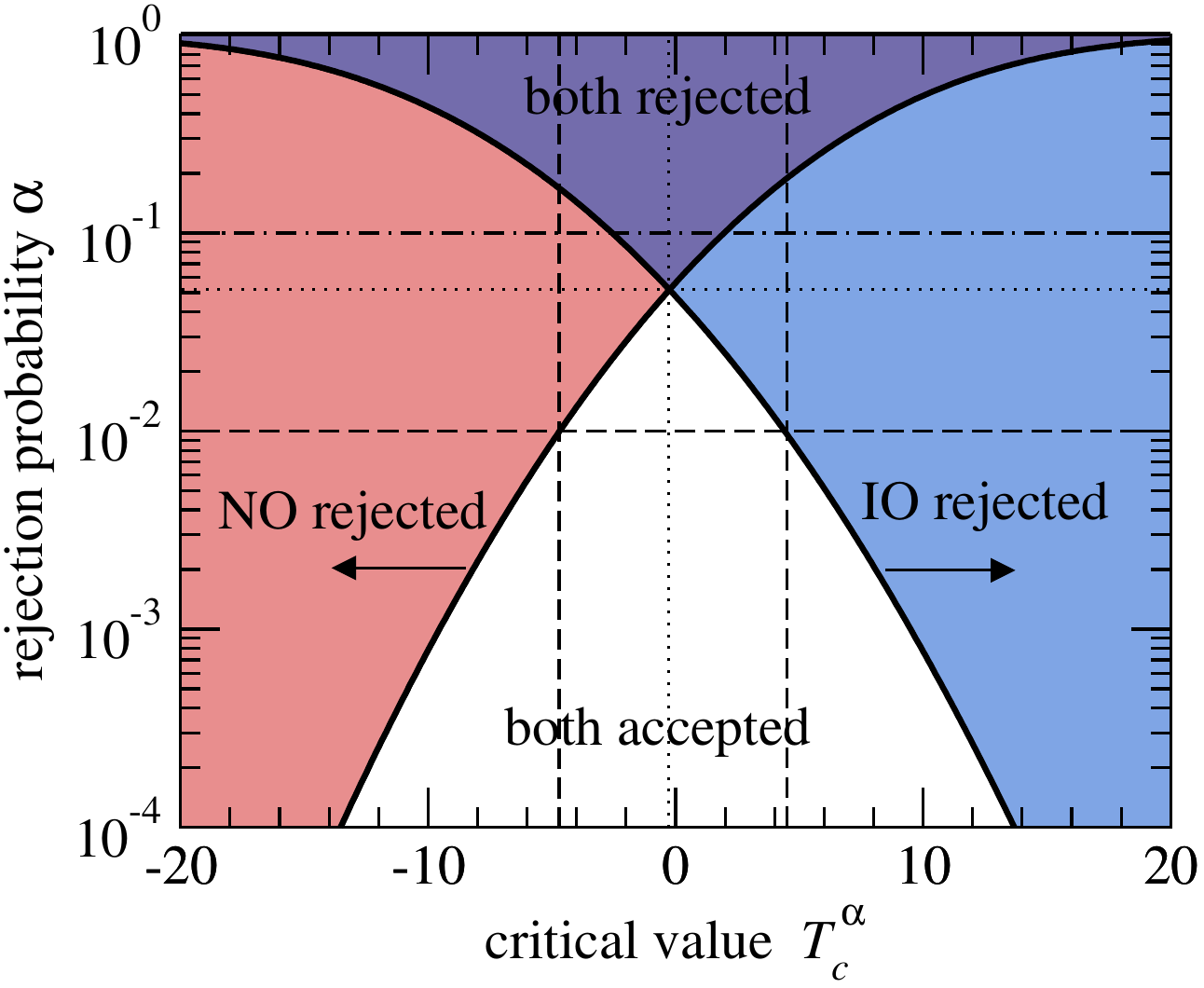}}
  \caption{Left: Distribution of the test statistic $T$ for our
  default configuration of the JUNO reactor experiment discussed in
  Sec.~\ref{sec:juno}. Histograms show the results of the MC
  simulation based on $10^5$ simulated experiments and black curves
  correspond to the Gaussian approximation discussed in
  Sec.~\ref{sec:gauss}. Right: The value of $\alpha$ as a function of
  the critical value $T_{c}^\alpha$ required for rejecting inverted
  (blue) and normal (red) ordering for the JUNO reactor experiment. In
  the purple region both mass orderings are rejected at the CL
  ($1-\alpha$), in the white region both orderings are consistent with
  data at the CL ($1-\alpha$). The dashed lines in both panels
  indicate $T_c^\alpha$ for $\alpha = 0.01$ for both orderings. The
  dotted lines indicate the crossing point $T_c^\NO =
  T_c^\IO$. The dot-dashed line in the right panel shows an example (for $\alpha=0.1$)  
  in which $T_{c,\textrm{IO}}^\alpha < T_{c,\textrm{NO}}^\alpha$. } \label{fig:Tc}
\end{center}
\end{figure}

It is important to note that within a frequentist approach, rejecting
one hypothesis at a given $\alpha$ does not automatically imply that
the other hypothesis could not also be rejected using the same
data. Instead, the only statement we can make is to either reject an
ordering or not. The value of $T = 0$ therefore does not a priori play
a crucial role in the analysis.  Let us illustrate this point at an
example.  In the left panel of Fig.~\ref{fig:Tc}, we show the
distributions of the test statistics $T$ for both mass orderings
obtained from the simulation of a particular configuration of the JUNO
reactor experiment. Experimental details will be discussed later in
Sec.~\ref{sec:juno}. In the right panel we show the corresponding
critical values $T_c^\alpha$ for testing both orderings and how they
depend on the chosen confidence level $1-\alpha$. The curves for
testing the different orderings cross around $\alpha = 5.2\%$,
indicated by the dotted lines. This represents the unique confidence
level for which the experiment in question will rule
out \emph{exactly} one of the orderings, regardless of the
experimental outcome. If, for instance, we would choose to test
whether either ordering can be rejected at a confidence level of 90\%,
then there is a possibility of an experimental outcome $T$ with
$T_{c,\IO}^{0.1} < T < T_{c,\NO}^{0.1}$, implying that \emph{both}
orderings could be rejected at the 90\%~CL. This situation is
indicated by the dash-dotted line in the right panel of
Fig.~\ref{fig:Tc} and applies to the purple region. Thus, in order to
claim a discovery of the mass ordering, it will not be sufficient to
test one of the orderings. If both orderings were rejected at high
confidence, it would mean either having obtained a very unlikely
statistical fluctuation, underestimating the experimental errors, or
neither ordering being a good description due to some new
physics. Conversely, if we would choose $\alpha = 0.01 < 0.052$ (dashed line in
both panels, white region in right panel), then there is the
possibility of obtaining $T_{c,\NO}^{0.01} < T < T_{c,\IO}^{0.01}$,
meaning that neither ordering can be excluded at the 99\%~CL.

The CL corresponding to the crossing condition $T_{c,\NO}^\alpha =
T_{c,\IO}^\alpha$ provides a possible sensitivity measure of a given
experiment. We will refer to it as ``crossing sensitivity'' below.\footnote{In the
  case of composite hypotheses, where the distribution of $T$ depends on
  the true values of some parameters (\eg, the CP phase in the case of
  long-baseline experiments), we define $T_{c,\NO}^\alpha$ and
  $T_{c,\IO}^\alpha$ in analogy to Eq.~\eqref{eq:Tc-composite}, \ie, we
  chose the largest or smallest value of $T_c^\alpha(\theta)$, depending
  on the mass ordering. Hence, the crossing sensitivity is
  independent of the true values of the parameters.} 
If $T_{c,\NO}^\alpha \approx - T_{c,\IO}^\alpha$ (as it is the case for the example shown
in Fig.~\ref{fig:Tc}), this is equivalent to testing the sign of
$T$. This test has been discussed also in~Ref.~\cite{Qian:2012zn, Ciuffoli:2013rza}. From
the definition of the sensitivity of an average experiment which we
are going to give in the next subsection it will be clear that the
crossing sensitivity is rather different from the median
sensitivity, which is typically what is intended by ``sensitivity'' in the existing literature. 
It should also be noted that the critical values for the
different orderings, as well as the crossing of the critical values, in
general are not symmetric with respect to $T=0$. The fact that
Fig.~\ref{fig:Tc} appears to be close to symmetric is a feature of the
particular experiment as well of the test statistic $T$. This would
not be the case for instance for the statistic $T'$ mentioned in
footnote~\ref{foot:Tprime}. Finally, note that Fig.~\ref{fig:Tc} is
only concerned with the critical value of $T$ and its dependence on
$\alpha$. As such, it does not tell us anything about the probability
of actually rejecting, for instance, inverted ordering if the normal ordering would
be the true one (power of the test). As discussed above, this
probability will typically also depend on the actual parameters within
the alternative ordering and can therefore not be given a particular
value. However, for the crossing point of the critical values, the
rejection power for the other ordering is at least $1-\alpha$.

\subsection{Median sensitivity or the sensitivity of an average experiment}
\label{sec:average}

Let us elaborate on how to compare such a statistical analysis to
previous sensitivity estimates massively employed in the literature,
in particular in the context of long-baseline oscillation
experiments. The most common performance indicator used for the mass
ordering determination is given by
\begin{equation}\label{eq:T0}
  T_0^\NO(\theta_0) = 
  \min_{\theta \in \IO} \sum_i \frac{[\mu_i^\NO(\theta_0) - \mu_i^\IO(\theta)]^2}{\sigma_i^2} 
\end{equation}
for testing normal ordering, with an analogous definition for inverted
ordering. This quantity corresponds to the test statistic $T$ defined
in Eq.~\eqref{eq:T} but the data $x_i$ are replaced by the predicted
observables $\mu_i(\theta_0)$ at true parameter values
$\theta_0$. Since no statistical fluctuations are included in this
definition it is implicitly assumed that it is representative for an
``average'' experiment. (This is sometimes referred to as the Asimov
data set \cite{Cowan:2010js}, and $T_0$ is sometimes denoted as
``$\overline{\Delta\chi^2}$'' \cite{Qian:2012zn}.)  $T_0$ is then
evaluated assuming a $\chi^2$ distribution with 1~dof in order to
quote a ``CL with which a given mass ordering can be identified''. In
the following, we will refer to this as the ``standard method'' or
``standard sensitivity''. Note that $T_0$ by itself is not a
statistic, since it does not depend on any random data. The
interpretation of assigning a $\chi^2$ distribution to it is not well
defined, and is motivated by the intuition based on nested hypothesis
testing (which is not applicable for the mass ordering question). In
the following we show that actually the relevant limiting distribution
for $T$ (but not for $T_0$) is Gaussian, not $\chi^2$.

The formalism described in section~\ref{sec:hypothesis-testing} allows
a more precise definition of an ``average'' experiment. One
possibility is to calculate the CL $(1-\alpha)$ at which a false
hypothesis can be rejected with a probability of 50\%, \ie, with a
rate for an error of the second kind of $\beta = 0.5$. In other
words, the CL $(1-\alpha)$ for $\beta = 0.5$ is the CL at which an
experiment will reject the wrong mass ordering with a probability of
50\%. We will call the probability $\alpha(\beta=0.5)$ the
``sensitivity of an average experiment'' or ``median sensitivity''. This
is the definition we are going to use in the following for comparing
our simulations of the various experiments to the corresponding
sensitivities derived from the standard method.

Let us note that the median sensitivity defined in this way is not the only relevant quantity in order to design an experiment,
since in practice one would like to be more certain than 50\% for
being able to reject a wrong hypothesis. Under the Gaussian
approximation to be discussed in the next section it is easy to
calculate the sensitivity $\alpha$ for any desired $\beta$, once the
median sensitivity is known.

\section{The Gaussian case for the test statistic $T$}
\label{sec:gauss}

A crucial point in evaluating a statistical test is to know the
distribution of the test statistic. In general this has to be
estimated by explicit Monte Carlo simulations, an exercise which we
are going to report on for a number of experiments later in this
paper. However, under certain conditions the distribution of the
statistic $T$ defined in Eq.~\eqref{eq:T} can be derived analytically
and corresponds to a normal distribution \cite{Qian:2012zn}:
\begin{equation}\label{eq:Tgauss1}
 T = \mathcal{N}( \pm T_0, 2\sqrt{T_0}) \,,
\end{equation}
where $\mathcal{N}(\mu,\sigma)$ denotes the normal distribution with
mean $\mu$ and standard deviation $\sigma$ and the + ($-$) sign holds
for true NO (IO).\footnote{Note that $T_0^\NO$ and $T_0^\IO$ are always
defined to be positive according to Eq.~\eqref{eq:T0}, while $T$ can
take both signs, see Eq.~\eqref{eq:T}.}  In general $T_0^\NO$ and
$T_0^\IO$ may depend on model parameters $\theta$. In that case the
distribution of $T$ will depend on the true parameter values and we
have to consider the rules for composite hypothesis testing as
outlined in section~\ref{sec:hypothesis-testing}. We provide a
derivation of Eq.~\eqref{eq:Tgauss1} in App.~\ref{app:Tgauss},
where we also discuss the conditions that need to be fulfilled for
this to hold in some detail.  In addition to assumptions similar to
the ones necessary for Wilk's theorem to hold, Eq.~\eqref{eq:Tgauss1}
applies if
\begin{itemize}
\item we are dealing with simple hypotheses, or consider composite hypotheses 
at fixed parameter values, or
\item if close to the respective $\chi^2$ minima the two hypotheses depend on 
the parameters ``in the same way'' (a precise definition is given via
Eq.~\eqref{eq:condV} in the appendix), or
\item if $T_0$ is large compared to the number of relevant parameters of the hypotheses.
\end{itemize}

\subsection{Simple hypotheses}
\label{sec:gauss-simple}

Let us now study the properties of the hypothesis test for the mass
ordering based on the statistic $T$ under the assumption that it
indeed follows a normal distribution as in
Eq.~\eqref{eq:Tgauss1}. First we consider simple hypotheses, \ie,
$T_0$ does not depend on any free parameters. As we shall see below,
this situation applies with good accuracy to the medium-baseline
reactor experiment JUNO.

\begin{figure}
\begin{center}
\includegraphics[height=0.45\textwidth]{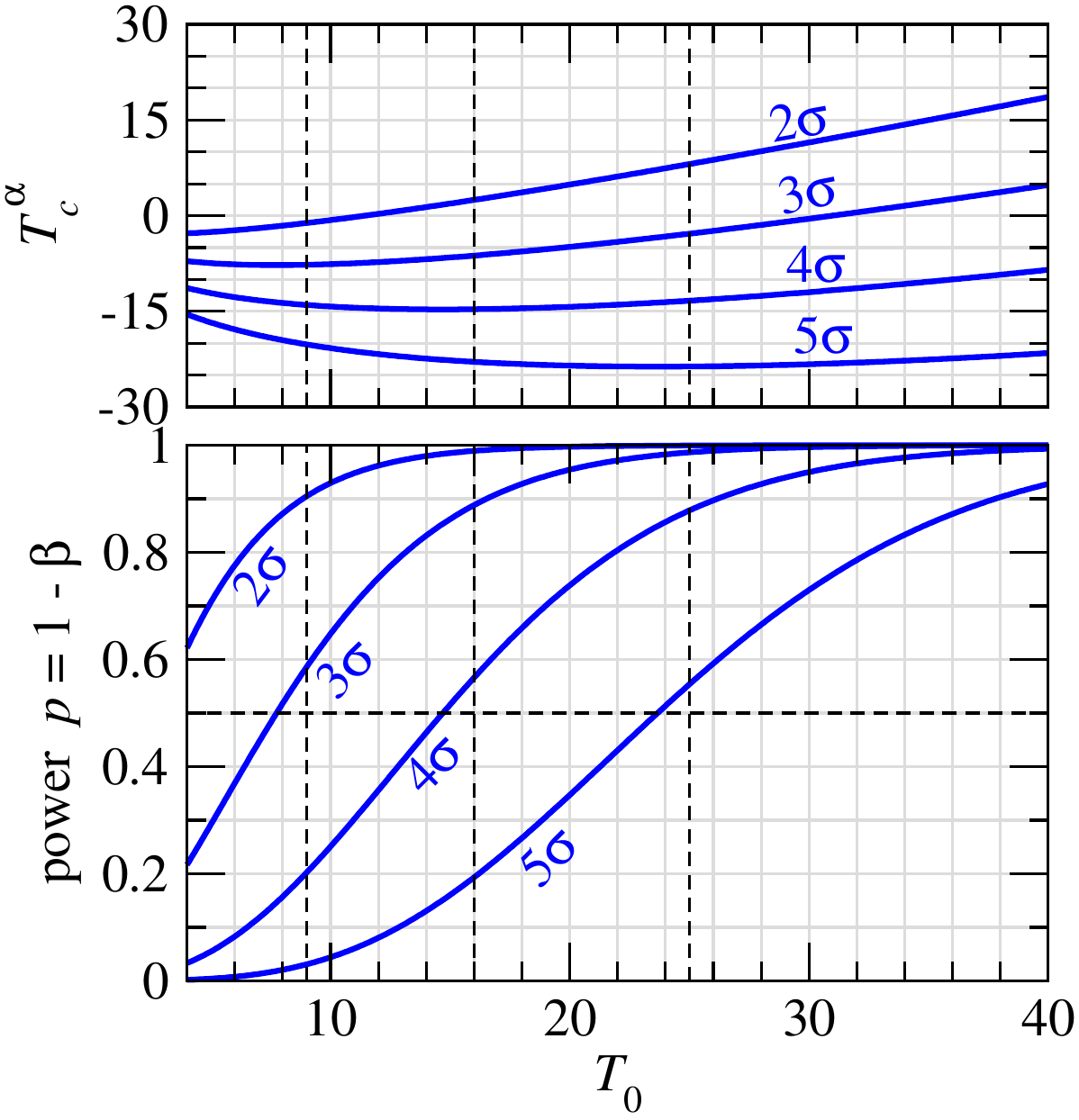}\quad
\includegraphics[height=0.45\textwidth]{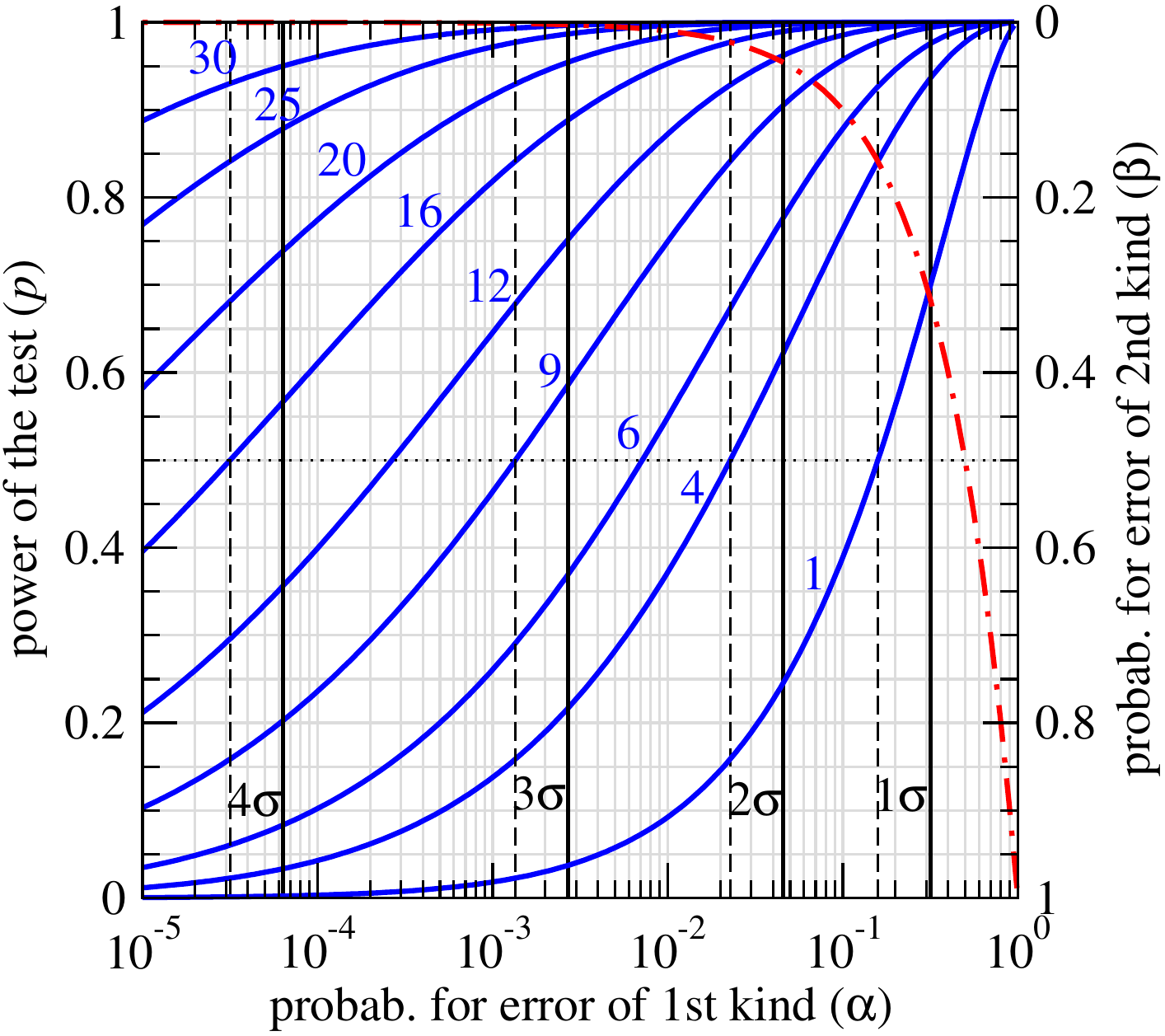}
  \caption{Gaussian approximation for the test statistics $T$.  Left
  upper panel: critical values for rejecting normal ordering as a
  function of $T_0$, see Eq.~\eqref{eq:Tca-gauss}, for different
  values of $\alpha$ as labeled in the plot. Left lower panel: power
  of the test as a function of $T_0$ for different values of $\alpha$,
  see Eq.~\eqref{eq:beta-gauss}.  Right panel: power of the test (left vertical axis) and the
  rate for an error of the second kind (right vertical axis)
  versus the CL ($1-\alpha$) for rejecting a given mass ordering for
  different values of $T_0$ as labeled in the plot. The vertical lines
  indicate the number of standard deviations, where we have used our
  standard convention Eq.~\eqref{eq:sigma-alpha} based on a 2-sided
  Gaussian for the solid lines and Eq.~\eqref{eq:sigma-alpha-1sided}
  based on a 1-sided Gaussian limit for the dashed lines. The dash-dotted red
  curve indicates $\alpha = \beta$, which follows in the Gaussian case
  from the condition $T_c^\NO = T_c^\IO$.
\label{fig:Gaussiansensitivity}}
\end{center}
\end{figure}

For definiteness we construct a test for $H_\NO$; the one for $H_\IO$ is
obtained analogously. Since large values of the test statistic favor
$H_\NO$ over the alternative hypothesis $H_\IO$, we would reject
$H_\NO$ for too small values of $T$. Hence, we need to find a critical
value $T_c^\alpha$ such that $P(T < T_c^\alpha) = \alpha$ if $H_\NO$
is correct. Since $H_\NO$ predicts $T = \mathcal{N}(T_0^\NO,2\sqrt{T_0^\NO})$,
we obtain
\begin{equation}\label{eq:Tca-gauss}
\alpha = \frac 12 \erfc\left(\frac{T_0^\NO-T_c^\alpha}{\sqrt{8 T_0^\NO}}\right) \quad \Leftrightarrow \quad T_c^\alpha = T_0^\NO - \sqrt{8T_0^\NO} \erfc^{-1}\left(2\alpha\right).
\end{equation}
The critical values $T_c^\alpha$ as a function of $T_0$ are shown for
several values of $\alpha$ in the upper left panel of
Fig.~\ref{fig:Gaussiansensitivity}. The labels in the left panel of
the figure in units of $\sigma$ are based on our default
convention based on the 2-sided Gaussian, Eq.~\eqref{eq:sigma-alpha}.

Let us now compute the power $p$ of the test, \ie, the
probability $p$ with which we can reject $H_\NO$ at the CL
$(1-\alpha)$ if the alternative hypothesis $H_\IO$ is true. As
mentioned above, $p$ is related to the rate for an error
of the second kind, $\beta$, since $p = 1 - \beta$. This probability is given by 
$\beta = P(T>T_c^\alpha)$ for true IO, where
$T_c^\alpha$ is given in Eq.~\eqref{eq:Tca-gauss}.
If $H_\IO$ is true we have $T = \mathcal{N}(-T_0^\IO, 2\sqrt{T_0^\IO})$ and hence
\begin{equation}\label{eq:beta-gauss}
\beta = \frac 12 \erfc\left(\frac{T_0^\IO+T_c^\alpha}{\sqrt{8 T_0^\IO}}\right)
 \approx \frac 12 \erfc\left(\sqrt{\frac{T_0}{2}} - \erfc^{-1}(2\alpha)  \right) \,,
\end{equation}
where the last approximation assumes $T_0 \equiv T_0^\NO \approx
T_0^\IO$, a situation we are going to encounter for instance in the
case of JUNO below. We shown $p = 1 - \beta$ as a function of $T_0$
for several values of $\alpha$ in the lower left panel of
Fig.~\ref{fig:Gaussiansensitivity}.

Equation~\eqref{eq:beta-gauss} (or the lower left panel of
Fig.~\ref{fig:Gaussiansensitivity}) contains all the information needed to
quantify the sensitivity of an experiment. In particular, it allows to
address the question of how likely it is that the wrong mass ordering will be
rejected at a given CL. For example, let us consider an experiment
with a median sensitivity of $4\sigma$, which implies $T_0 \approx
14.7$. If we now demand that we want to reject the wrong mass ordering
with a probability of 90\% ($\beta = 0.1$), then this experiment will
be able to do this only at slightly more than 99\%~CL. In the right panel
of Fig.~\ref{fig:Gaussiansensitivity} we show $\beta$ as a function of
$\alpha$ for several fixed values of $T_0$ using
Eq.~\eqref{eq:beta-gauss}. This plot allows a well defined
interpretation of the ``$\Delta\chi^2$'' used in the standard method
({\it i.e., $T_0$}) under the Gaussian approximation. For a given
$T_0$ and a chosen sensitivity $\alpha$ we can read off the
probability with which the experiment will be able to reject the wrong
ordering at the ($1-\alpha$) CL.

\begin{figure}
\begin{center}
\includegraphics[height=0.45\textwidth]{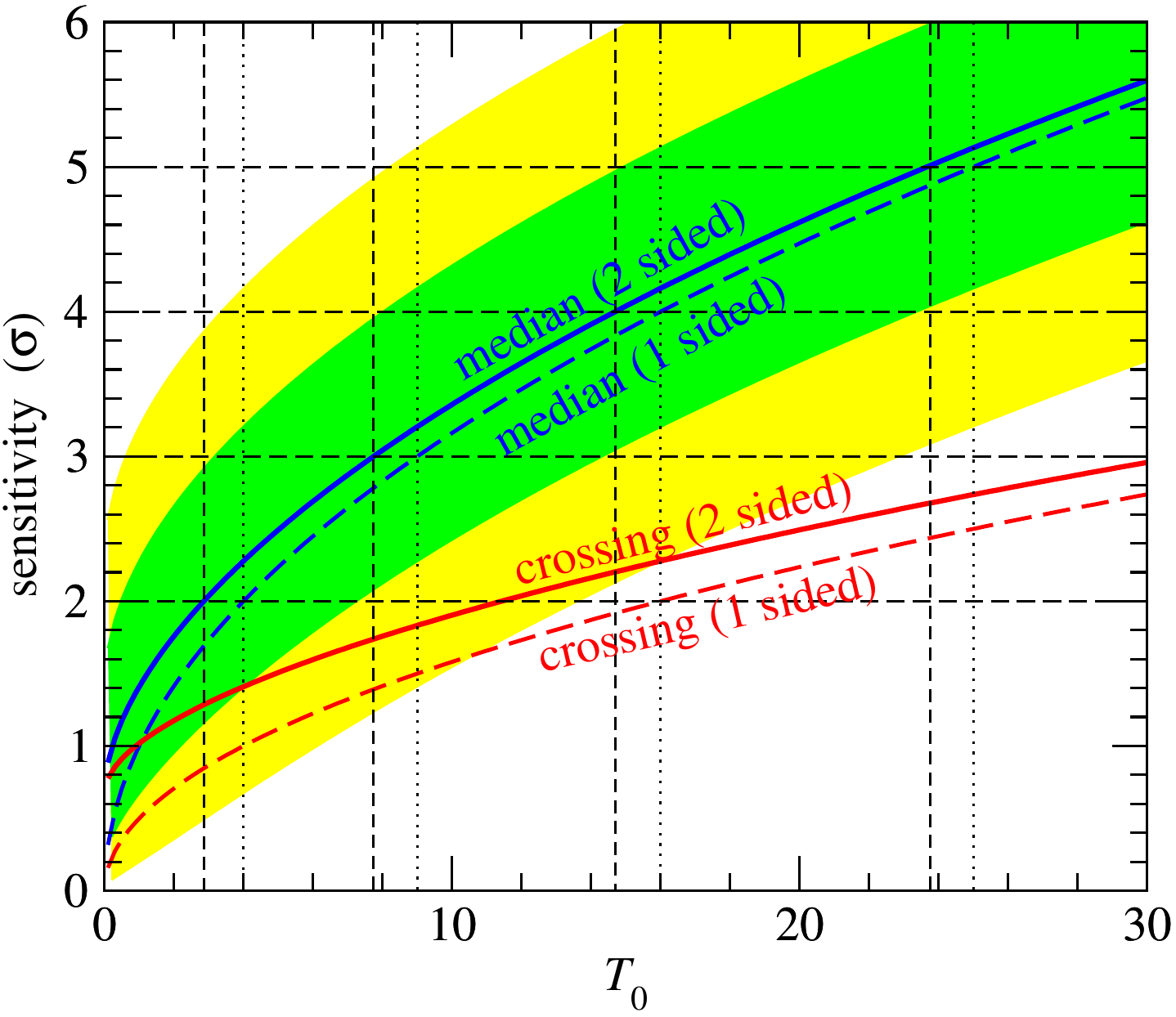}
\end{center} 
\caption{Median sensitivity ($\beta=0.5$) as a function of $T_0$, see
  Eq.~\eqref{eq:median-sens-gauss}.  The curves labeled ``crossing''
  show the sensitivity corresponding to the condition $T_c^\NO =
  T_c^\IO$ according to Eq.~\eqref{eq:alpha-crossing}. The solid
  curves use the 2-sided Gaussian to convert $\alpha$ into $n\sigma$,
  Eq.~\eqref{eq:sigma-alpha}, whereas the dashed curves are based on
  the 1-sided test, Eq.~\eqref{eq:sigma-alpha-1sided}. The latter
  correspond to the ``standard sensitivity'' of $n = \sqrt{T_0}$ and
  $n = \sqrt{T_0}/2$ for the crossing sensitivity.  The edges of the green and 
  yellow bands are obtained from the conditions on the rate for an error 
  of the second kind $\beta = 1/2 \pm 0.6827/2$ and $\beta = 1/2 \pm 
  0.9545/2$, respectively.
   \label{fig:sens-gauss}}
\end{figure}

Now it is also straight forward to compute the median sensitivity, which
we have defined in section~\ref{sec:average} as the $\alpha$ for which $\beta =
0.5$. From Eq.~\eqref{eq:beta-gauss} we obtain
\begin{equation}\label{eq:median-alpha-gauss}
 \alpha = 
\frac 12 \erfc\left(\frac{T_0^\IO + T_0^\NO}{\sqrt{8 T_0^\IO}}  \right)
\approx 
\frac 12 \erfc\left(\sqrt{\frac{T_0}2}\right) \qquad \text{(median sensitivity)}.
\end{equation}
Using our standard convention Eq.~\eqref{eq:sigma-alpha} to convert
$\alpha$ into standard deviations the median sensitivity is $n\sigma$,
with
\begin{equation}\label{eq:median-sens-gauss}
 n = \sqrt{2} \erfc^{-1}\left[\frac 12 \erfc\left(\sqrt{\frac{T_0}2}\right) \right]
 \qquad \text{(median sensitivity)}.
\end{equation}
We show $n(T_0)$ in Fig.~\ref{fig:sens-gauss}. This curve corresponds
to a section of the lower left panel (or right panel) of
Fig.~\ref{fig:Gaussiansensitivity} at $p = 0.5$. The green and yellow
shaded bands indicate the CL at which we expect being able to reject
NO if IO is true with a probability of 68.27\% and 95.45\%,
respectively. The edges of the bands are obtained from the conditions
$\beta = 1/2 \pm 0.6827/2$ and $\beta = 1/2 \pm 0.9545/2$,
respectively. They indicate the range of obtained rejection confidence
levels which emerge from values of $T$ within $1\sigma$ and $2\sigma$
from its mean assuming true IO.

Note that if we had used the 1-sided Gaussian rule from
Eq.~\eqref{eq:sigma-alpha-1sided} to convert the probability
Eq.~\eqref{eq:median-alpha-gauss} we would have obtained $n
= \sqrt{T_0}$ for the median sensitivity. Indeed, this corresponds
exactly to the ``standard sensitivity'' as defined in
section~\ref{sec:average}.\footnote{We would have obtained the result
$n = \sqrt{T_0}$ also when using a 2-sided test to calculate $\alpha$
from the distribution of $T$ combined with the 2-sided convention to
convert it into standard deviations. Note, however, that for the
purpose of rejecting a hypothesis clearly a 1-sided test for $T$
should be used, and therefore we do not consider this possibility
further.} We show this case for illustration as dashed curve in 
Fig.~\ref{fig:sens-gauss}. The dashed vertical
lines in the right panel of Fig.~\ref{fig:Gaussiansensitivity} show explicitly that using
this convention we obtain $\beta = 0.5$ at $n\sigma$ exactly for $T_0
= n^2$. Note that with our default convention we actually obtain
an \emph{increase} in the sensitivity compared to $\sqrt{T_0}$ used in
the ``standard method''. The exponential nature of $\erfc$
implies that the difference will not be large, in particular for large
$T_0$, see \fig~\ref{fig:sens-gauss}.  For
instance, the values of $T_0$ corresponding to a median sensitivity of
$2 \sigma$, $3 \sigma$, $4 \sigma$ according to
Eq.~\eqref{eq:median-sens-gauss} are 2.86, 7.74, 14.7, respectively,
which should be compared to the standard case of $T_0 = n^2$, \ie, 4,
9, 16. In summary, we obtain the first important result of this paper:
\emph{the sensitivity obtained by using the standard method is
very close to the median sensitivity within the Gaussian
approximation.}

Before concluding this section let us also mention the sensitivity
defined by the crossing point $T_c^\NO = T_c^\IO$ discussed at the end
of section~\ref{sec:MO}. This is the sensitivity $\alpha$ for which
the critical values are the same for both orderings, which implies
that regardless of the outcome of the experiment exactly one of the
two hypotheses can be rejected at that CL. In the Gaussian
approximation this implies that $\alpha = \beta$, i.e., the
rates for errors of the first and second kinds are the
same. Using Eq.~\eqref{eq:Tca-gauss} and the analog expression for IO
we obtain by imposing $T_c^\NO = T_c^\IO$
the probability
\begin{equation}\label{eq:alpha-crossing}
 \alpha = \frac{1}{2}\erfc\left(\frac{T_0^\NO + T_0^\IO}
{\sqrt{8T_0^\NO} + \sqrt{8T_0^\IO}} \right)\approx
\frac{1}{2}\erfc\left(\frac{1}{2}\sqrt{\frac{T_0}{2}}\right) \qquad
 (T_c^\NO = T_c^\IO) \,.
\end{equation}
The corresponding sensitivity is shown as red solid curve in
Fig.~\ref{fig:sens-gauss}. For this curve we use our default
convention to convert $\alpha$ into $\sigma$ according to
Eq.~\eqref{eq:sigma-alpha}. If we instead had used the 1-sided
Gaussian convention from Eq.~\eqref{eq:sigma-alpha-1sided} to convert
the probability Eq.~\eqref{eq:alpha-crossing} we would have obtained
the simple rule $n = \sqrt{T_0}/2$ (dashed red curve). This can be
seen also in the right panel of Fig.~\ref{fig:Gaussiansensitivity}, where the red dash-dotted
curve indicates the condition $\alpha = \beta$. For a given $T_0$ the
probability $\alpha$ for $T_c^\NO = T_c^\IO$ can be read off from the
section of the corresponding blue curve with the red curve. By
considering the dashed vertical lines we observe the rule $n
= \sqrt{T_0}/2$ from the 1-sided conversion of $\alpha$ into
$n\sigma$.  For our default conversion it turns out that the
sensitivity from the condition $T_c^\NO = T_c^\IO$ is always more than
half of the median sensitivity in units of $\sigma$. From the 68.27\%
and 95.45\% bands in Fig.~\ref{fig:sens-gauss} one can see that
for a ``typical'' experimental outcome the sensitivity will be
significantly better than the one given by the crossing condition.

\subsection{Composite hypotheses}
\label{sec:gauss-comp}

Let us now generalize the discussion to the case where $T_0$ depends on
parameters. This will be typically the situation for long-baseline
experiments, where event rates depend significantly on the (unknown)
value of the CP phase $\delta$. It is straight forward to apply the
rules discussed in section~\ref{sec:hypothesis-testing} assuming that
$T = \mathcal{N}(T_0^\NO(\theta), 2\sqrt{T_0^\NO(\theta)})$ for normal ordering
and $T = \mathcal{N}(-T_0^\IO(\theta), 2\sqrt{T_0^\IO(\theta)})$ for inverted
ordering.

First we must ensure that we can reject NO for all possible values of
$\theta$ at $(1-\alpha)$ confidence. Hence, Eq.~\eqref{eq:Tca-gauss} becomes,
\begin{equation}
(T_c^\alpha)_{\rm min} = 
\min_{\theta\in\NO}\left[T_0^\NO(\theta) - \sqrt{8T_0^\NO(\theta)} \erfc^{-1}\left(2\alpha\right) \right] \,,
\end{equation}
\ie, we have to choose the smallest possible $T_c^\alpha$. Considering $T_c^\alpha$ from 
Eq.~\eqref{eq:Tca-gauss} as a function of $T_0$, we see that
$T_c^\alpha$ has a minimum at $T_0 = 2[\erfc^{-1}(2\alpha)]^2$, and the
value at the minimum is $-2[\erfc^{-1}(2\alpha)]^2$. This minimum is
also visible in Fig.~\ref{fig:Gaussiansensitivity} (upper left panel). Hence, we have
\begin{equation}\label{eq:Tca-gauss-comp}
(T_c^\alpha)_{\rm min} = \left\{
\begin{array}{l@{\quad}c@{\quad}l}
 -2[\erfc^{-1}(2\alpha)]^2 & \text{if} & \hat T_0^\NO <  2[\erfc^{-1}(2\alpha)]^2 \\
\hat T_0^\NO - \sqrt{8\hat T_0^\NO} \erfc^{-1}\left(2\alpha\right) 
& \text{if} & \hat T_0^\NO >  2[\erfc^{-1}(2\alpha)]^2 
\end{array}
\right.
\end{equation}
where $\hat T_0^\NO$ is the minimum of $T_0^\NO(\theta)$ with respect
to the parameters $\theta$.

The expression for the rate for an error of the second kind, Eq.~\eqref{eq:beta-gauss} will now depend on the true values of $\theta$ in the alternative hypothesis:
\begin{equation}\label{eq:beta-gauss-comp}
\beta(\theta) = \frac 12 \erfc\left[
\frac{T_0^\IO(\theta) + (T_c^\alpha)_{\rm min}}{\sqrt{8 T_0^\IO(\theta)}} \right] \,.
\end{equation}
The median sensitivity is obtained by setting $\beta(\theta) = 0.5$. This leads to the equation
$T_0^\IO(\theta) = - (T_c^\alpha)_{\rm min}$ which has to be solved for $\alpha$.
Note that this is a recursive definition, since which case 
in Eq.~\eqref{eq:Tca-gauss-comp} to be used
can only be decided after $\alpha$ is computed. However, it turns out 
that in situations of interest the first case applies. In this
case we have $T_0^\IO(\theta) = 2[\erfc^{-1}(2\alpha)]^2$. Typically
it also holds that $\hat T_0^\IO \approx \hat T_0^\NO$ and therefore $\hat
T_0^\NO < T_0^\IO(\theta)$ and $\hat T_0^\NO <
2[\erfc^{-1}(2\alpha)]^2$ for $\alpha$ corresponding to the median
sensitivity. Hence, we obtain the result that 
\begin{equation}\label{eq:median-gauss-comp}
\alpha(\theta) \approx
 \frac 12 \erfc\sqrt{\frac{T_0^\IO(\theta)}{2}} \qquad \text{(median sensitivity)}
\end{equation}
is a useful expression for estimating the median sensitivity for
composite hypotheses in the Gaussian approximation. We will confirm
this later on by comparing it to the full Monte Carlo simulations of
long-baseline experiments. Also note the similarity with the
expression in case of simple hypotheses (see Eq.~\eqref{eq:median-alpha-gauss}).

Finally we can also calculate the ``crossing sensitivity'' by
requiring $(T_c^\alpha)_{\rm min}^\NO = (T_c^\alpha)_{\rm min}^\IO$,
for which exactly one hypothesis can be rejected. Again this is a
recursive definition, however, if $\hat T_0^\NO \simeq \hat T_0^\IO$
it turns out that only the second case 
in Eq.~\eqref{eq:Tca-gauss-comp} is relevant. This leads to
\begin{equation}\label{eq:crossing-composite}
  \alpha = \frac 12 \erfc\left(\frac{1}{\sqrt{8}} 
  \frac{\hat T_0^\NO + \hat T_0^\IO}{\sqrt{\hat T_0^\NO} + \sqrt{\hat T_0^\IO}}
\right) \approx
\frac 12 \erfc\left(\frac 12 \sqrt{\frac{\hat T_0}{2}} \right) 
\qquad (T_c^\NO = T_c^\IO) \,,
\end{equation}
where the last relation holds for $\hat T_0 \equiv \hat
T_0^\NO \approx \hat T_0^\IO$, which again is very similar to the case
for simple hypotheses, Eq.~\eqref{eq:alpha-crossing}.

\section{Monte Carlo simulations of experimental setups}
\label{sec:numericalresults}

Let us now apply the methods presented above to realistic experimental
configurations.  We have performed Monte Carlo (MC) studies to determine
the sensitivity to the neutrino mass ordering for three different
types of experiments, each of which obtains their sensitivity through
the observation of different phenomena: ($a$) JUNO~\cite{JUNO}:
interference (in the vacuum regime) between the solar and atmospheric
oscillation amplitudes at a medium baseline reactor neutrino
oscillation experiment; ($b$) PINGU~\cite{Aartsen:2013aaa} and
INO~\cite{INO}: matter effects in atmospheric neutrino oscillations;
($c$) \NOvA~\cite{Ayres:2004js} and LBNE~\cite{Adams:2013qkq}: matter
effects in a long baseline neutrino beam experiment. In each case we
have followed closely the information given in the respective
proposals or design reports, and we adopted bench mark setups which
under same assumptions reproduce standard sensitivities in the
literature reasonably well. The specific details that have been used
to simulate each experiment are summarized in
App.~\ref{app:simulation}.

\subsection{Medium-baseline reactor experiment: JUNO}
\label{sec:juno}

For the simulations in this paper
we adopt an experimental configuration for the JUNO reactor experiment based
on Refs.~\cite{JUNO, DB2-WWang, Li:2013zyd}, following the analysis
described in Ref.~\cite{Blennow:2013vta}. A 20~\kt\ liquid scintillator
detector is considered at a distance of approximately 52~km from 10
reactors with a total power of 36~GW, with an exposure of 6
years, \ie, $4320 \rm\, \kt\, GW \, yr$. The energy resolution is
assumed to be $3\%\sqrt{1\,{\rm MeV}/E}$. For further details see
App.~\ref{app:juno}.

The unique feature of this setup is that the sensitivity to the mass
ordering is rather insensitive to the true values of the oscillation
parameters within their uncertainties. Being a $\bar\nu_e$
disappearance experiment, the survival probability depends neither on
$\theta_{23}$ nor on the CP phase $\delta$, and all the other
oscillation parameters are known (or will be known at the time of the
data analysis of the experiment) with sufficient precision such that
the mass ordering sensitivity is barely affected. Therefore we are
effectively very close to the situation of simple hypotheses for this
setup. Note that although the mass ordering sensitivity is insensitive
to the true values, the $\chi^2$ minimization with respect to
oscillation parameters, especially $|\Delta m^2_{31}|$, is crucial
when calculating the value of the test statistic $T$.

In the left panel of Fig.~\ref{fig:Tc} we show the distribution of the test statistic
$T$ from a Monte Carlo simulation of $10^5$ data sets for our default
JUNO configuration. For each true mass ordering we compare those
results to the normal distributions expected under the Gaussian
approximation, namely $\mathcal{N}(T_0^\NO, 2\sqrt{T_0^\NO})$ for normal
ordering and $\mathcal{N}(-T_0^\IO, 2\sqrt{T_0^\IO})$ for inverted ordering,
where $T_0^\NO$ and $T_0^\IO$ are the values of the test statistic
without statistical fluctuation (Asimov data set). For the considered
setup we find $T_0^\NO = 10.1$ and $T_0^\IO = 11.1$, and we observe
excellent agreement of the Gaussian approximation with the Monte Carlo
simulation, see also, \eg, Ref.~\cite{Qian:2012xh}.

\begin{table}
\begin{tabular}{lc@{\quad}c@{\quad}c@{\quad}c}
\hline\hline
energy resolution & \multicolumn{2}{c}{$3\%\sqrt{1\,{\rm MeV}/E}$} 
                  & \multicolumn{2}{c}{$3.5\%\sqrt{1\,{\rm MeV}/E}$} \\
\hline 
                  & normal & inverted       & normal & inverted  \\
\hline
$T_0 \, (\sqrt{T_0}\sigma)$ &   10.1 ($3.2\sigma$) & 11.1 ($3.3\sigma$) 
                            &   5.4 ($2.3\sigma$)  & 5.9 ($2.4\sigma$)\\
median sens.  &   $7.3\times 10^{-4} \,(3.4\sigma)$ & $4.3\times 10^{-4} \,(3.5\sigma)$ 
              &   $1.0\times 10^{-2} \,(2.5\sigma)$ & $7.5\times 10^{-3} \,(2.7\sigma)$ \\ 
crossing sens.&  \multicolumn{2}{c}{$5.2\% \,(1.9\sigma)$}
              &  \multicolumn{2}{c}{$12\% \,(1.6\sigma)$}\\
\hline\hline 
\end{tabular}
  \caption{Sensitivity of the JUNO reactor experiment for $4320 \rm\,
  \kt\, GW \, yr$ exposure for two different assumptions on the energy
  resolution. We give the value of the test statistic without
  statistical fluctuation, $T_0$, and the ``standard sensitivity''
  $\sqrt{T_0}\sigma$. The median sensitivity is calculated according
  to Eq.~\eqref{eq:median-alpha-gauss}. The ``crossing sensitivity''
  corresponds to the CL where exactly one mass ordering can be
  rejected regardless of the outcome, which is calculated according to
  Eq.~\eqref{eq:alpha-crossing}.
\label{tab:juno}}
\end{table}

Hence we can apply the formalism developed in
section~\ref{sec:gauss} directly to evaluate the sensitivity of the experiment
in terms of $T_0^\NO$ and $T_0^\IO$. For instance,
Eq.~\eqref{eq:median-alpha-gauss} gives for the median sensitivity
$\alpha = 7.3\,(4.3) \times 10^{-4}$ for testing normal (inverted)
ordering, which corresponds to 3.4$\sigma$ (3.5$\sigma$). As discussed
in section~\ref{sec:gauss} those numbers are rather close to the
``standard sensitivity'' based on $n=\sqrt{T_0}$, which would give
3.2$\sigma$ (3.3$\sigma$). For the given values of $T_0^\NO$ and $T_0^\IO$
we can now use Fig.~\ref{fig:Gaussiansensitivity} to obtain the probability to reject an 
ordering if it is false (\ie, the power of the test) for any desired
confidence level $(1-\alpha)$. The confidence level at which exactly
one mass ordering can be rejected (crossing point $T_c^\NO = T_c^\IO$)
is obtained from Eq.~\eqref{eq:alpha-crossing} as $\alpha = 5.2\%$ or
1.9$\sigma$, see also Fig.~\ref{fig:Tc}. Those numbers are summarized in
Tab.~\ref{tab:juno}. There we give also the corresponding results for
the same setup but with a slightly worse energy resolution of
$3.5\%\sqrt{1\,{\rm MeV}/E}$, in which case significantly reduced
sensitivities are obtained, highlighting once more the importance to
achieve excellent energy reconstruction abilities. We have checked
that also in this case the distribution of $T$ is very close to the
Gaussian approximation.

\subsection{Atmospheric neutrinos: PINGU and INO}
\label{sec:atm}

We now move to atmospheric neutrino experiments, which try to
determine the mass ordering by looking for the imprint of the matter
resonance in the angular and energy distribution of neutrino induced
muons. The resonance will occur for neutrinos (antineutrinos) in the
case of normal (inverted) ordering. The INO experiment~\cite{INO} uses
a magnetized iron calorimeter which is able to separate neutrino and
antineutrino induced events with high efficiency, which provides
sensitivity to the mass ordering with an exposure of around 500~\mbox{\kt\,yr}
(10 year operation of a 50~\kt\ detector). Alternatively, the
PINGU~\cite{Aartsen:2013aaa} experiment, being a low-energy extension
of the IceCube detector, is not able to separate neutrino and
antineutrino induced muons on an event-by-event basis. This leads to a
dilution of the effect of changing the mass ordering, which has to be
compensated by exposures exceeding 10~\mbox{Mt\,yr}, which can be achieved
for a few years of running time. In both cases the ability to reconstruct
neutrino energy and direction will be crucial to determining the mass ordering.

\begin{table}
\begin{tabular}{l@{\quad}c@{\quad}c@{\quad}c@{\quad}c@{\quad}c}
\hline\hline
& $\sigma_{E_\nu}$ & $\sigma_{\theta_\nu}$ & exposure & $T_0^\NO$ (med.\ sens.) & $T_0^\IO$ (med.\ sens.)\\
\hline
INO & $0.1 E_\nu$ & $10^\circ$ & 10 yr $\times$ 50 \kt & 5.5 ($2.6\sigma$) & 5.4 ($2.6\sigma$)\\
PINGU & $0.2 E_\nu$ & $29^\circ / \sqrt{E_\nu / \rm GeV}$ & 5 yr & 12.5 ($3.7\sigma$) & 12.0 ($3.6\sigma$)\\
\hline\hline
\end{tabular}
  \caption{Main characteristics of our default setups for INO and
  PINGU. We give energy resolutions for neutrino energy and direction
  reconstruction and default exposure. For PINGU we assume an energy
  dependent effective detector mass. The last two columns give the
  value of $T_0$ and the median sensitivity using
  Eq.~\eqref{eq:median-sens-gauss} for the two orderings, assuming
  $\theta_{23} = 45^\circ$. \label{tab:atm}}
\end{table}

Our simulations for the INO and PINGU experiments are based on
Refs.~\cite{Blennow:2012gj} and \cite{Blennow:2013vta},
respectively. We summarize the main characteristics of our default
setups in Tab.~\ref{tab:atm}, further technical details and references
are given in App.~\ref{app:atm}. Let us stress that the
sensitivity of this type of experiments crucially depends on
experimental parameters such a systematic uncertainties, efficiencies,
particle identification, and especially the ability to reconstruct
neutrino energy and direction. Those parameters are still not settled,
in particular for the PINGU experiment, and final sensitivities may
vary by few sigmas, see for instance Refs.~\cite{Aartsen:2013aaa, Winter:2013ema}. Our
setups should serve as representative examples in order to study the
statistical properties of the resulting sensitivities. While the final
numerical answer will depend strongly on to be defined experimental
parameters, we do not expect that the statistical behavior will be
affected significantly.

\begin{figure}
\begin{center}
\includegraphics[width=0.4\textwidth]{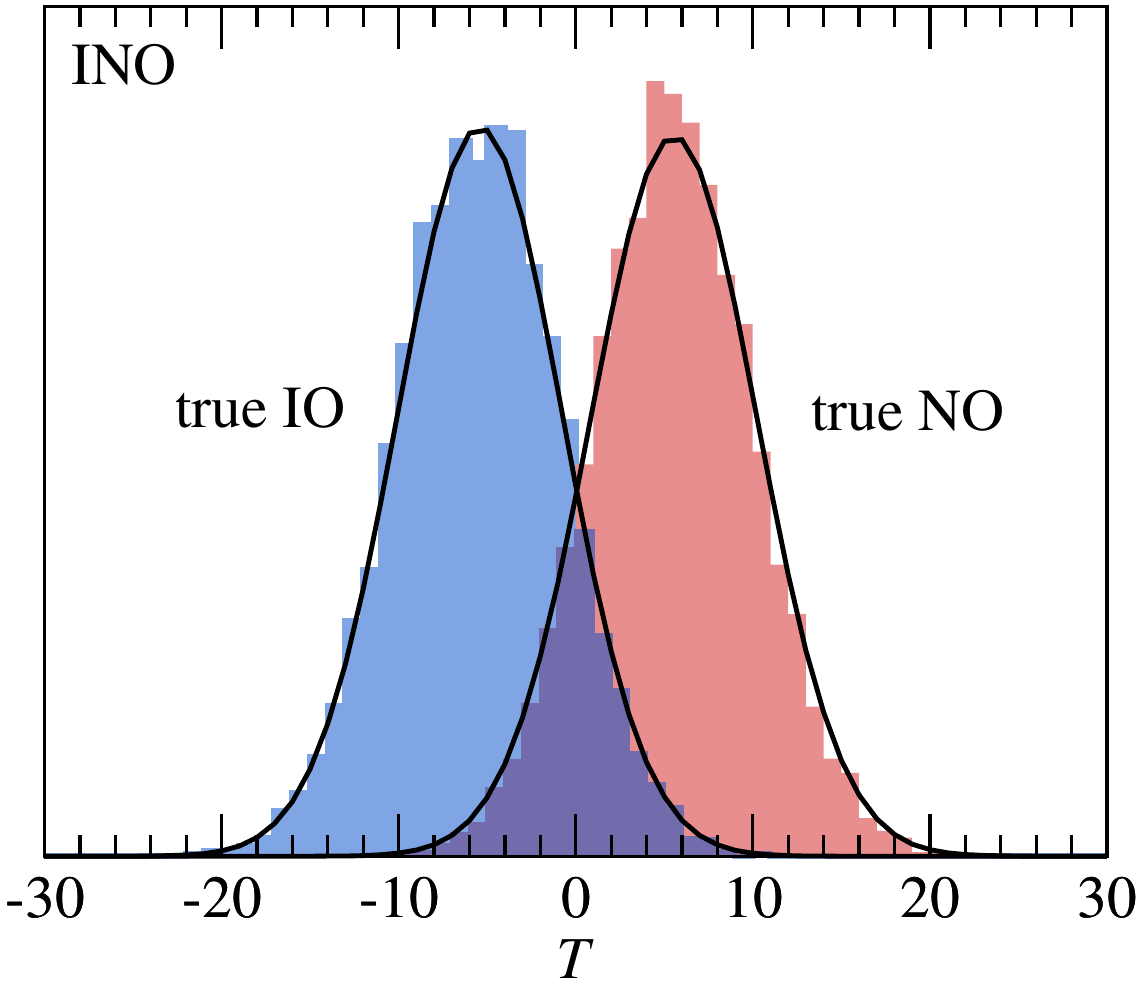} 
\caption{Simulated distributions of the test statistic $T$ in the INO experiment. We use our default setup as defined in Tab.~\ref{tab:atm} and assume $\theta_{23} = 45^\circ$. Solid curves show the Gaussian approximation from Eq.~\eqref{eq:Tgauss1}.}
\label{fig:Tdist-ino}
\end{center}
\end{figure}

\begin{figure}
\begin{center}
\includegraphics[width=0.31\textwidth]{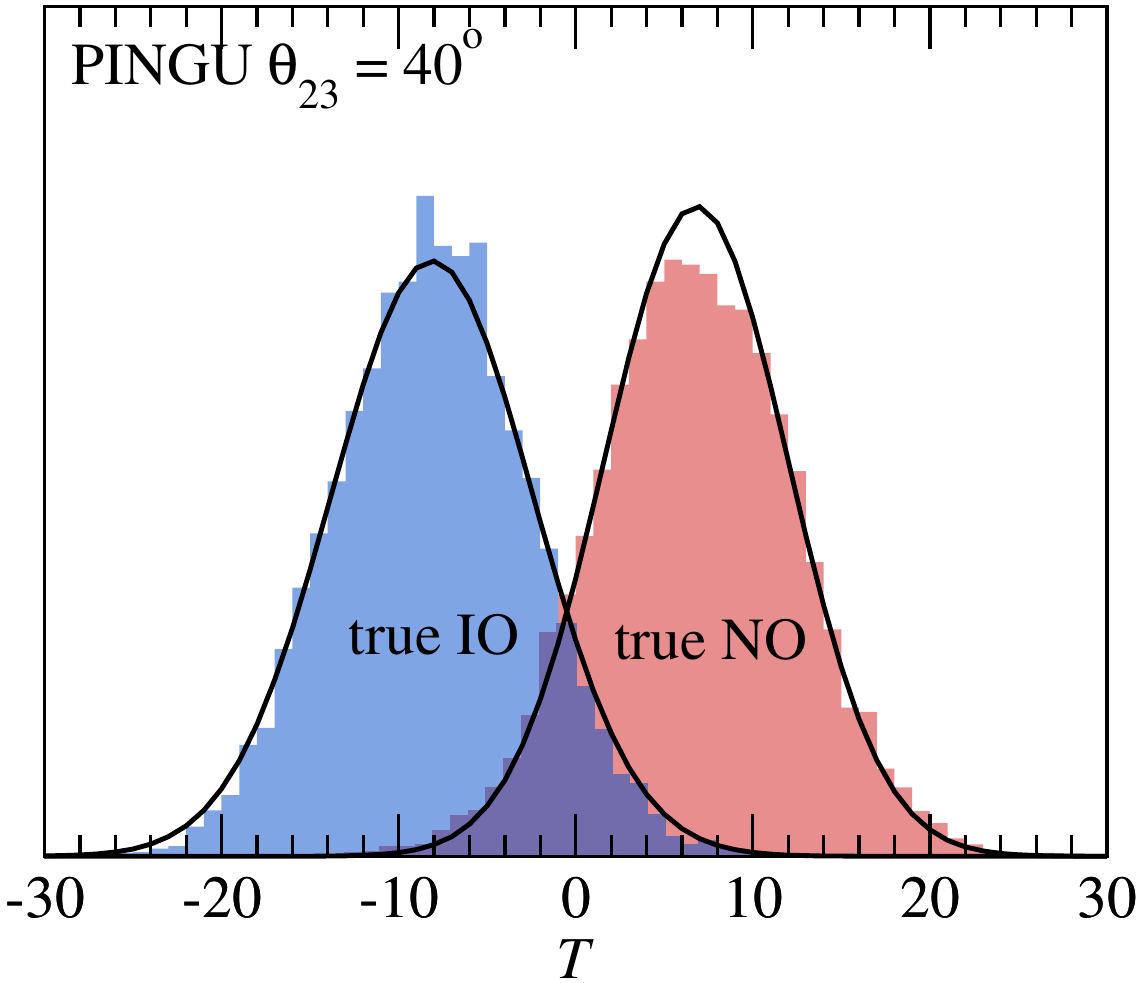} 
\includegraphics[width=0.31\textwidth]{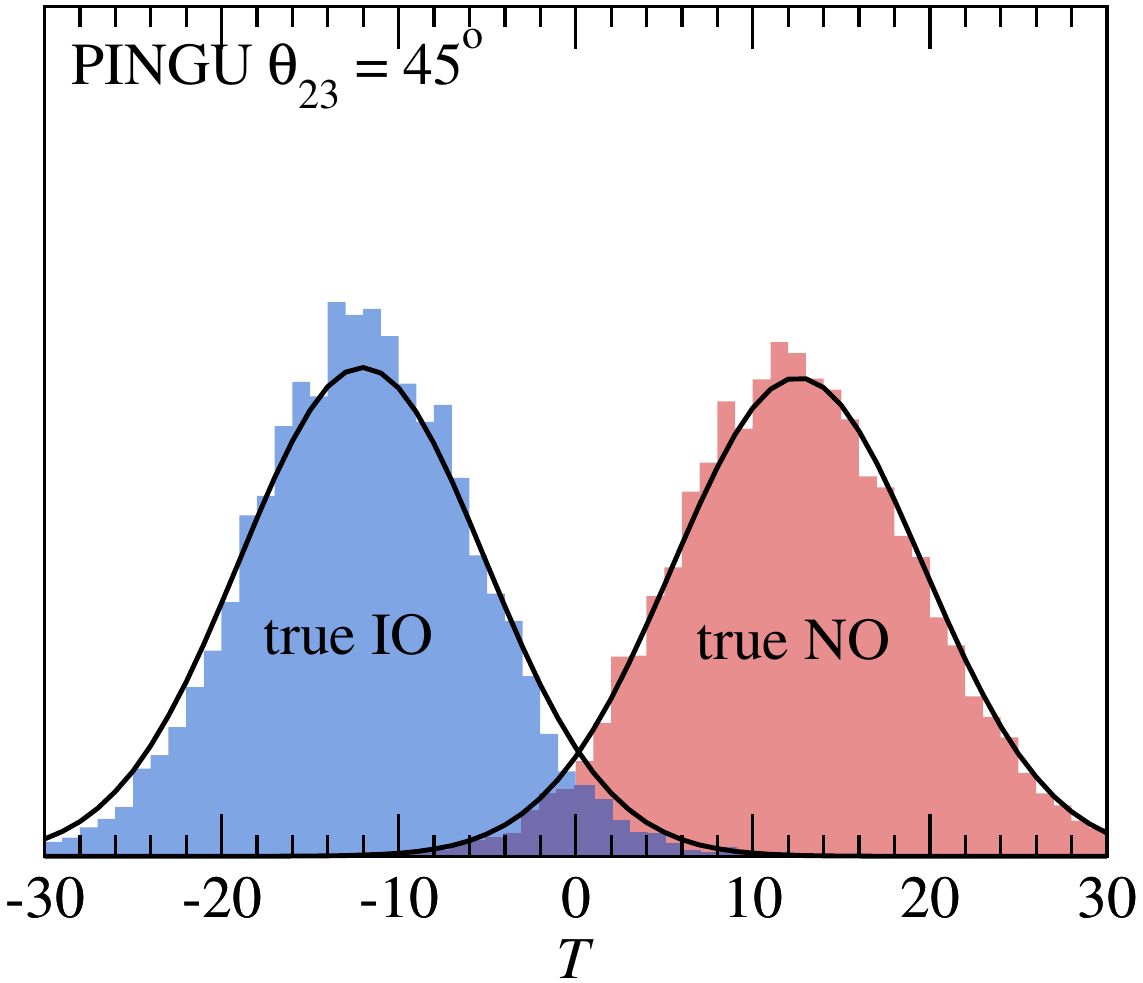} 
\includegraphics[width=0.31\textwidth]{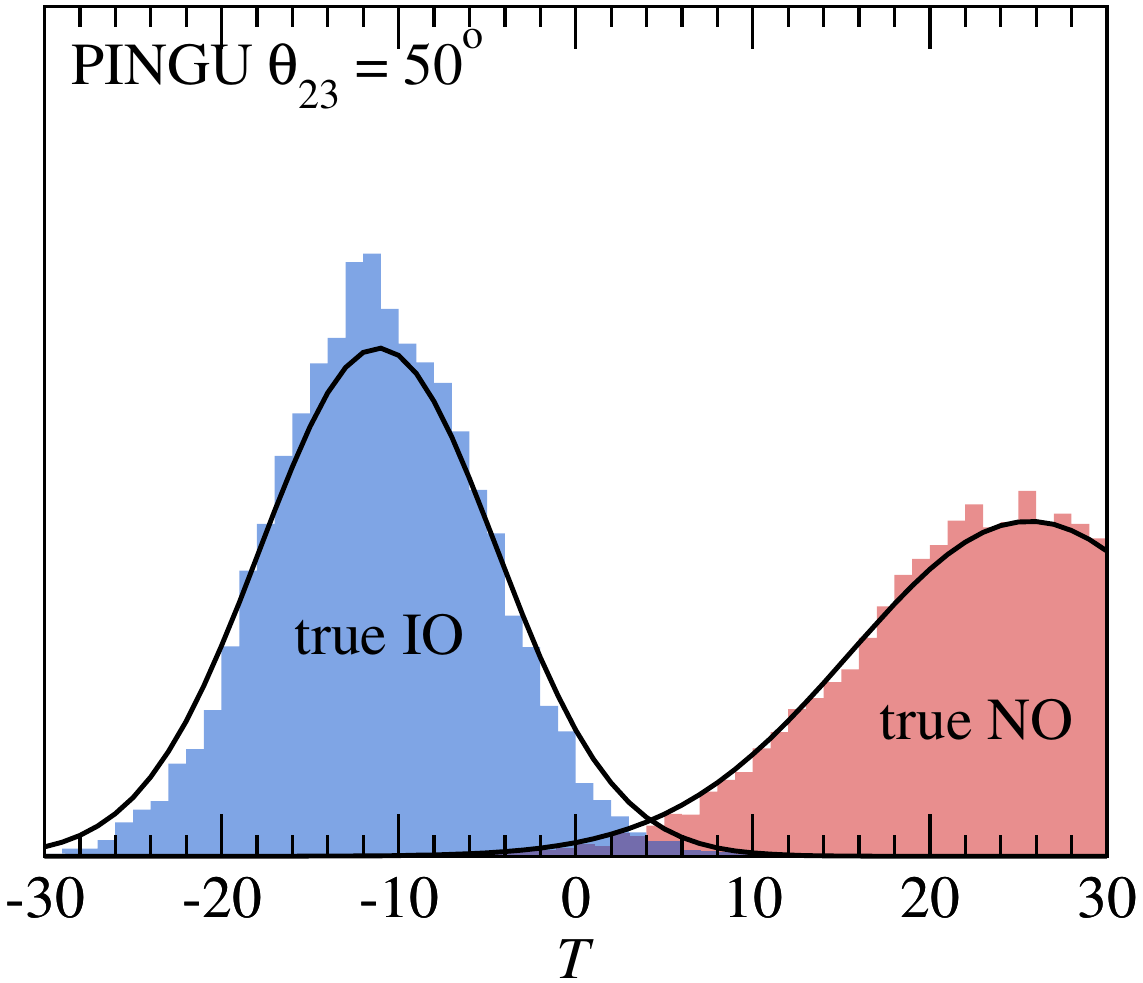} 
\caption{Simulated distributions of the test statistic $T$ in the
  PINGU experiment with $\theta_{23} = 40^\circ, \, 45^\circ, \,
  50^\circ$ for the left, middle, right panel, respectively.  We use
  our default setup as defined in Tab.~\ref{tab:atm}. Solid curves
  show the Gaussian approximation from Eq.~\eqref{eq:Tgauss1}.}
\label{fig:Tdist-pingu}
\end{center}
\end{figure}

In Figs.~\ref{fig:Tdist-ino} and \ref{fig:Tdist-pingu} we show the
distributions of the test statistic $T$ for the INO and PINGU
experiments, respectively, obtained from a sample of $10^4$ simulated
data sets for each mass ordering, using the default setups from
Tab.~\ref{tab:atm}.  We observe good
agreement with the Gaussian approximation (see also
Ref.~\cite{Franco:2013in} for a simulation in the context of
PINGU). Those results justify the use of the simple expressions from
section~\ref{sec:gauss} also for INO and PINGU in order to calculate
median sensitivities or rates for errors of the first and
second kind.

\begin{figure}
\begin{center}
\includegraphics[width=0.4\textwidth]{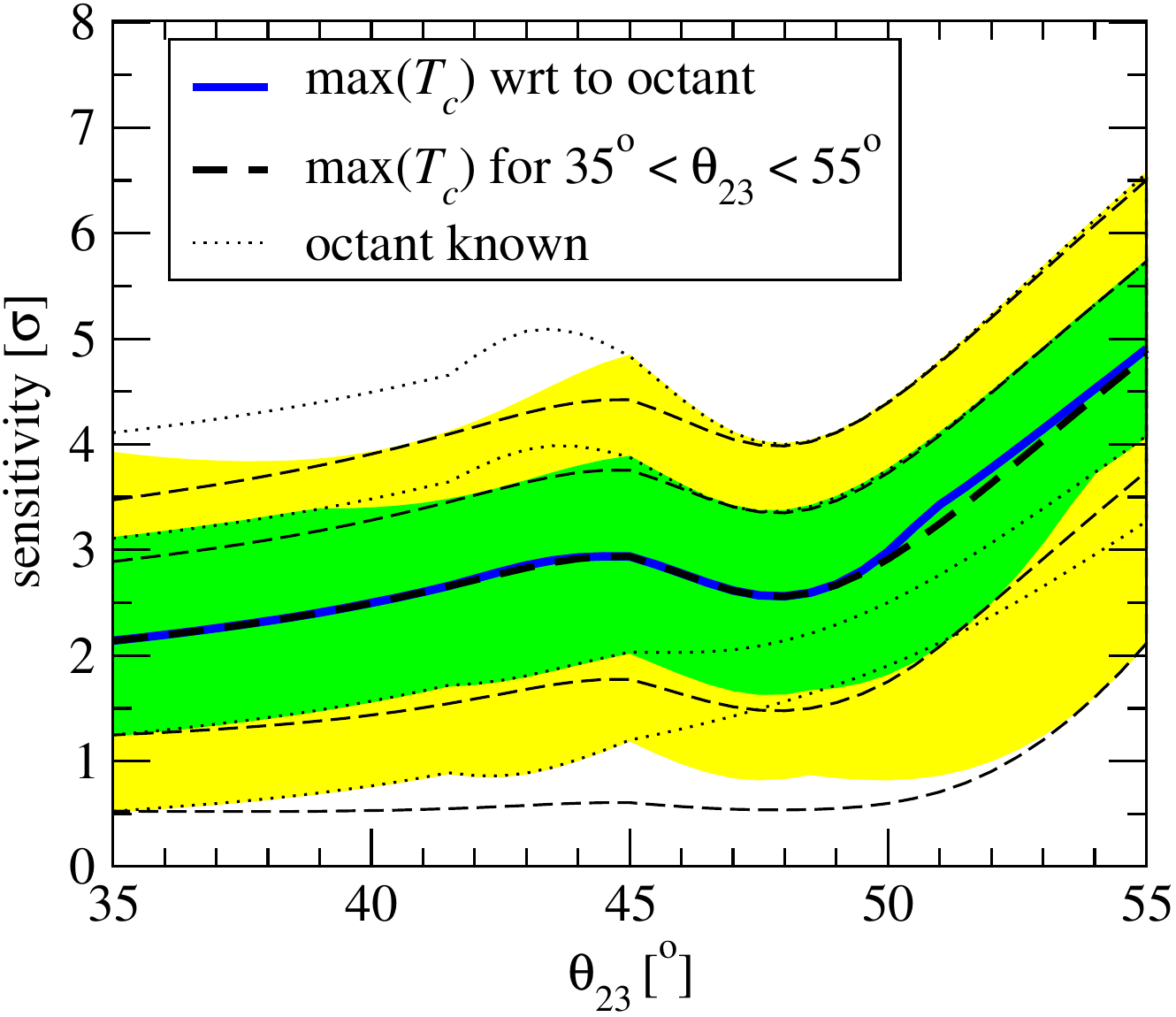} \qquad 
\includegraphics[width=0.4\textwidth]{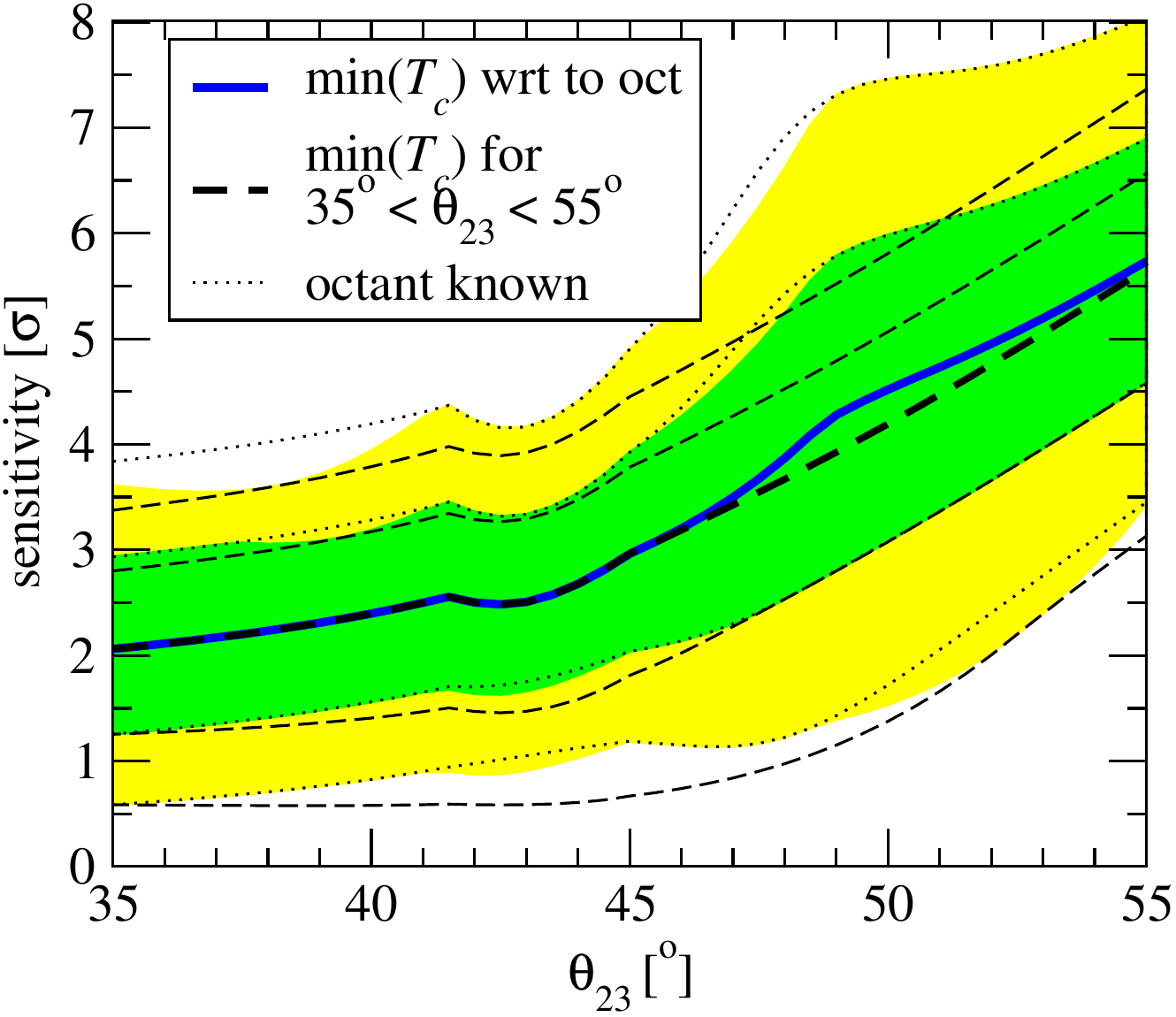} 
  \caption{Median sensitivity for PINGU after 3~years data taking as a
  function of the true value of $\theta_{23}$. Left (right) panel
  shows a test for NO (IO), which means that the true ordering is
  inverted (normal). For the thick black dashed curve we consider the range
  $35^\circ < \theta_{23} < 55^\circ$ for the true value of
  $\theta_{23}$ when calculating the critical value for the test
  statistic ($T_c^\alpha$), and the thin dashed curves indicate the
  corresponding 68.27\% and 95.45\% probability ranges of obtained
  rejection significances. For the blue solid curve and the
  corresponding green (68.27\%) and yellow (95.45\%) probability bands
  we assume that $\theta_{23}$ is known up to its octant when
  calculating $T_c^\alpha$. The dotted curves show the 68.27\% and 95.45\% probability ranges
  assuming that $\theta_{23}$ including its octant is known (simple hypothesis test).}
\label{fig:sens_pingu}
\end{center}
\end{figure}

In Fig.~\ref{fig:Tdist-pingu} we illustrate the dependence of the
distributions for PINGU on the true value of $\theta_{23}$. From this
figure it is clear that the true value of $\theta_{23}$ plays an
important role for the sensitivity to the mass ordering, with better
sensitivity for large values of $\theta_{23}$ (a similar dependence is
holds also for INO, see, \eg, Refs.~\cite{Petcov:2005rv, Blennow:2012gj}).
The dependence on other parameters is rather weak (taking into account
that, at the time of the experiment, $\theta_{13}$ will be known even
better than today).  Let us discuss the $\theta_{23}$ dependence in
more detail for the case of PINGU, where from now on we use the
Gaussian approximation.  The problem arises when calculating the
critical value for the test statistic $T$ in order to reject the
null-hypothesis at a given CL. If we follow our rule for composite
hypothesis, Eq.~\eqref{eq:Tc-composite}, and minimize (for NO) or
maximize (for IO) $T_c^\alpha(\theta_{23})$ over $\theta_{23}$ in the
range $35^\circ$ to $55^\circ$ we obtain the black dashed curves in
Fig.~\ref{fig:sens_pingu}. This is equivalent to using
Eq.~\eqref{eq:median-gauss-comp}. The chosen range for $\theta_{23}$
corresponds roughly to the $3\sigma$ range obtained from current data
\cite{GonzalezGarcia:2012sz}. However, this may be too conservative,
since at the time of the experiment T2K and \NOvA\ will provide a very
accurate determination of $\sin^22\theta_{23}$. Hence, $\theta_{23}$
will be known with good precision up to its octant, see for instance
Fig.~5 of Ref.~\cite{Huber:2009cw}.  If we minimize (maximize)
$T_c^\alpha(\theta_{23})$ only over the two discrete values
$\theta_{23}^{\rm true}$ and $90^\circ - \theta_{23}^{\rm true}$ we
obtain the blue solid curves in Fig.~\ref{fig:sens_pingu}. The green
and yellow bands indicate the corresponding 68.27\% and 95.45\%
probability ranges of expected rejection significances. The dotted
curves show the corresponding information but using only the true
value of $\theta_{23}$ when calculating $T_c^\alpha$. This last case
corresponds to the ideal situation of perfectly knowing $\theta_{23}$
(including its octant), in which case NO and IO become simple
hypotheses.  The median sensitivity for known $\theta_{23}$ is not
shown in the figure for clarity, but it is very similar to the blue
solid curves.

We obtain the pleasant result that all three methods give very similar
values for the median sensitivity, ranging from $2\sigma$ at
$\theta_{23} \simeq 35^\circ$ up to $5\sigma$ ($6\sigma$) rejection of
NO (IO) at $\theta_{23} \simeq 55^\circ$. Only for the NO test and
$\theta_{23} \simeq 50^\circ$ we find that taking the
octant degeneracy into account leads to a larger spread of the 68.27\% and 95.45\%
probability ranges for the sensitivity, implying a higher risk of
obtaining a rather weak rejection. Actually, this region of parameter
space (true IO and $\theta_{23} > 45^\circ$) is the only one where the
octant degeneracy severely affects the sensitivity to the mass
ordering \cite{Winter:2013ema}. Let us emphasize that the octant
degeneracy is always fully taken into account when minimizing the
$\chi^2$. Here we are instead concerned with the dependence of the
critical value $T_c^\alpha$ on $\theta_{23}$.

\subsection{Long-baseline appearance experiments: \NOvAt\ and LBNE}

Long-baseline neutrino beam experiments try to identify the neutrino
mass ordering by exploring the matter effect in the
$\nu_\mu \to \nu_e$ appearance channel. Whether the resonance occurs for neutrinos or for
antineutrinos will determine the mass ordering. A crucial feature in
this case is that the appearance probability, and therefore also the
event rates, depend significantly on the unknown value of the CP phase
$\delta$. Most likely $\delta$ will remain unknown even at the time
the mass ordering measurement will be performed, and therefore taking
the $\delta$ dependence into account is essential. In the nomenclature of
sections~\ref{sec:statanalysis} and \ref{sec:gauss} we are dealing
with composite hypothesis testing. In this work we consider three
representative experimental configurations to study the statistical
properties of the mass ordering sensitivity, namely
\NOvA~\cite{Ayres:2004js}, \LBNE{10}, and
\LBNE{34}~\cite{Adams:2013qkq}, which provide increasing sensitivity
to the mass ordering. Tab~\ref{tab:LBLsetups} summarizes their main features, 
while further details are given in App.~\ref{app:lbl}.

\begin{table}

\begin{tabular}{l@{\quad}c@{\quad}c@{\quad}c@{\quad}c@{\quad}c@{\quad}c}
\hline\hline
   & L (km) & Off-axis angle & $\nu$ flux peak & Detector & M(\kt) & Years $(\nu,\bar\nu)$  \\
   \hline
 \NOvA & 810 & 14 mrad & 2 GeV  & TASD & 13~\kt & (3,3)    \\
 \LBNE{10(34)} & 1290 & -- & 2.5 GeV & LAr  & 10(34)~\kt & (5,5)     \\ \hline\hline
\end{tabular}
\caption{Main characteristics of the long baseline setups considered in this work. 
In both cases the beam power is 700~kW. The \NOvA\ detector is a Totally Active Scintillator Detector (TASD), while for LBNE a Liquid Argon (LAr) detector is considered.  
\label{tab:LBLsetups}}
\end{table}

\begin{figure}
\begin{center}
\includegraphics[width=0.4\textwidth]{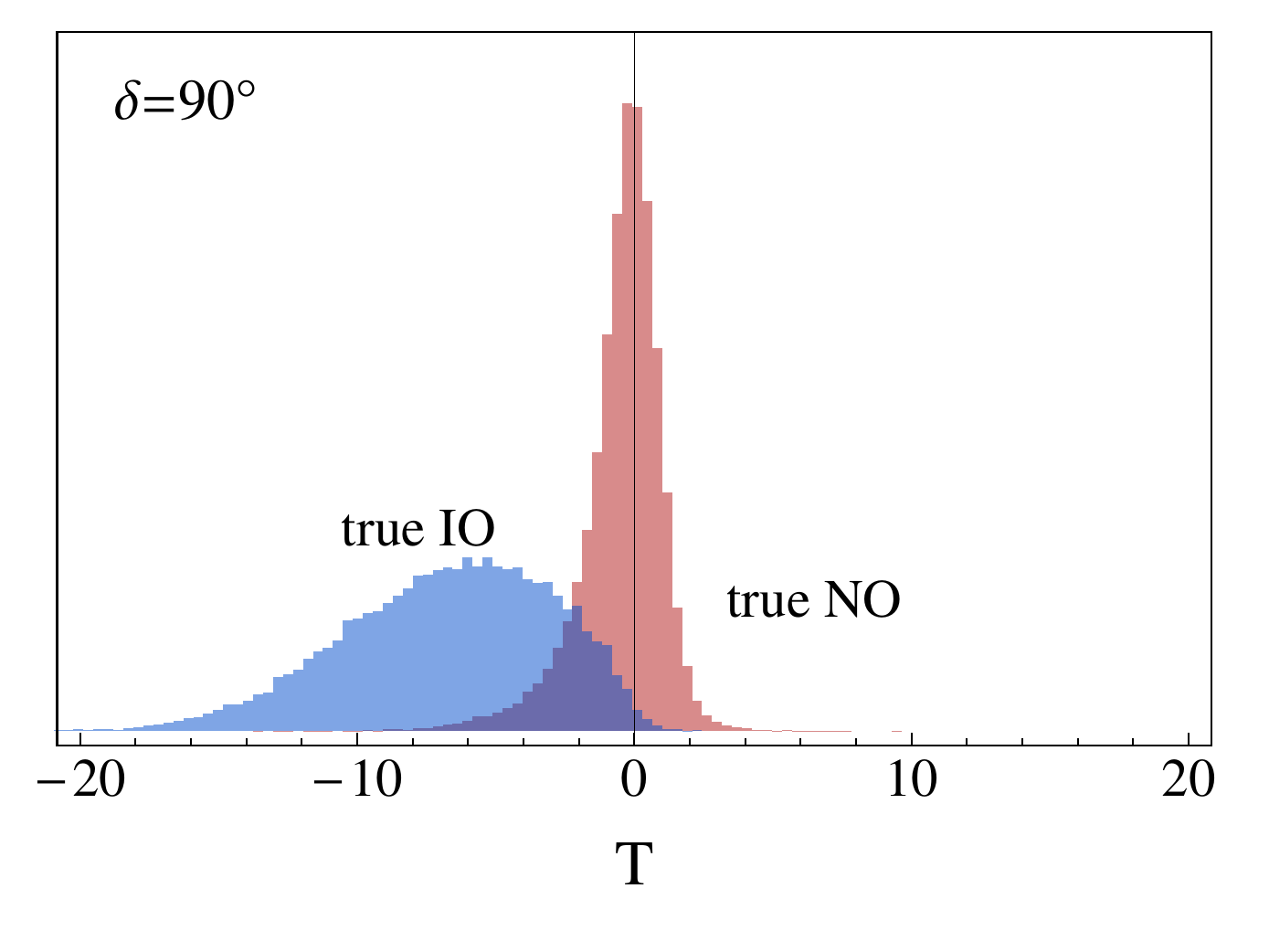} 
\includegraphics[width=0.4\textwidth]{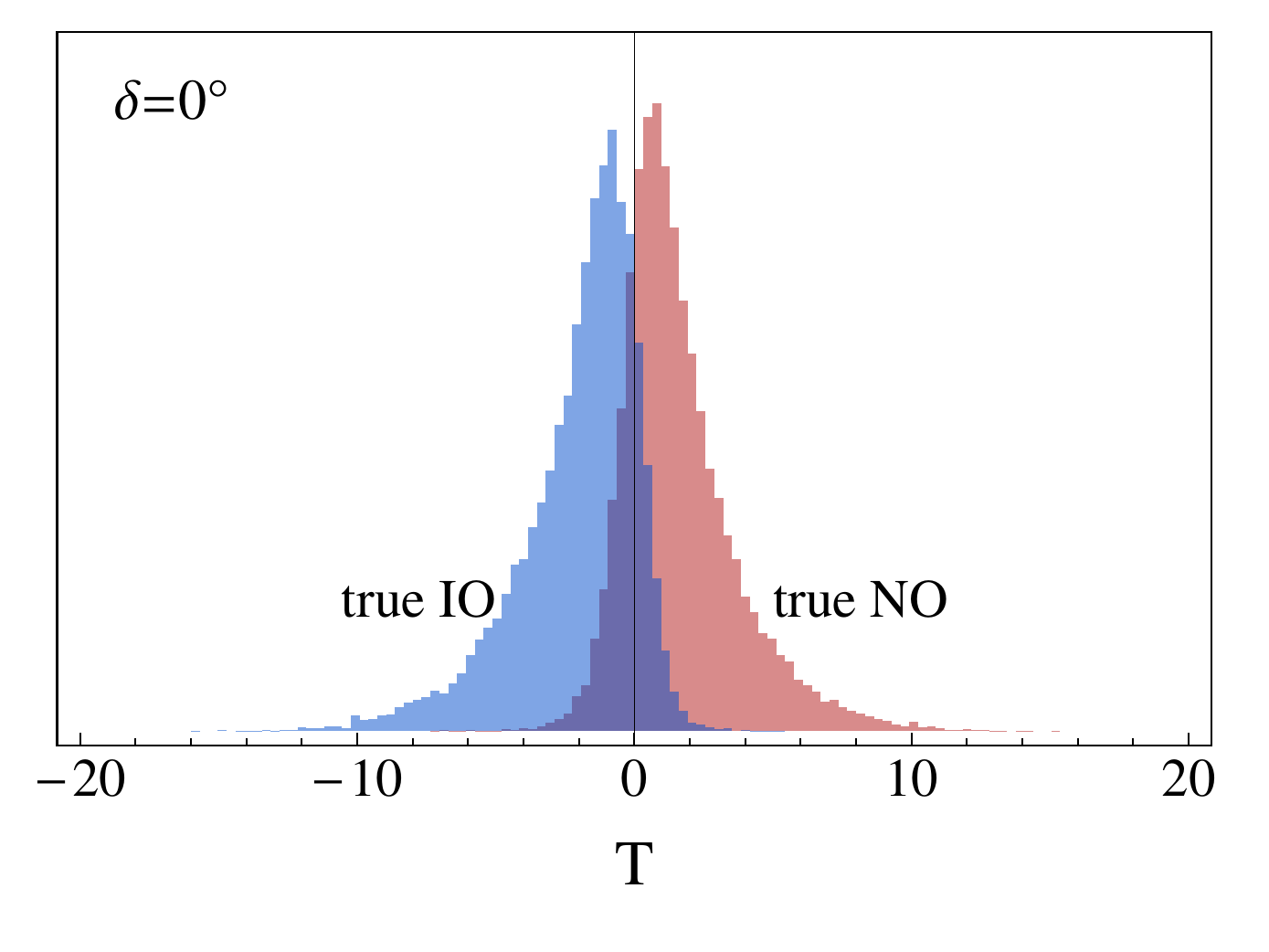} \\
\includegraphics[width=0.4\textwidth]{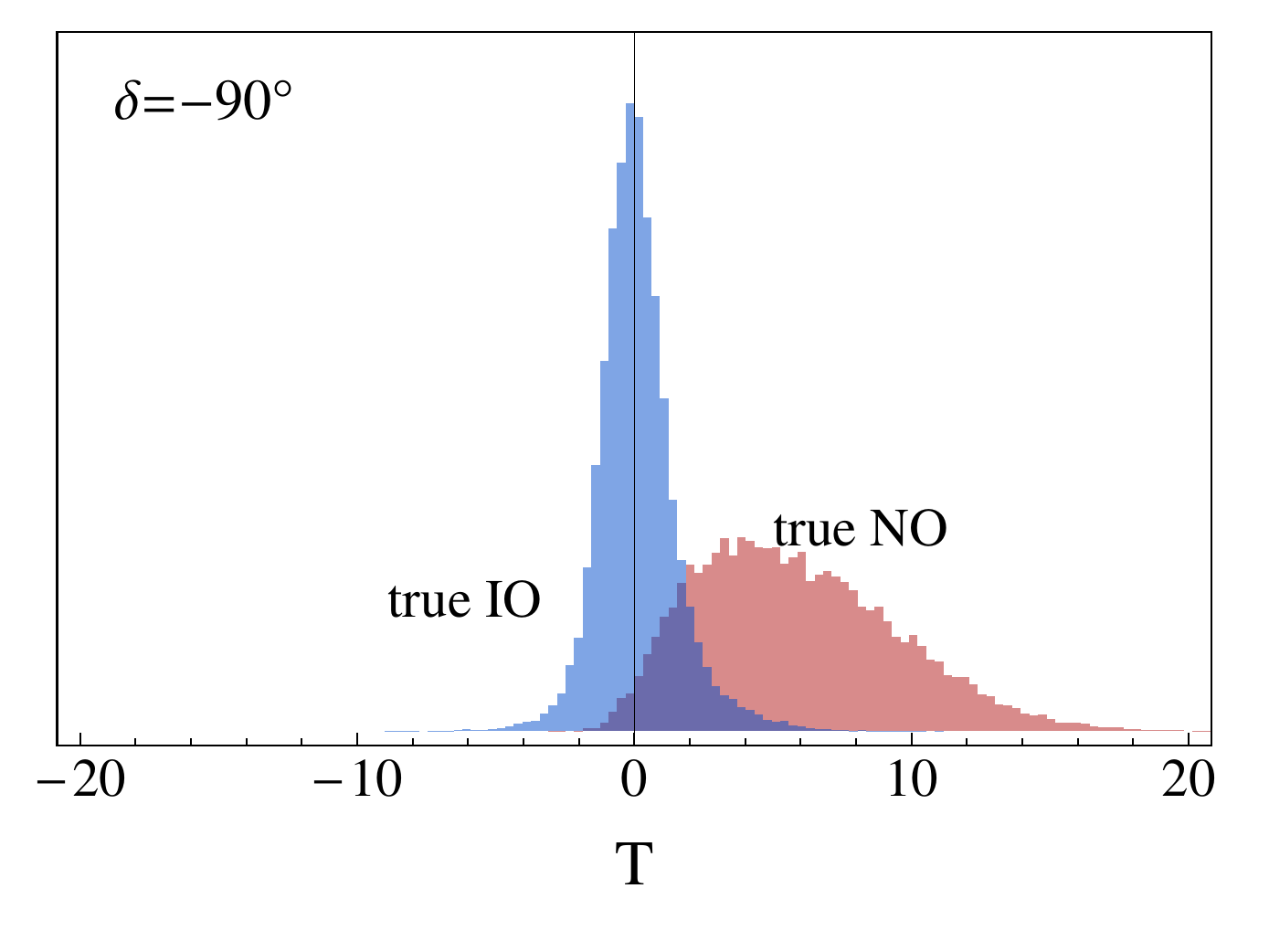}
\includegraphics[width=0.4\textwidth]{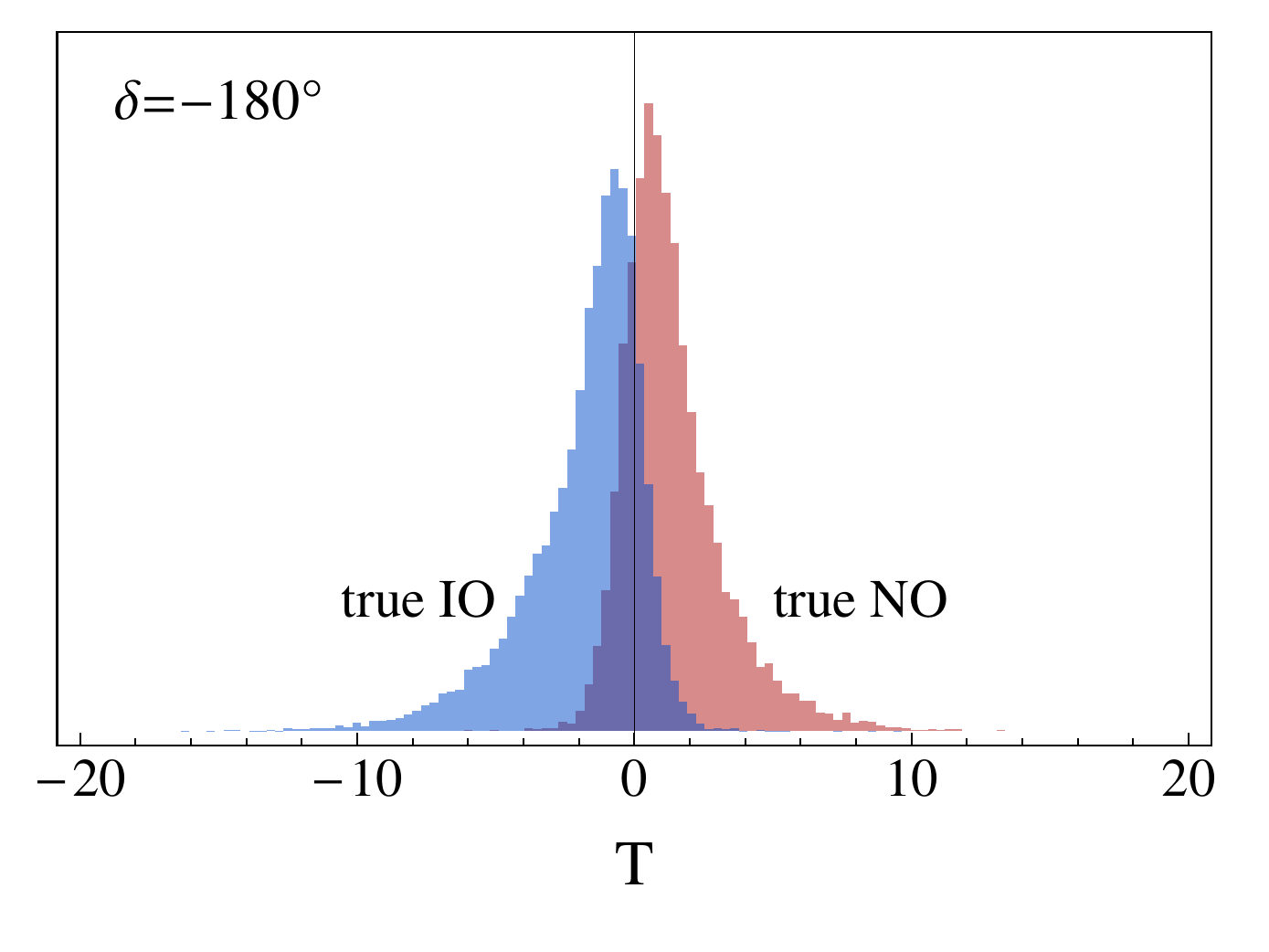}
\caption{The simulated distributions of the test statistic $T$ in the \NOvA\ experiment for different true values of $\delta$, as indicated by the labels. The red (blue) distributions assume a true normal (inverted) ordering.}
\label{fig:novadistributions}
\end{center}
\end{figure}
\begin{figure}
\begin{center}
\includegraphics[width=0.4\textwidth]{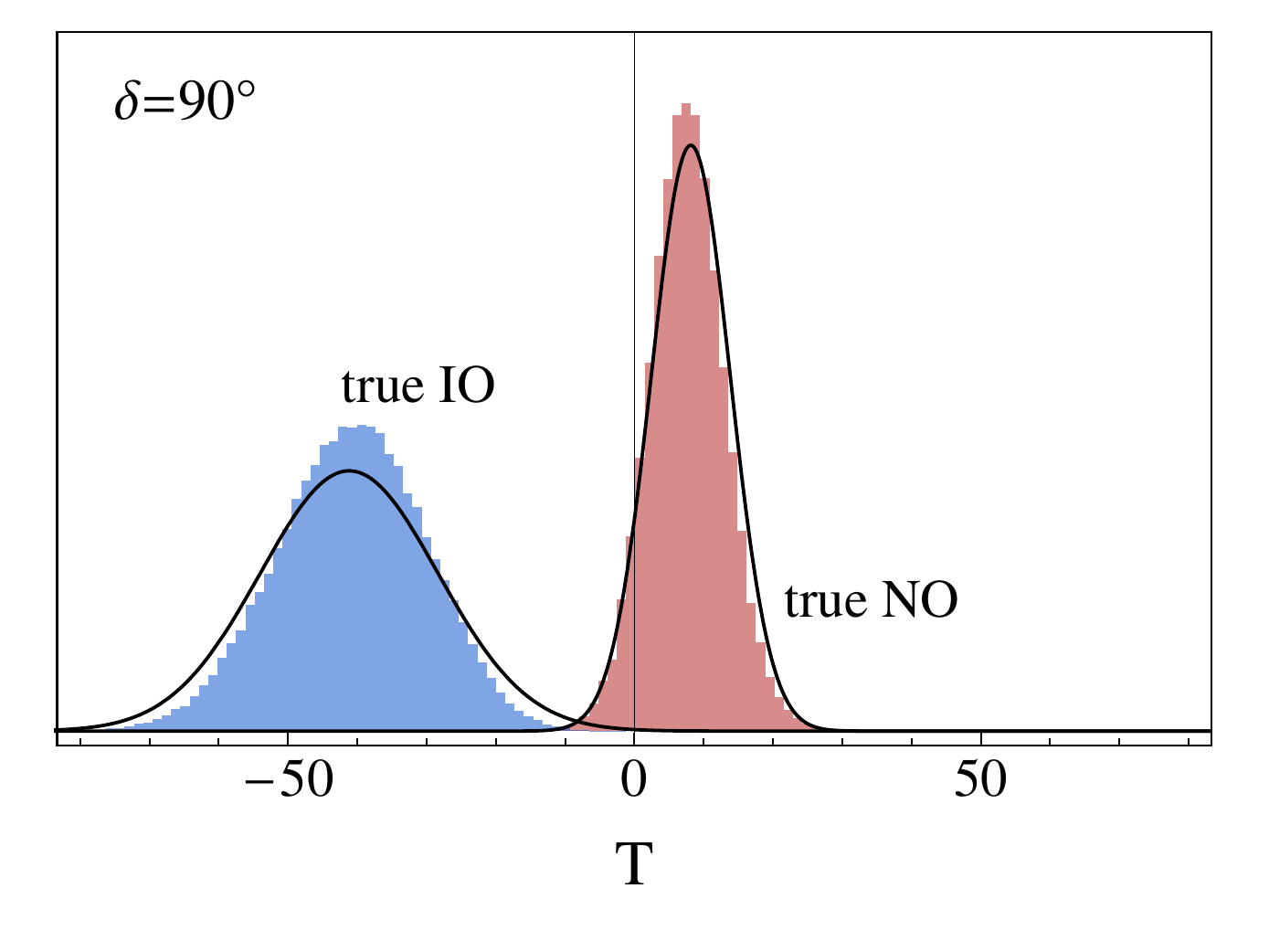}
\includegraphics[width=0.4\textwidth]{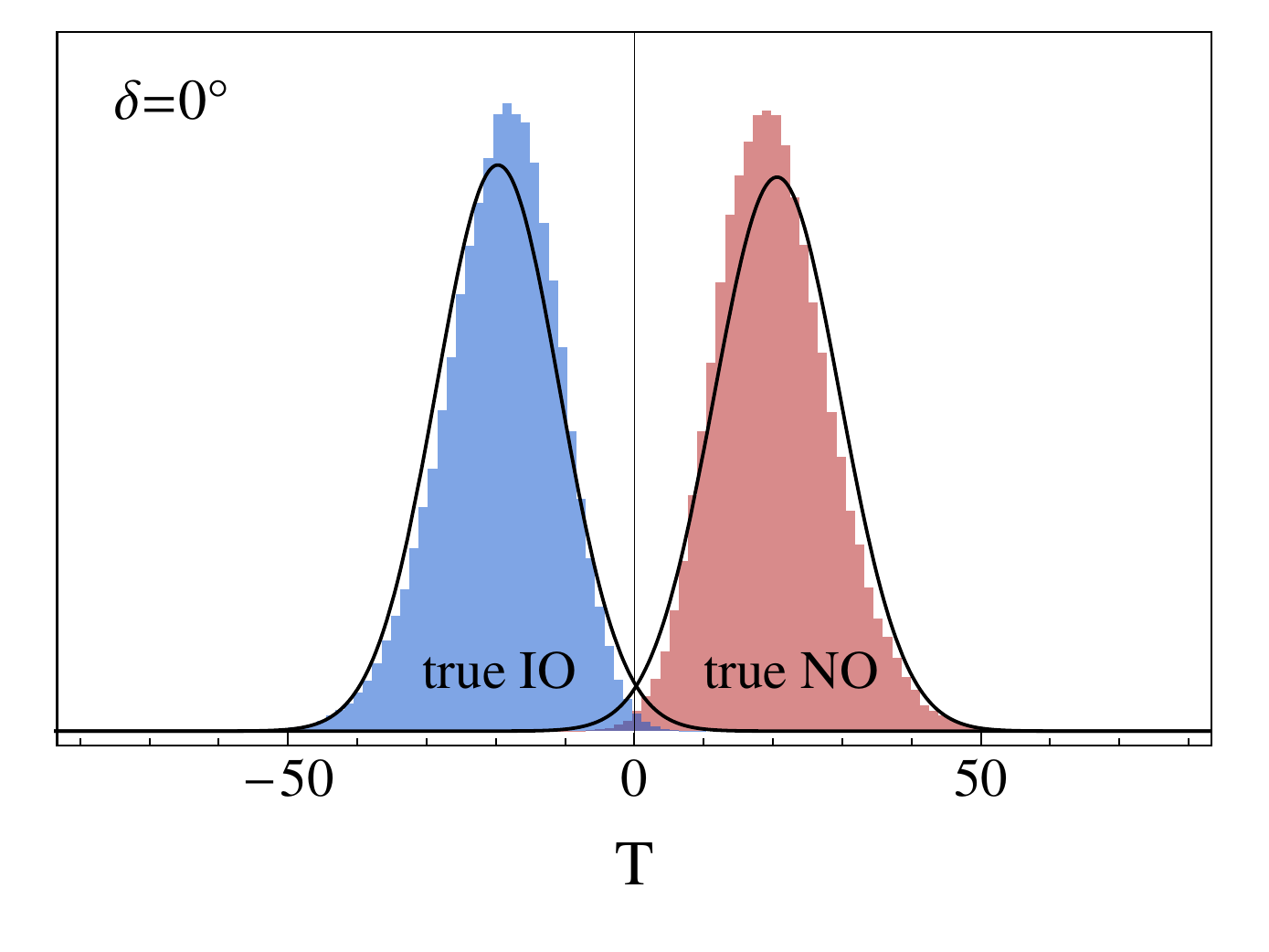} \\
\includegraphics[width=0.4\textwidth]{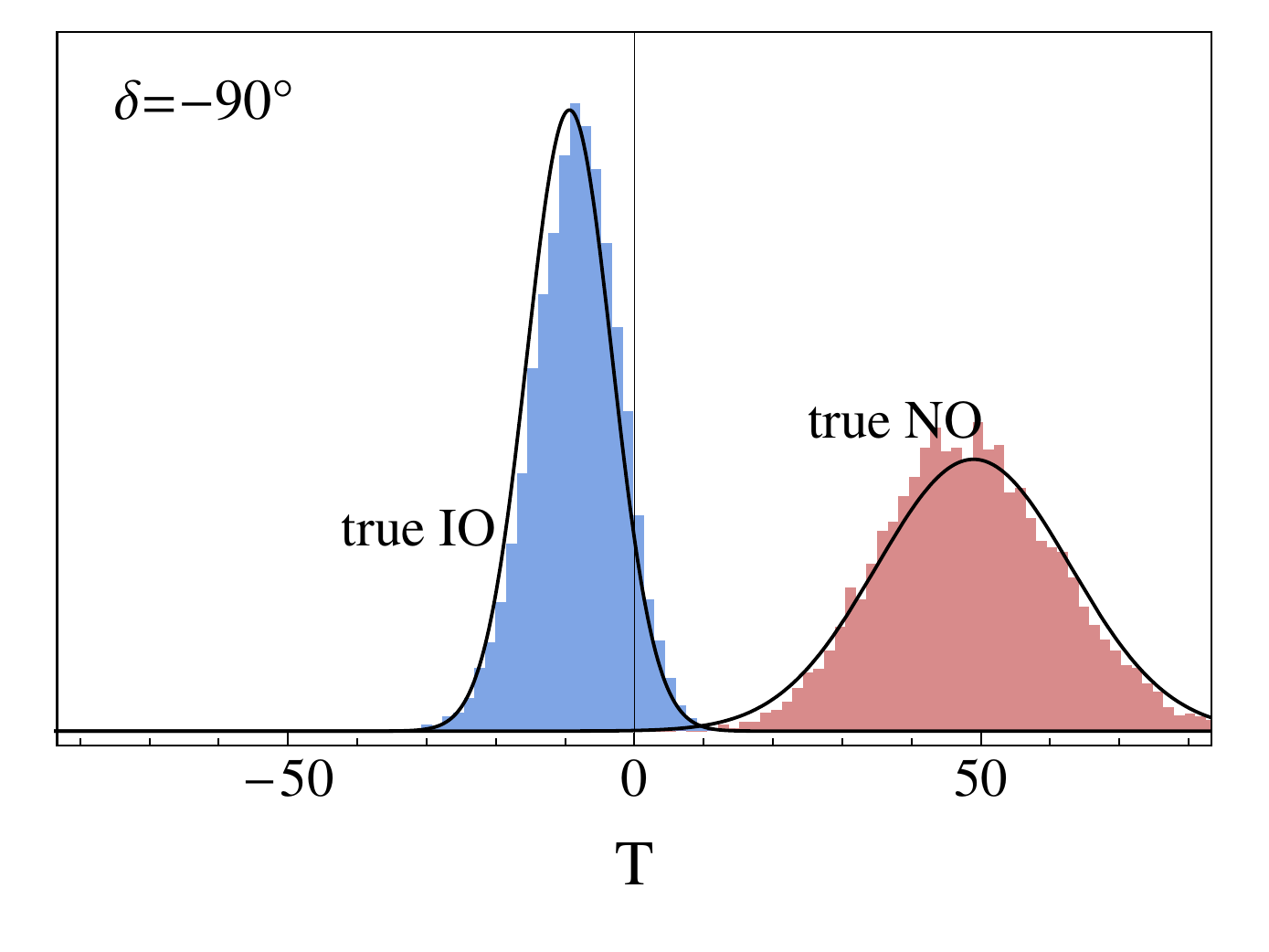}
\includegraphics[width=0.4\textwidth]{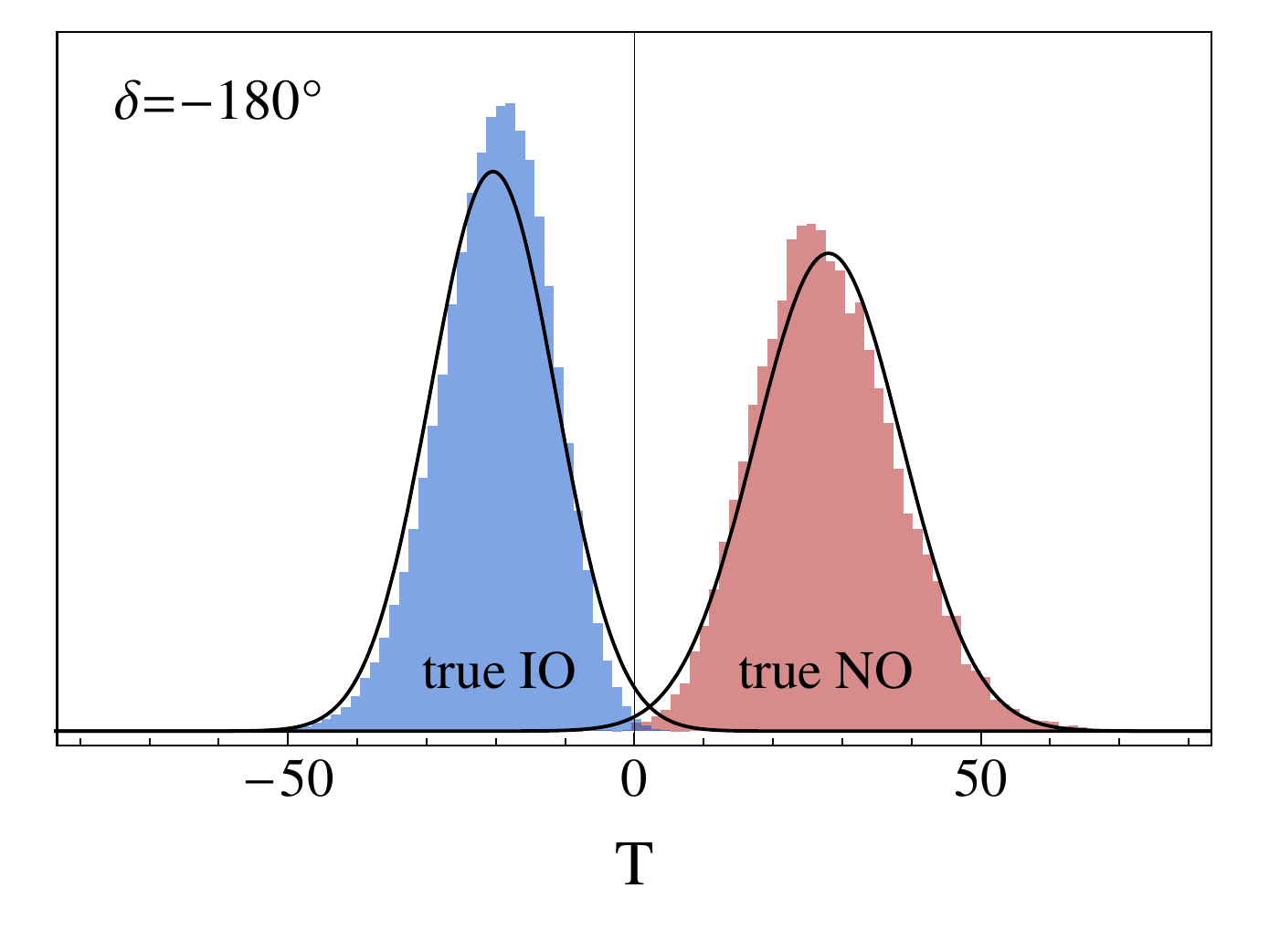}
\caption{The simulated distributions of the test statistic $T$ in the \LBNE{10} experiment for different true values of $\delta$, as indicated by the labels. The red (blue) distributions assume a true normal (inverted) ordering. Solid curves indicate the Gaussian approximation for $T$
from Eq.~\eqref{eq:Tgauss1}.}
\label{fig:lbnedistributions}
\end{center}
\end{figure}

Figs.~\ref{fig:novadistributions} and~\ref{fig:lbnedistributions} show
the probability distributions for the test statistic $T$ defined in
Eq.~\eqref{eq:T}, for the \NOvA\ and \LBNE{10} setups,
respectively. The distributions are shown for both mass orderings, and
for different values of $\delta$, as indicated in each panel. Our results are based on a sample of
$6\times 10^5$ simulations for \NOvA\ and $4\times 10^5$ for \LBNE{10} per value of
$\delta$, and we scan $\delta$ in steps of $10^\circ$. As can be
seen from the figures, both the shape and mean of the distributions
present large variations with the value of $\delta$. From the
comparison between the two figures it is clear that the \NOvA\
experiment will achieve very limited sensitivity to the mass ordering.
On the other hand, for the \LBNE{10}
setup the situation is much better: the overlapping region is reduced,
and is only sizable for certain combinations of values of $\delta$ in
the two mass orderings.

We also note that for \NOvA\ there are clear deviations from the
Gaussian shape for the $T$ distributions, while for the \LBNE{10}
experiment they are close to the Gaussian approximation discussed in
section~\ref{sec:gauss}, namely $T = \mathcal{N}(\pm T_0(\theta),
2\sqrt{T_0(\theta)})$. For comparison, in
Fig.~\ref{fig:lbnedistributions} the Gaussian approximation is
overlaid on the histograms from the Monte Carlo. Those results are in
agreement with the considerations of App.~\ref{app:Tgauss}. As
discussed there, one expects that the median of the $T$ distribution
should remain around $\pm T_0$, even if corrections to the shape of the
distribution are significant. We have checked that this does indeed hold
for \NOvA. Furthermore, assuming that there is only one relevant
parameter ($\delta$ in this case), Eq.~\eqref{eq:cond-1param} implies
that deviations from Gaussianity can be expected if $T_0 \sim 1$,
which is the case for \NOvA, whereas for $T_0 \gg 1$ (such as for
LBNE) one expects close to Gaussian distributions for $T$.

One can also notice in Figs.~\ref{fig:novadistributions} and~\ref{fig:lbnedistributions}
that the shape of the distributions for a given value of $\delta$ in
one ordering is rather similar to the mirrored image of the
distribution corresponding to the other mass ordering and
$-\delta$. The reason for this is the well-known fact that the
standard mass ordering sensitivity is symmetric between changing the
true ordering and $\delta\to -\delta$, \ie, $T_0^\NO(\delta) \approx
T_0^\IO(-\delta)$, see \eg, Figs.~8 and 9 of Ref.~\cite{Patterson:2012zs}
and Fig.~4-13 of Ref.~\cite{Adams:2013qkq}.\footnote{This can be understood
by considering the expressions for the oscillation probabilities,
taking into account the fact that, if matter effects are sufficiently strong, the $\chi^2$ minimum in the wrong
ordering tends to take place close to $\delta = \pm\pi/2$. }  Furthermore, using the formalism in
App.~\ref{app:Tgauss}, in particular Eq.~\eqref{eq:cond-1param},
one can show that also the deviations from the Gaussian distribution
will obey the same symmetry.  Below we will show that despite the
deviations from Gaussianity for \NOvA, the final sensitivities obtained
from the Monte Carlo will be surprisingly close to the Gaussian
expectation. As expected, this will be even more true for \LBNE{10}.

\begin{figure}
\begin{center}
\includegraphics[width=0.4\textwidth]{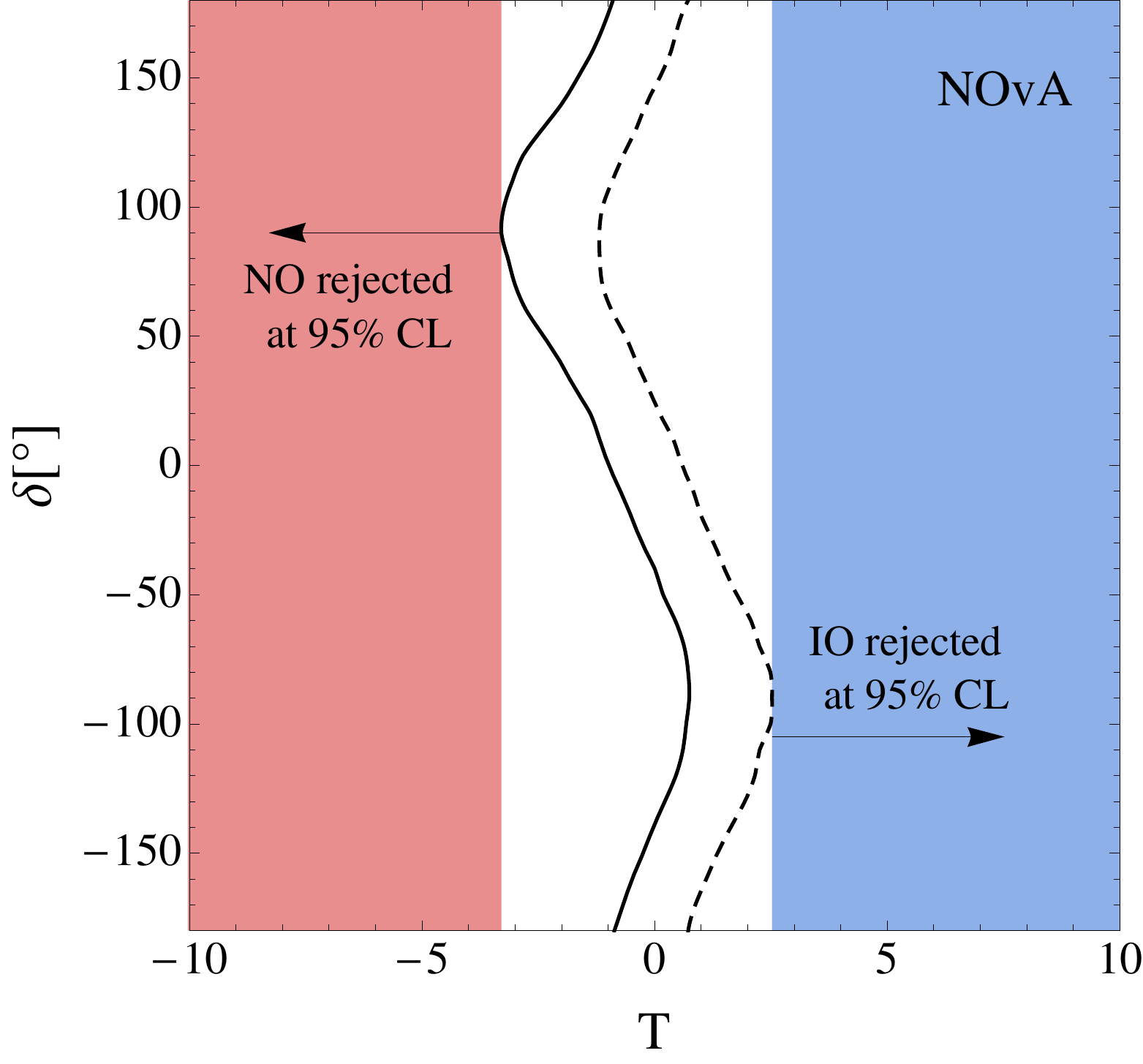}
\includegraphics[width=0.4\textwidth]{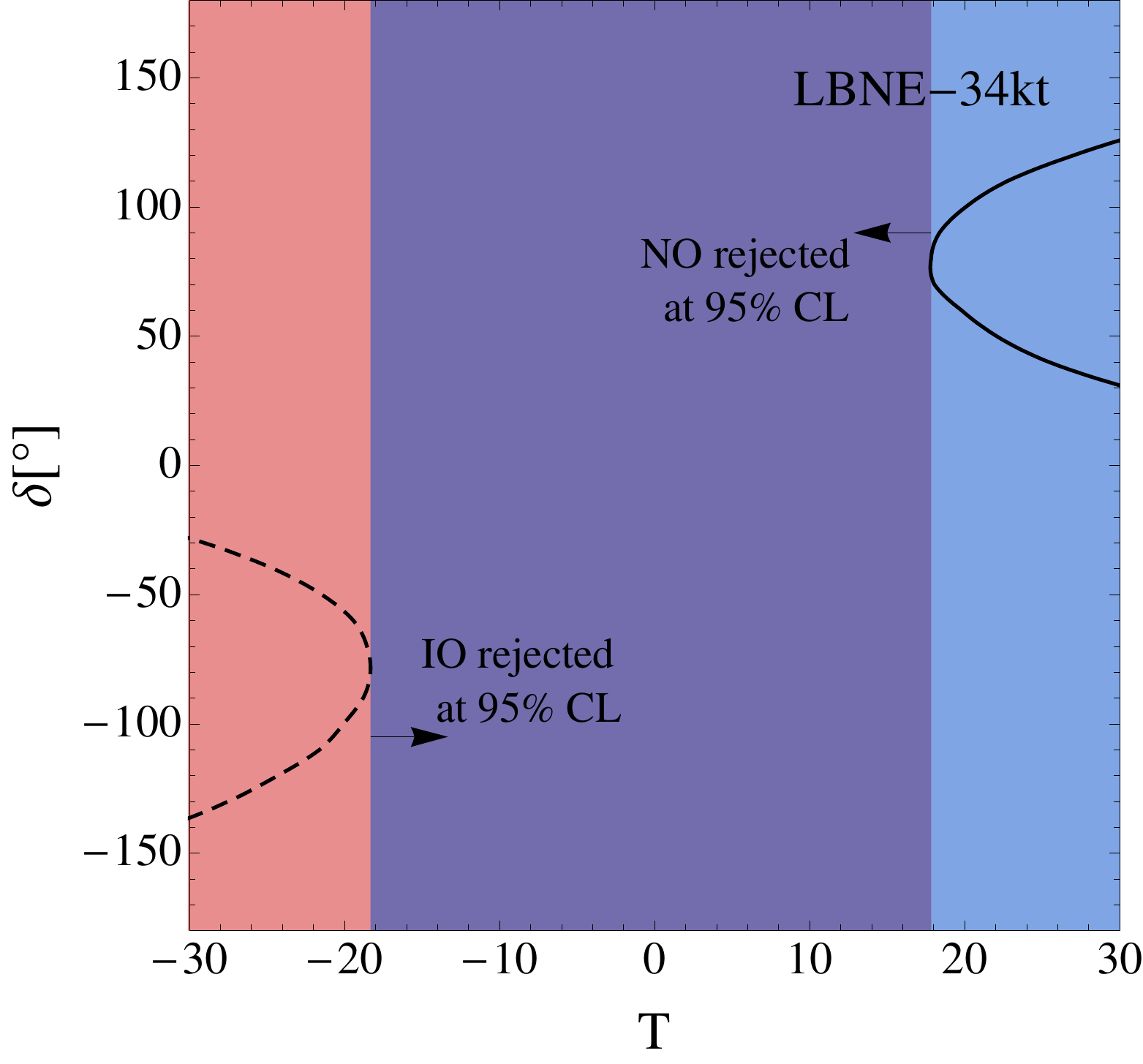}
\caption{The critical value $T_c$ corresponding to 95\% confidence level as a function of the CP-violating phase $\delta$ for \NOvA~(left panel) and \LBNE{34}~(right panel). The solid (dashed) lines correspond to testing the normal (inverted) ordering. The red (blue) region corresponds to values of $T$ which would reject all parameter values in the normal (inverted) ordering and thereby reject normal (inverted) ordering at 95\% confidence level. In the white region, there are parameter values in both orderings which are allowed, while in the purple region none of the two orderings would be compatible with data at 95\%~CL.}
\label{fig:Tc-vs-delta}
\end{center}
\end{figure}

Due to the strong dependence on the CP phase $\delta$ we need to choose
the critical value $T_c^\alpha$ such that the null hypothesis can be
rejected at $(1-\alpha)$~CL for all possible values of $\delta$, see
discussion in sections~\ref{sec:hypothesis-testing}
and \ref{sec:gauss-comp}. This is illustrated in
Fig.~\ref{fig:Tc-vs-delta}, which is analogous to Fig.~\ref{fig:Tc} (right panel)
for a fixed CL. The continuous (dashed) black curves in
Fig.~\ref{fig:Tc-vs-delta} show the values of $T_c^\alpha$ that lead
to the probability of 5\% to find a smaller (larger) value of $T$
under the hypothesis of a true normal (inverted) ordering as a
function of the true value of $\delta$. The left panel shows the
result for \NO$\nu$A, while the right panel corresponds to \LBNE{34}. 
The number of data sets simulated for \LBNE{34} in this case is $10^5$ 
per value of $\delta$, which is again scanned in steps of $10^\circ$.
As discussed in Sec.~\ref{sec:hypothesis-testing}, a composite null
hypothesis can only be rejected if we can reject all parameter sets
$\theta \in H$. In our case, this would imply rejecting the hypothesis
for all values of $\delta$. Therefore, in order to guarantee a CL
equal to $(1-\alpha)$, the most conservative value of $T_c^\alpha$
will have to be chosen. This automatically defines two values
$T^\alpha_c (\NO)$ and $T_c^\alpha (\IO)$, which are the values which
guarantee that a given hypothesis can be rejected at the
95\%~CL. These values will generally be different, and are indicated
in the figures by the arrows. In Fig.~\ref{fig:Tc-vs-delta} we
encounter the two situations already discussed in
Sec.~\ref{sec:statanalysis} (cf.~Fig.~\ref{fig:Tc}):
\begin{itemize}
 \item $T^\alpha_c(\IO) > T^\alpha_c(\NO)$: this is the case of NO$\nu$A, left panel. There is an intermediate region (shown in white) in which none of the hypotheses would be rejected at $(1-\alpha)$~CL. The reason why this intermediate region appears is because the experiment is not sensitive enough to the observable we want to measure, and a measurement at the chosen CL may not be reached.
 \item $T^\alpha_c(\IO) < T^\alpha_c(\NO)$: this is the case of \LBNE{34}, right panel. There is an overlap region (shown in purple) in which both hierarchies would be rejected at $(1-\alpha)$~CL. A statistical fluctuation may bring the result of the experiment into this region,
although this would typically not be expected.
 \end{itemize}
The intermediate case $T^\alpha_c(\IO) = T^\alpha_c(\NO)$ would correspond
to the ``crossing point'' discussed in Sec.~\ref{sec:statanalysis}, 
Fig.~\ref{fig:Tc}, which defines the CL at which exactly one of the
hypotheses can be excluded. 

\begin{figure}
\begin{center}
\includegraphics[width=0.49\textwidth]{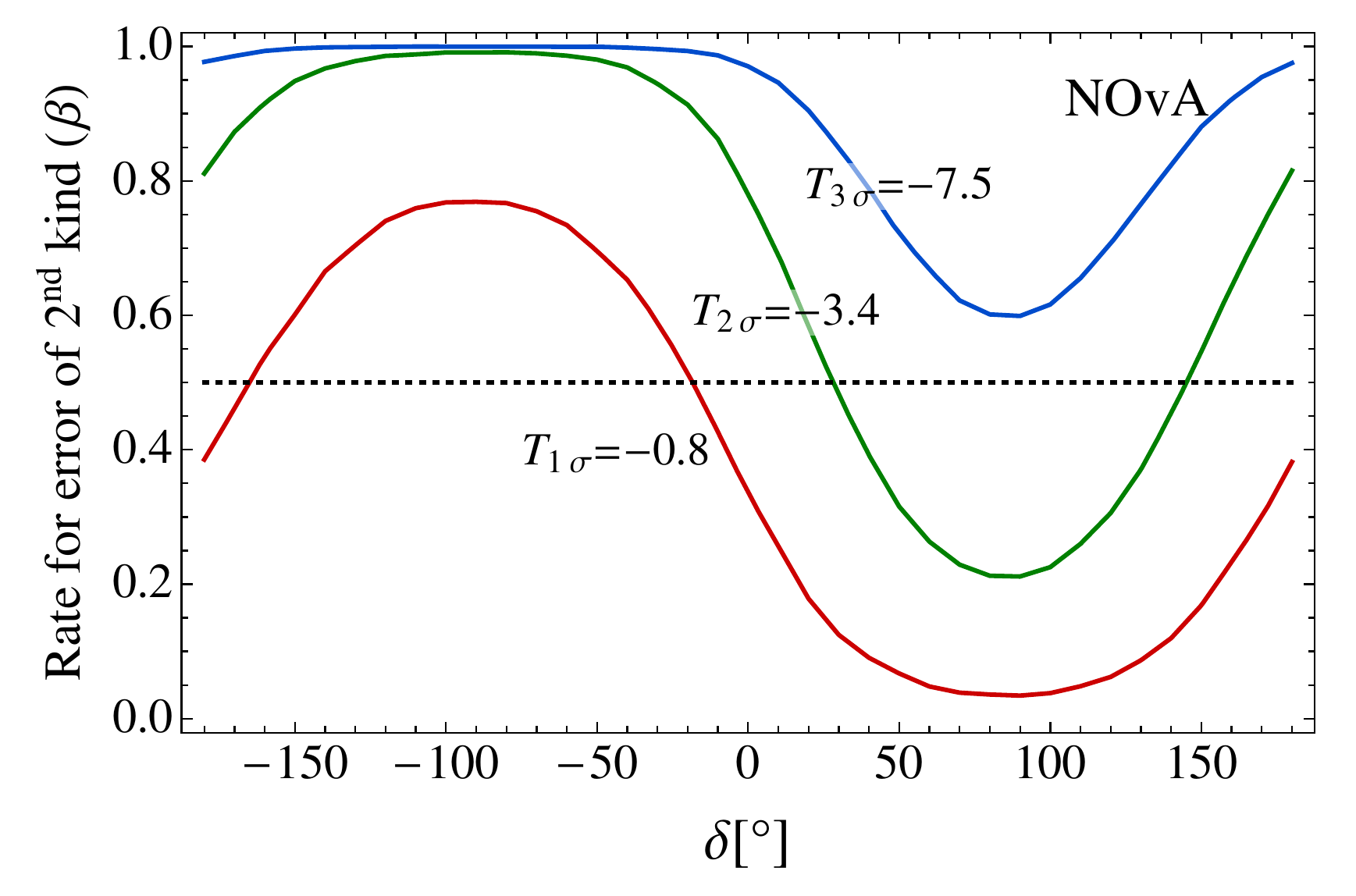}
\includegraphics[width=0.49\textwidth]{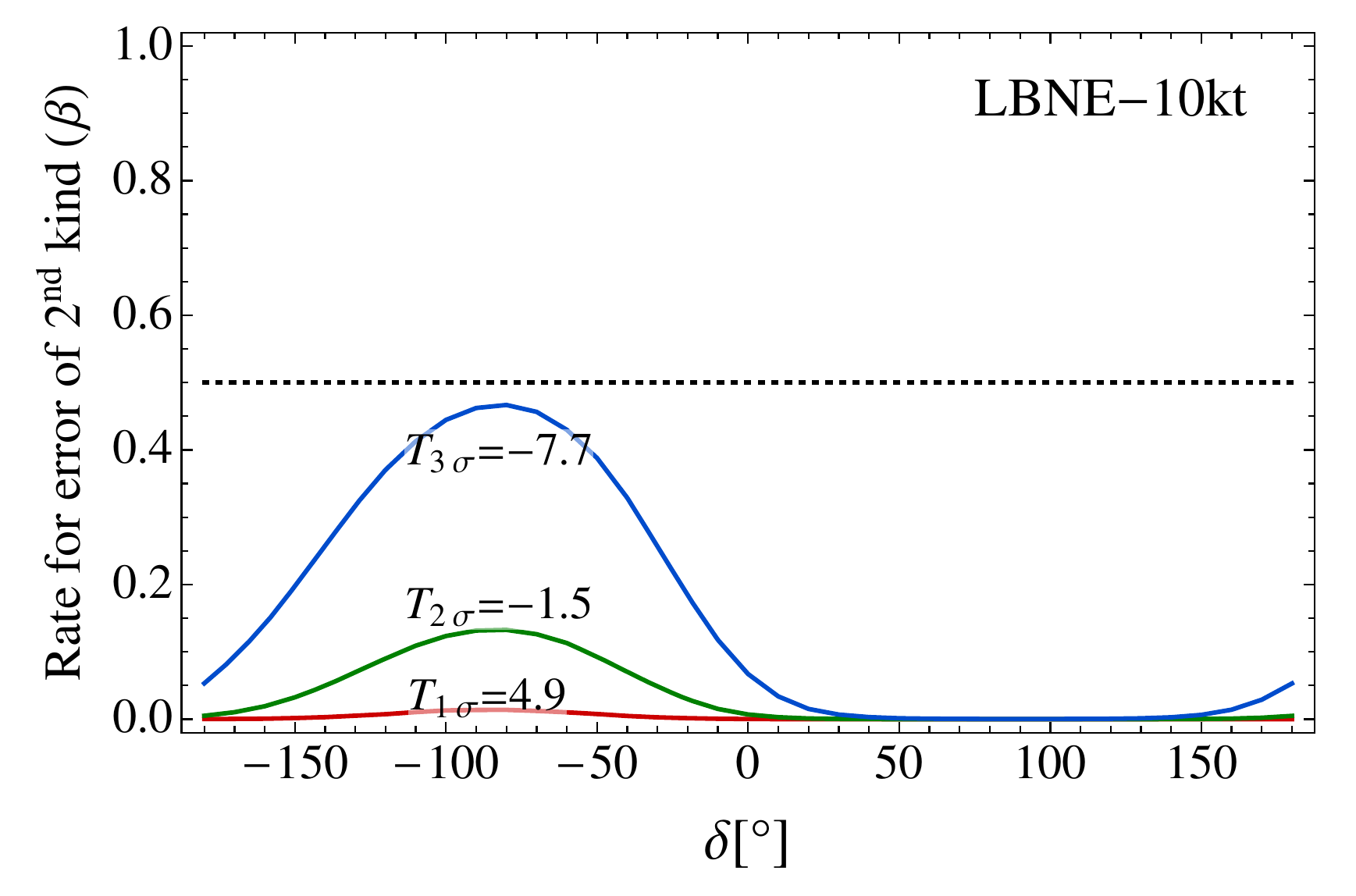}
\caption{Probability of accepting normal ordering if inverted ordering is true 
(\ie, rate for an error of the second kind) as a function of the true $\delta$ in
IO for the \NOvA~(left panel) and \LBNE{10}~(right panel)
experiments. The different curves correspond to tests at 1$\sigma$,
2$\sigma$, 3$\sigma$ confidence level, as labeled in the
plot. Furthermore the corresponding critical values $T^\alpha_c$ are
given. The horizontal dotted lines indicate the median experiment, $\beta=0.5$. }
\label{fig:beta-vs-delta}
\end{center}
\end{figure}

Let us now evaluate the rate for an error of the second kind corresponding to a
given value of $\alpha$.  After the value of $T_c^\alpha$ is
determined for a given hypothesis and $\alpha$, we can compute the
rate for an error of the second kind, $\beta$, as a function
of the true value of $\delta$, as discussed in
Sec.~\ref{sec:statanalysis}. We show this probability in
Fig.~\ref{fig:beta-vs-delta} for the \NOvA\ and the \LBNE{10}
experiments in the left- and right-hand panels, respectively. To be
explicit, we show the probability of accepting normal ordering at 
1$\sigma$, 2$\sigma$, 3$\sigma$ CL, \ie, $\alpha=32\%, 4.55\%, 0.27\%$, (regardless of
the value of $\delta$ in the NO) although the true ordering is
inverted. This probability depends on the true value of $\delta$ in
the IO, which is shown on the horizontal axis. By doing a cut at $\beta=0.5$ on the left-hand panel (indicated by the dotted line), we can get an idea on the median sensitivity that will be obtained for \NOvA: for $\delta=-90^\circ$ it will be around $1\sigma$, while for $\delta=90^\circ$ it will reach almost the $3\sigma$ level. This seems to be roughly consistent with the expected standard
sensitivities usually reported in the literature, see for instance
Ref.~\cite{Patterson:2012zs}. Similarly, for \LBNE{10}, we expect that the sensitivity for the median experiment will be around $3\sigma$ for $\delta=-90^\circ$, while for other values of $\delta$ we expect it to be much larger. This is also in agreement with the results from Ref.~\cite{Adams:2013qkq}, for instance.

\begin{figure}
\begin{center}
\includegraphics[width=0.49\textwidth]{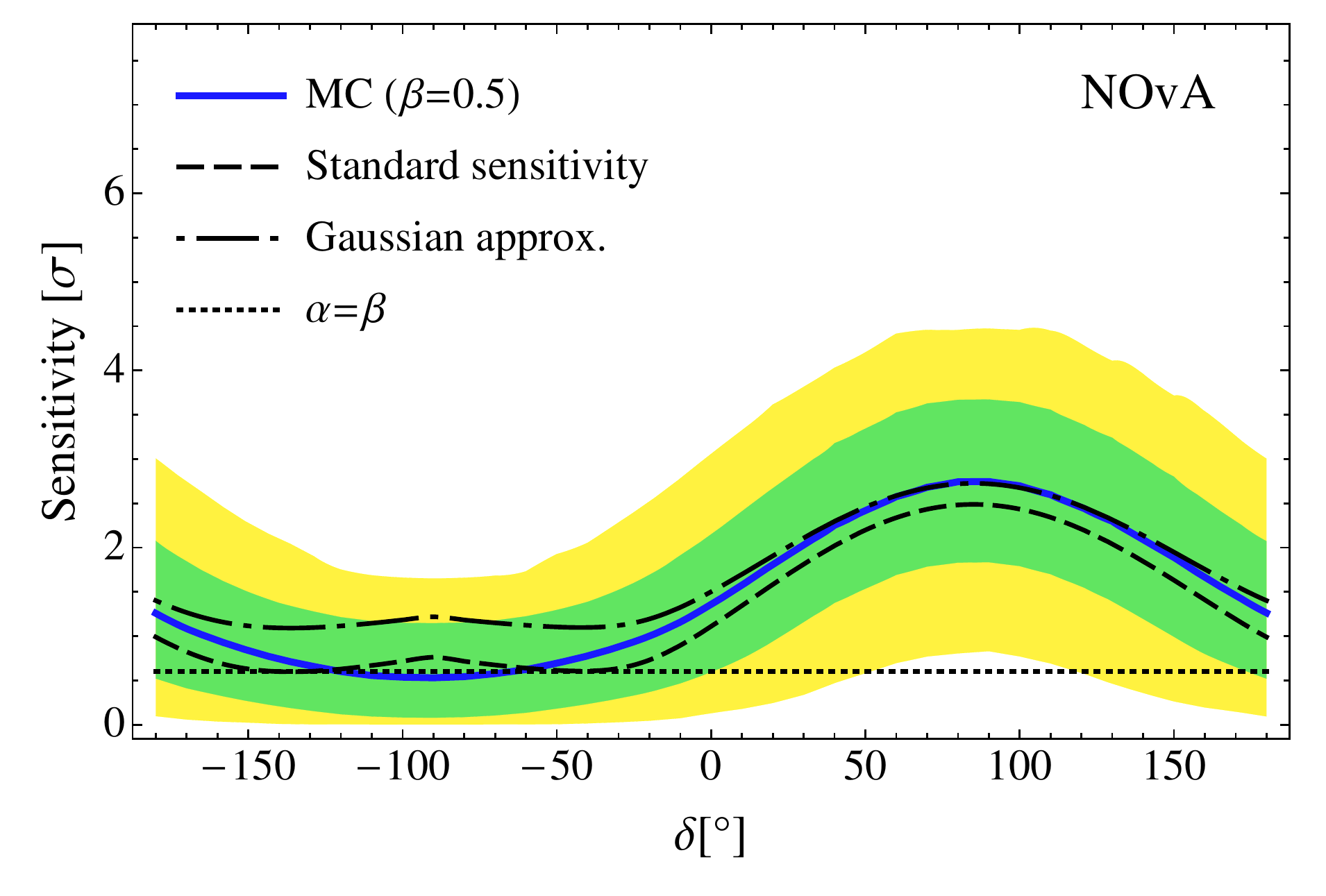}
\includegraphics[width=0.49\textwidth]{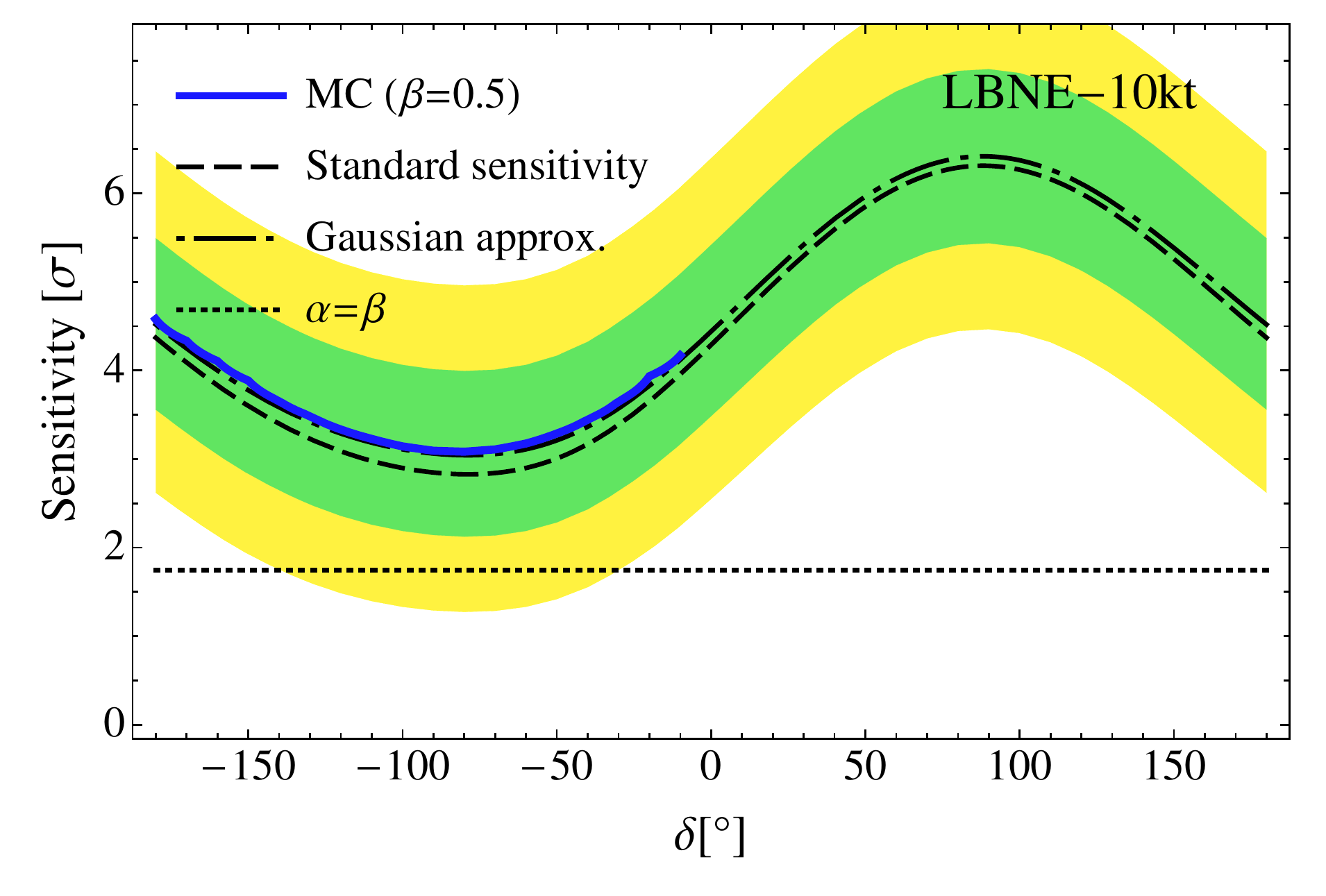}
  \caption{Comparison of the median sensitivities based on a full MC
  simulation to the results based on the Gaussian
  approximation Eq.~\eqref{eq:median-gauss-comp}. The number of sigmas
  at which the normal mass ordering can be rejected with a probability
  of 50\% are shown as a function of the true value of $\delta$ in the
  inverted ordering for \NOvA~(left panel) and \LBNE{10}~(right
  panel). The results obtained by a full MC simulation are shown by the solid thick lines. 
  The results for the Gaussian approximation are shown by
  the dot-dashed curves while the dashed curves correspond to
  the ``standard sensitivity'', \ie, $n = \sqrt{T_0}$.
  The dotted horizontal lines
  show the sensitivity corresponding to the ``crossing point''
  defined in Sec.~\ref{sec:statanalysis}, which guarantees that $\beta \lesssim \alpha$. 
   The missing points in the curve for the MC
  results for \LBNE{10} require a number of simulations above
  $4\times10^5$ (per value of $\delta$) and are therefore not computed
  here. The green (yellow) band shows the range of $\sigma$ with which a false null hypothesis 
  will be rejected in 68.27\% and 95.45\% of the experiments. }
\label{fig:sigmas-vs-std}
\end{center}
\end{figure}

Let us now investigate in detail how our median sensitivity compares
to the ``standard sensitivities'' widely used in the literature. In
Fig.~\ref{fig:sigmas-vs-std} the solid thick curves show the
results for the median sensitivity derived from full MC
simulations. The shaded green and yellow bands are analogous to those
shown in Fig.~\ref{fig:sens-gauss}, and show the range in the number of sigmas with
which we expect to be able to reject NO if IO is true in
68.27\% and 95.45\% of the experiments, respectively.  We also show how
these results compare to the Gaussian approximation discussed in
section~\ref{sec:gauss}. The value of the $\chi^2$ is computed without
taking statistical fluctuations into account (what is called $T_0$ in
Sec.~\ref{sec:statanalysis}). We then use
Eq.~\eqref{eq:median-gauss-comp} to compute the confidence level
$(1-\alpha)$ at which the normal ordering can be rejected with a
probability of 50\% if the inverted ordering is true, as a function of
the true value of $\delta$ in the IO. Then, for the dot-dashed curves
we use a 2-sided Gaussian to convert $\alpha$ into number of
$\sigma$, \ie, Eq.~\eqref{eq:sigma-alpha}, the same prescription is
also used for the MC result. We observe good agreement, in particular
for LBNE. This indicates that, for the high-statistics data from LBNE,
we are very close to the Gaussian limit, whereas from the smaller data
sample (and smaller values of $T_0$) in \NOvA\ deviations are visible,
but not dramatic. We also show the results using a 1-sided Gaussian,
Eq.~\eqref{eq:sigma-alpha-1sided}, to convert $\alpha$ into number of
sigmas, which leads to $n = \sqrt{T_0}$, \ie, the standard
sensitivity. This is shown by the dashed lines. As discussed in
Sec.~\ref{sec:statanalysis} we observe that the standard sensitivity
slightly under-estimates the true sensitivity.\footnote{Note that
traditionally the ``standard sensitivity for IO'' denotes the case
when IO is true and refers to the sensitivity to reject NO. In the
language of the present paper we call this a ``test for NO''. This is
also consistent with the formula in the Gaussian approximation,
Eq.~\eqref{eq:median-gauss-comp}, which contains $T_0^\IO$ when
considering a test for NO. This has to be taken into account when
comparing \eg, Fig.~\ref{fig:sigmas-vs-std} (corresponding to a test
for NO) to similar curves in the literature.}  Finally, the dotted
horizontal line in Fig.~\ref{fig:sigmas-vs-std} corresponds to the
significance of the crossing point $T_c^\NO = T_c^\IO$ defined in
Sec.~\ref{sec:statanalysis}, \ie, the confidence level at which
exactly one hypothesis can be excluded regardless of the outcome of
the experiment. The results are independent of the value of $\delta$,
and guarantee that the rate for an error of the second kind $\beta$ is
at most equal to $\alpha$, unlike for the median experiment where
$\beta =0.5$. The results for the crossing point are also consistent
with the Gaussian expectation Eq.~\eqref{eq:crossing-composite}.

\section{Comparison between facilities: future prospects}
\label{sec:comparison}

In this section we give a quantitative comparison between the
different experiments that have been considered in this paper. We do a
careful simulation of all the facilities using the details available
in the literature from the different collaborations, see
App.~\ref{app:simulation} for details.  We have checked that our
standard sensitivities are in good agreement with the respective
proposals or design reports. Nevertheless, we do not explore in which
way the assumptions made in the literature towards efficiencies,
energy resolution, angular resolution, systematics, etc may affect the
results, with the only exception of JUNO, as we explain below.  Since
we are mainly interested in the statistical method for determining the
mass ordering, such analysis is beyond the scope of this paper. Our
results will be shown as a function of the date, taking the starting
points from the official statements of each collaboration. Obviously, such projections always are subject to large
uncertainties.

\begin{figure}
\begin{center}
\includegraphics[width=0.49\textwidth]{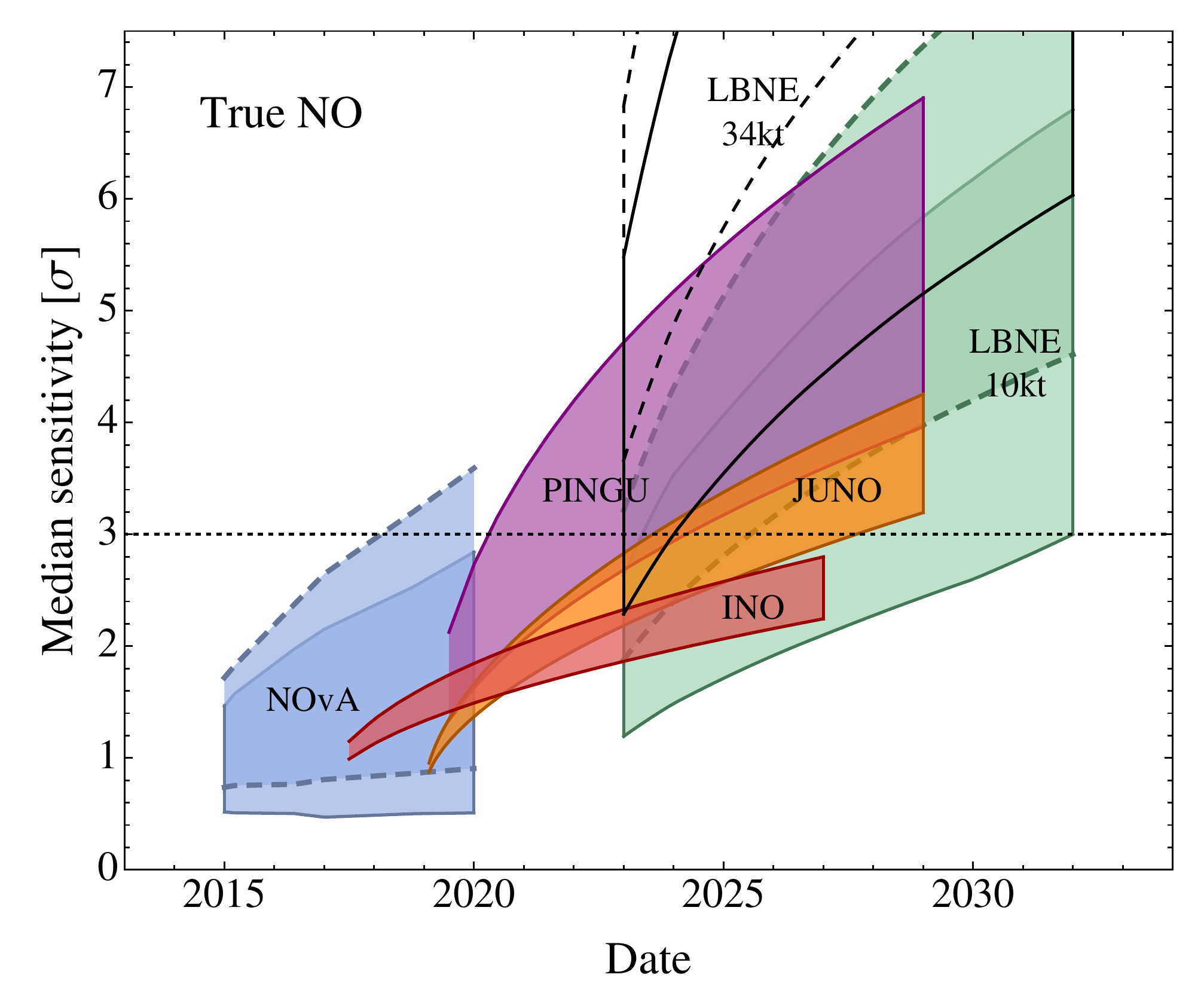}
\includegraphics[width=0.49\textwidth]{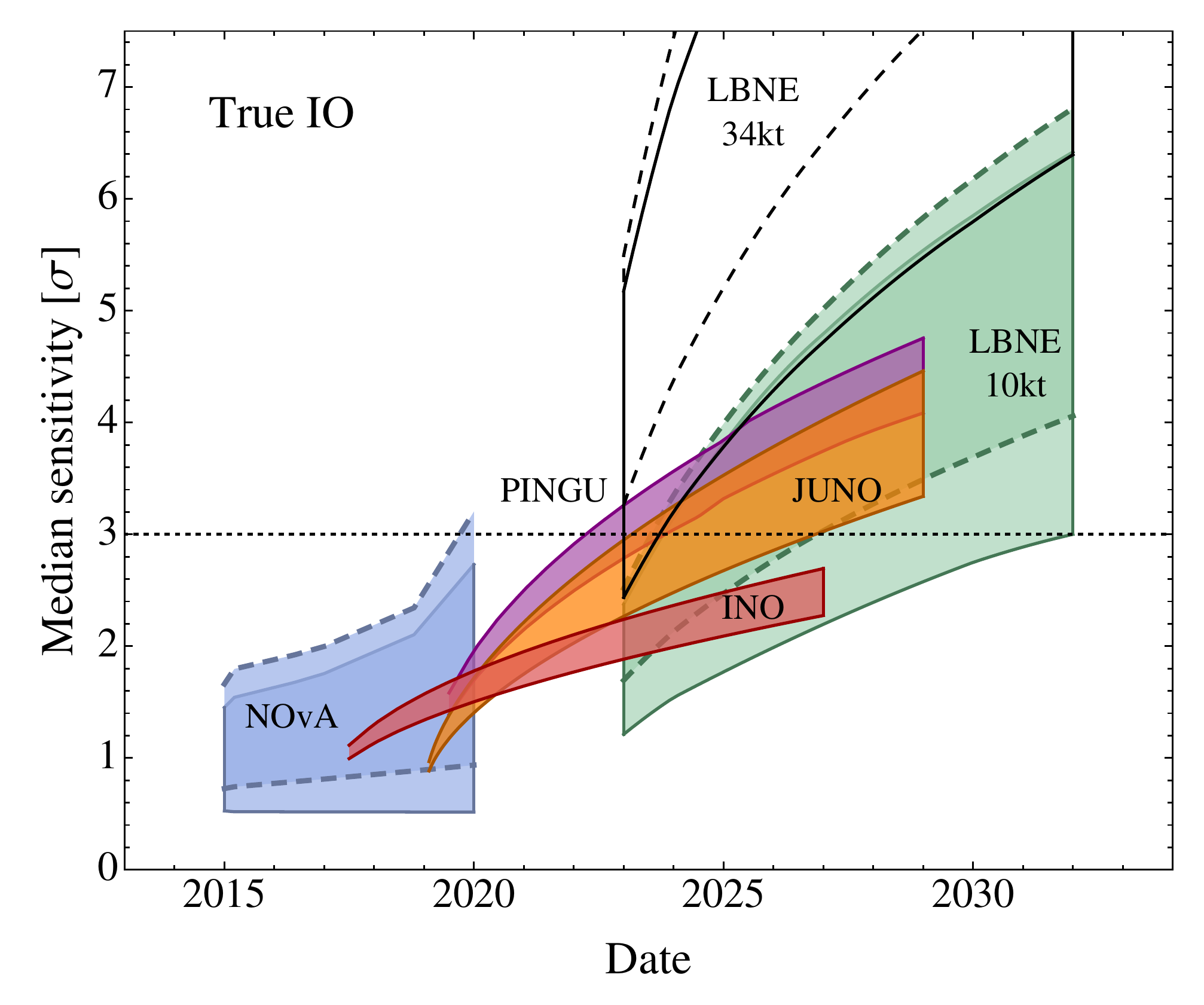}
  \caption{The left (right) panel shows the median sensitivity in
    number of sigmas for rejecting the IO (NO) if the NO (IO) is true
    for different facilities as a function of the date. The width of
    the bands correspond to different true values of the CP phase
    $\delta$ for \NOvA\ and LBNE, different true values of
    $\theta_{23}$ between $40^\circ$ and $50^\circ$ for INO and PINGU,
    and energy resolution between $3\%\sqrt{1~\textrm{MeV}/E}$ and
    $3.5\%\sqrt{1~\textrm{MeV}/E}$ for JUNO. For the long baseline experiments, the bands with solid (dashed) contours correspond to a true value for $\theta_{23}$ of $ 40^\circ$ ($ 50^\circ$). In all cases, octant degeneracies are fully searched for. }
\label{fig:timeplot-median}
\end{center}
\end{figure}

\begin{figure}
\begin{center}
\includegraphics[width=0.49\textwidth]{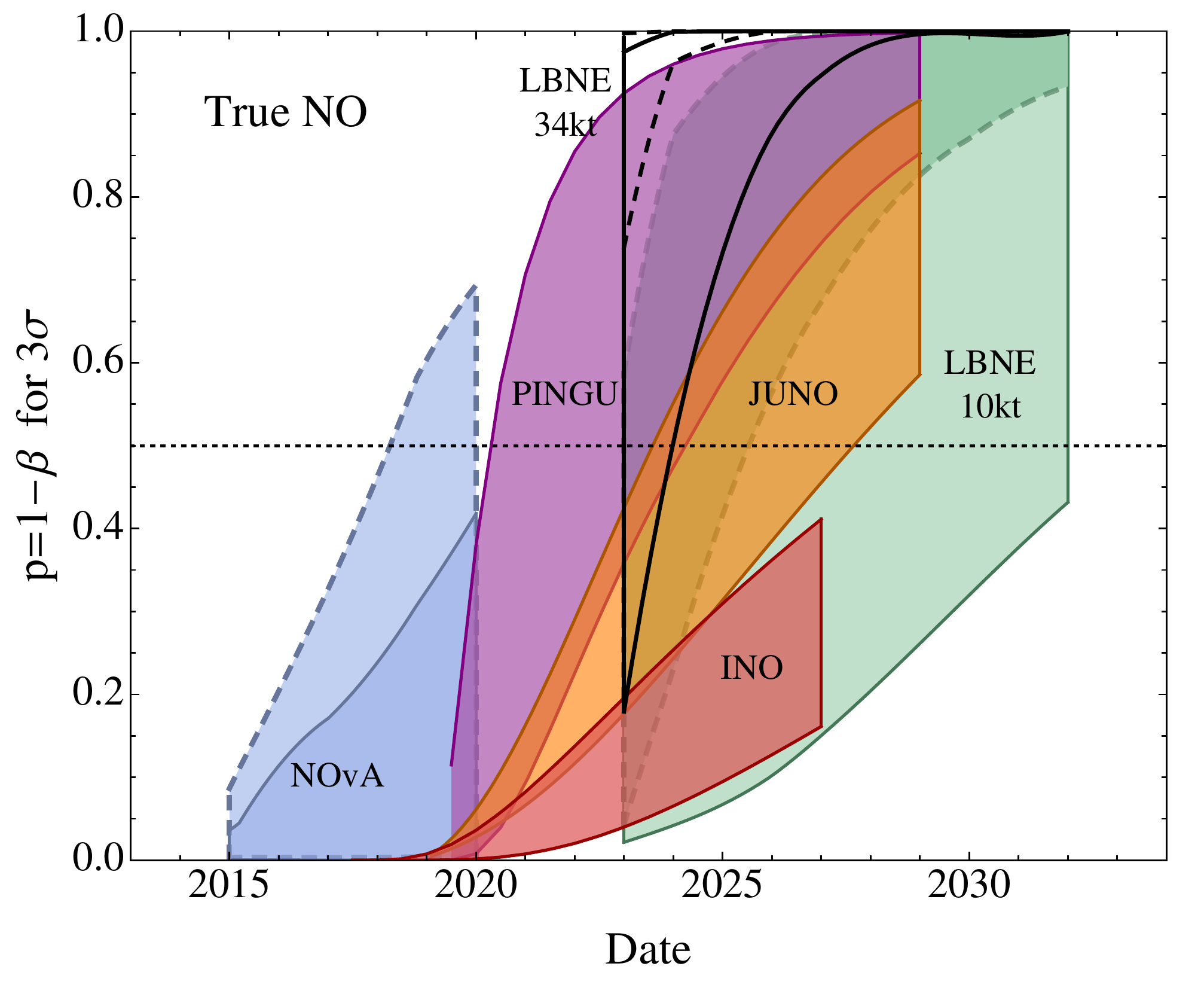}
\includegraphics[width=0.49\textwidth]{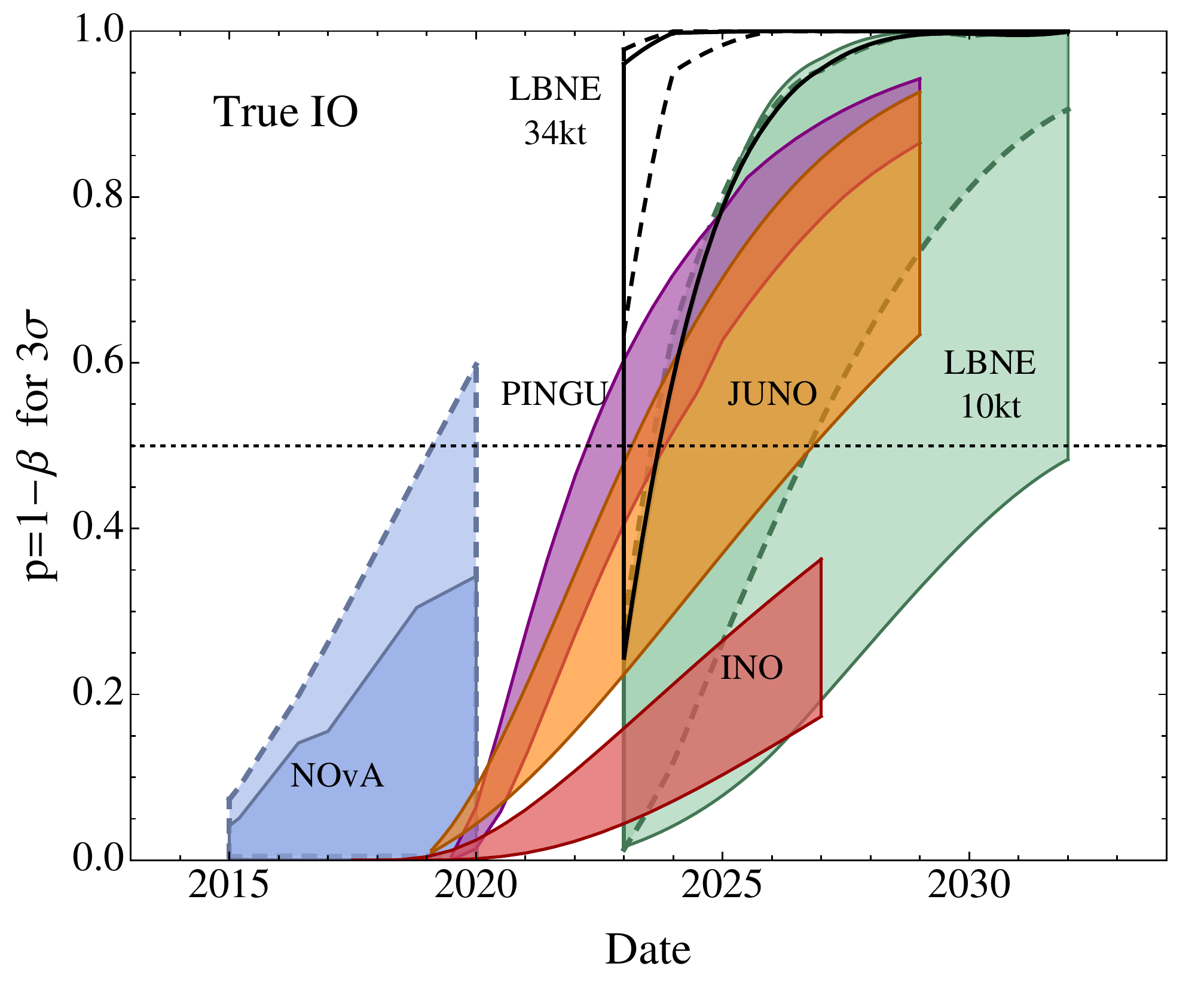}
  \caption{Probability that the wrong ordering can be rejected at
    $3\sigma$ (99.73\%~CL) for a true NO (left) and IO (right) for
    different facilities as a function of the date. The width of the
    bands has the same origin as in Fig.~\ref{fig:timeplot-median}.
    The dotted horizontal line indicates the median experiment
    ($\beta=0.5$).}
\label{fig:timeplot-beta}
\end{center}
\end{figure}

Fig.~\ref{fig:timeplot-median} shows the median sensitivities for the
various experiments, \ie, the number of sigmas with which an ``average
experiment'' for each facility can rejected a given mass ordering if
it is false. In some sense this is similar to the standard sensitivity
of $\sqrt{T_0}$ commonly applied in the literature. A different
question is answered in Fig.~\ref{fig:timeplot-beta}, namely: what is the
probability that the wrong mass ordering can be rejected at a confidence
level of $3\sigma$? The confidence level has been chosen arbitrarily
to $3\sigma$, based on the convention that this would correspond
to ``evidence'' that the wrong ordering is false. Below we discuss
those plots in some detail.

In order to keep the number of MC simulations down to a feasible
level, we use the Gaussian approximation whenever it is reasonably
justified. As we have shown in Sec.~\ref{sec:numericalresults}, this
is indeed the case for PINGU, INO, and JUNO. With respect to the LBL
experiments, even though we have seen that the agreement with the
Gaussian case is actually quite good (see
Fig.~\ref{fig:sigmas-vs-std}), there are still some deviations, in
particular in the case of \NOvA. Consequently, in this case we have
decided to use the results from the full MC simulation whenever
possible. The results for the \NOvA\ experiment are always obtained
using MC simulations, while in the case of \LBNE{10} the results from
a full MC are used whenever the number of simulations does not have to
exceed $4\times10^5$ (per value of $\delta$). As was mentioned in
the caption of Fig.~\ref{fig:sigmas-vs-std}, this means that, in order
to reach sensitivities above $\sim 4\sigma$ (for the median
experiment), results from the full MC cannot be used. In these cases,
we will compute our results using the Gaussian approximation
instead. As mentioned in App.~\ref{app:Tgauss}, the approximation is
expected to be quite accurate precisely for large values of
$T_0$. Finally, for \LBNE{34}, all the results have to be computed
using the Gaussian approximation, since the median sensitivity for
this experiment reaches the $4\sigma$ bound already for one year of
exposure only, even for the most unfavorable values of $\delta$.

For each experiment, we have determined the parameter that has the largest impact on the results, and we draw a band according to it to show the range of sensitivities that should be expected in each case. Therefore, we want to stress that the meaning of each band may be different, depending on the particular experiment that is considered. In the case of long baseline experiments (\NOvA, \LBNE{10} and \LBNE{34}), the results mainly depend on the value of the CP-violating phase $\delta$. In this case, we do a composite hypothesis test as described in Secs.~\ref{sec:hypothesis-testing} and~\ref{sec:gauss-comp}, and we draw the edges of the band using the values of true $\delta$ in the true ordering that give the worst and the best results for each setup. Nevertheless, since for these experiments the impact due to the true value of $\theta_{23}$ is also relevant, we show two results, corresponding to values of $\theta_{23}$ in the first and second octant. In all cases, the octant degeneracy is fully searched for (see App.~\ref{app:lbl} for details). In the case of PINGU and INO, the most relevant parameter is $\theta_{23}$. We find that, depending on the combination of true ordering and $\theta_{23}$ the results will be very different. Therefore, in this case we also do a composite hypothesis test, using $\theta_{23}$ as an extra parameter. Finally, the case of JUNO is somewhat different. In this case, the uncertainties on the oscillation parameters do not have a big impact on the results. Instead, the energy resolution is the parameter which is expected to have the greatest impact, see for instance Ref.~\cite{Kettell:2013eos} for a detailed discussion. Therefore, in this case the width of the band shows the change on the results when the energy resolution is changed between $3\%\sqrt{1~\textrm{MeV}/E}$ and $3.5\%\sqrt{1~\textrm{MeV}/E}$. For JUNO we do a simple hypothesis test, as described in Sec.~\ref{sec:gauss-simple}. 

The starting dates assumed for each experiment are: 2017 for
INO~\cite{INOstartingdate}, 2019 for PINGU~\cite{Aartsen:2013aaa} and
JUNO~\cite{JUNO} and 2022 for LBNE~\cite{LBNEhomepage}. Note that the
official running times for PINGU and JUNO are 5 and 6 years,
respectively. For illustrative purposes we extend the time in the
plots to 10 years, in order to see how sensitivities would evolve
under the adopted assumptions about systematics. For the \NOvA\
experiment, we assume that the nominal luminosity will be achieved by
2014~\cite{Patterson:2012zs} and we consider 6 years of data taking
from that moment on. 

From the comparison of Figs.~\ref{fig:timeplot-median}
and \ref{fig:timeplot-beta} one can see that, even though the median
sensitivity for INO would stay below the $3\sigma$
CL, there may be a sizable probability (up to $\sim40\%$) that a
statistical fluctuation will bring the result up to $3\sigma$. For \NOvA, 
such probability could even go up to a 60\%, depending on the 
combination of $\theta_{23}$, $\delta$ and the true mass ordering. In the
case of LBNE, the dependence on the true value of $\delta$ is
remarkable, in particular for the power of the test. We clearly
observe the superior performance of the 34~\kt\ configuration over the
10~\kt\ one. For 34~\kt\ a $3\sigma$ result can be obtained at very high
probability for all values of $\delta$, and for some values of
$\delta$ a much higher rejection significance of the wrong ordering is
achieved with high probability.

For the atmospheric neutrino experiments INO and PINGU we show the
effect of changing the true value of $\theta_{23}$ from $40^\circ$ to
$50^\circ$. The effect is particularly large for PINGU and a true
NO. As visible in Fig.~\ref{fig:sens_pingu}, for NO the sensitivity
changes significantly between $40^\circ$ and $50^\circ$, whereas for
IO they happen to be similar, as reflected by the width of the bands
in Figs.~\ref{fig:timeplot-median} and \ref{fig:timeplot-beta}. The
reason for this behavior is that for true IO and $\theta_{23} >
45^\circ$ the mass ordering sensitivity is reduced due to the octant
degeneracy \cite{Winter:2013ema}. In the context of PINGU, let us
stress that the precise experimental properties (in particular the
ability to reconstruct neutrino energy and direction) are still very
much under investigation \cite{Aartsen:2013aaa}. While we consider our
adopted configuration (see Sec.~\ref{sec:atm} and App.~\ref{app:atm}
for details) as a representative bench mark scenario, the real
sensitivity may be easily different by few standard deviations, once the
actual reconstruction abilities and other experimental parameters are
identified. To lesser extent this applies also to INO.

Let us also mention that in this work we only consider the sensitivity
of individual experiments, and did not combine different setups. It
has been pointed out in a number of studies that the sensitivity can
be significantly boosted in this way \cite{Huber:2009cw,
Blennow:2012gj, Ghosh:2012px, Winter:2013ema, Blennow:2013vta}. We
also expect that in this case, if the combined $T_0$ is sufficiently
large, the Gaussian approximation should hold. However, we stress that
a detailed investigation of this question is certainly worth pursuing
in future work.

\section{Discussion and summary}
\label{sec:summary}

The sensitivity of a statistical test is quantified by reporting two numbers:
\begin{enumerate}
\item 
the confidence level $(1-\alpha)$ at which we want to reject a given
hypothesis, which corresponds to a rate for an error
of the first kind, $\alpha$; and
\item
the probability $p$ with which a hypothesis can be rejected at CL
($1-\alpha$) if it is false (the power of the test), which is related
to the rate for an error of the second kind, $\beta = 1- p$ .
\end{enumerate}
In this work we have applied this standard approach to the
determination of the type of the neutrino mass ordering. With the help
of those concepts it is straight forward to quantify the sensitivity
of a given experimental configuration aiming to answer this important
question in neutrino physics. We consider a test statistic $T$ (see
Eq.~\eqref{eq:T}) in order to perform the test, which is based on the
ratio of the likelihood maxima under the two hypotheses normal and
inverted ordering. Under certain conditions, see
App.~\ref{app:Tgauss}, the statistic $T$ is normal distributed
(Gaussian approximation)~\cite{Qian:2012zn}. In the limit of no
statistical fluctuations (Asimov data set) the test statistic $T$
becomes the usual $\Delta \chi^2$ (up to a sign) massively used in
the literature for sensitivity calculations. In this work we denote
this quantity by $T_0$ (in Ref.~\cite{Qian:2012zn} it has been
denoted by $\overline{\Delta\chi^2}$). The sensitivity of an
average experiment (in the frequentist sense) can be defined as
the confidence level $(1-\alpha)$ at which a given hypothesis can be
rejected with a probability $\beta = 50\%$ (``median
sensitivity''). An important result of our work is the following:

\bigskip

\emph{The sensitivity obtained by using the standard method of taking the square-root of the $\Delta\chi^2$ without statistical fluctuations is
very close to the median sensitivity obtained within the Gaussian approximation
for the test statistic $T$.}

\bigskip

\begin{table}
\begin{tabular}{c@{\,\,}lc@{\,\,}lccccc}
\hline\hline
$T_0$ & \multicolumn{2}{c}{std.\ sens.} & \multicolumn{2}{c}{median sens.} & \phantom{l}crossing sens.\phantom{m} & $\beta$ for $3\sigma$ & \phantom{m}68.27\% range & \phantom{a}95.45\% range \\ 
\hline
9 & $99.73\%$ & $(3.0\sigma)$ & $99.87\%$ & $(3.2 \sigma)$ & $93.32\% \, (1.8 \sigma)$ & $0.41$ & $2.3 \sigma - 4.2\sigma$ & $1.4 \sigma - 5.1\sigma$ \\ 
16 & $99.9937\%$ & $(4.0\sigma)$ & $99.9968\%$ & $(4.2 \sigma)$ & $97.72\% \, (2.3 \sigma)$ & $0.11$ & $3.2 \sigma - 5.1\sigma$ & $2.3 \sigma - 6.1\sigma$ \\ 
25 & $99.999943\%$ & $(5.0\sigma)$ & $99.999971\%$ & $(5.1 \sigma)$ & $99.38\% \, (2.7 \sigma)$ & $0.013$ & $4.2 \sigma - 6.1\sigma$ & $3.2 \sigma - 7.1\sigma$ \\ 
\hline\hline
\end{tabular}
  \caption{Sensitivity measures for the neutrino mass ordering in the
    Gaussian approximation assuming $T_0^\NO = T_0^\IO$. The columns
    show $T_0$, the standard sensitivity $n=\sqrt{T_0}$, the median
    sensitivity (Eqs.~\eqref{eq:median-alpha-gauss},
    \eqref{eq:median-sens-gauss}), the crossing sensitivity where
    exactly one hypothesis is rejected (equivalent to testing the sign
    of $T$, Eq.~\eqref{eq:alpha-crossing}), the probability $\beta$ of
    accepting a mass ordering at the $3\sigma$ CL although it is false
    (rate for an error of the second kind, Eq.~\eqref{eq:beta-gauss}), and the range
    of rejection confidence levels obtained with a probability of
    68.27\% and 95.45\%. We convert CL into standard deviations using
    a 2-sided Gaussian. \label{tab:metrics}}
\end{table}

In section~\ref{sec:gauss} we provide simple formulas, based on the
Gaussian approximation, which allow quantification of the sensitivity in
terms of error rates of the first and second kind for a given $T_0$. For
instance, Eqs.~\eqref{eq:beta-gauss} and \eqref{eq:beta-gauss-comp}
contain simple expressions for the computation of $\beta$ for given
values of $\alpha$ and $T_0$, whereas Eq.~\eqref{eq:median-gauss-comp}
allows the computation of the median sensitivity in terms of $T_0$.
In Tab.~\ref{tab:metrics} we give a collection of sensitivity measures
based on the Gaussian approximation for the three example values $T_0
= 9,16,25$. The columns ``std.\ sens.'' and ``median sens.''
demonstrate explicitly the statement emphasized above, that the median
sensitivity is close to the $n=\sqrt{T_0}$ rule. The crossing
sensitivity corresponds to the CL at which exactly one of the two
hypotheses can be rejected. This is similar to testing the sign of the
test statistic $T$, a test which has been discussed
in Ref.~\cite{Ciuffoli:2013rza} and also mentioned
in Ref.~\cite{Qian:2012zn}. By construction, this test gives smaller confidence levels than the median sensitivity and 
is not necessarily connected to what would be expected from an experiment. We give in
the table also the probability for accepting a hypothesis at the
$3\sigma$ level although it is false (rate for an error of the second kind). The
last two columns in the table give the range of obtained rejection
significance with a probability of 68.27\% and 95.45\% (assuming that
the experiment would be repeated many times). Those are a few examples
of how to apply the equations from section~\ref{sec:gauss}. These
sensitivity measures provide different information and all serve to
quantify the sensitivity of an experiment within a frequentist framework.
They can be compared to similar sensitivity measures given
in Ref.~\cite{Qian:2012zn} in a Bayesian context (see, \eg, their Tab.~IV).

In the second part of the paper we report on the results from Monte
Carlo simulations for several experimental setups which aim to address
the neutrino mass ordering: the medium-baseline reactor experiment
JUNO, the atmospheric neutrino experiments INO and PINGU, and the
long-baseline beam experiments \NOvA\ and LBNE. In each case we have
checked by generating a large number of random data sets how well the
Gaussian approximation is satisfied. Our results indicate that the
Gaussian approximation is excellent for JUNO, INO, and
PINGU. For \NOvA\ the $T$ distributions deviate significantly from
Gaussian (strongly dependent on the true value of the CP phase
$\delta$), however the Gaussian expressions for the sensitivities
still provide a fair approximation to the results of the Monte
Carlo. For LBNE the Gaussian approximation is again fulfilled
reasonably well.  This is in agreement with our analytical
considerations on the validity of the Gaussian approximation given in
App.~\ref{app:Tgauss}, where we find that for experiments with
$T_0$ large compared to the number of relevant parameters Gaussiantiy
should hold. Hence, we expect that the Gaussian approximation should
hold to very good accuracy also for experiments with a high
sensitivity to the mass ordering, such as for instance a neutrino
factory~\cite{Choubey:2011zzq, Ballett:2012rz, Christensen:2013va} or the LBNO experiment~\cite{Stahl:2012exa},
when explicit Monte Carlo simulations become exceedingly unpractical
due to the very large number of data sets needed in order to explore
the high confidence levels.

In section~\ref{sec:comparison} we provide a comparison of the
sensitivities of the above mentioned facilities using the statistical
methods discussed in this paper. Figures~\ref{fig:timeplot-median}
and \ref{fig:timeplot-beta} illustrate how the median sensitivity
and the probability to reject the wrong mass ordering at $3\sigma$ CL for the various experiments,
respectively, could evolve as function of
time based on official statements of the collaborations. While this
type of plots is subject to large error bars on the time axis (typically asymmetric) as well as concerning actual experimental
parameters, our results indicate that it is likely that the wrong
mass ordering will be excluded at $3\sigma$ CL within the next 10 to 15
years.

\bigskip

{\bf Acknowledgments.} We thank Walter Winter for comments on the
PINGU sensitivity and Enrique Fernandez-Martinez for useful
discussions. This work was supported by the G\"oran Gustafsson
Foundation (M.B.) and by the U.S. Department of Energy under award
number \protect{DE-SC0003915} (P.C. and P.H.). T.S.\ acknowledges
partial support from the European Union FP7 ITN INVISIBLES (Marie
Curie Actions, PITN-GA-2011-289442).

\appendix

\section{The distribution of $T$}
\label{app:Tgauss}

Consider $N$ data points $x_i$, and the two hypotheses, $H$ and $H'$,
and we want to test whether one of them can be rejected by the
data. The theoretical predictions for the observed data under the two
hypotheses are denoted by $\mu_i$ and $\mu'_i$, respectively. The
prediction $\mu_i$ ($\mu'_i$) may depend on a set of $P$ ($P'$)
parameters $\theta_\alpha$ ($\theta'_\alpha$) which have to be
estimated from the data. For the case of the mass ordering we have
$P=P'$ and $H$ and $H'$ depend on the same set of parameters.
However, here we want to be more general.

Under $H$ the data $x_i$ will be distributed as
$\mathcal{N}(\mu_i(\theta_\alpha^0),\sigma_i)$, where
$\mathcal{N}(m,\sigma)$ denotes the normal distribution with mean $m$
and variance $\sigma^2$ and $\theta_\alpha^0$ are the unknown true
values of the parameters. If $H'$ is true $x_i$ will be distributed as
$\mathcal{N}(\mu'_i({\theta'}_\alpha^0),\sigma'_i)$. Once the experiment
has been performed one can build for each hypothesis a least-square
function:
\begin{align}
X^2(\theta_\alpha; H) & = \sum_i 
\left(\frac{\mu_i(\theta_\alpha) - x_i }{\sigma_i}\right)^2 \\
X^2(H) & = \min_{\theta_\alpha} \sum_i 
\left(\frac{\mu_i(\theta_\alpha) - x_i }{\sigma_i}\right)^2 
= 
\sum_i \left(\frac{\mu_i(\hat\theta_\alpha) - x_i }{\sigma_i}\right)^2
\label{eq:chisq}
\end{align}			       
and similar for $H'$. Here $\hat\theta_\alpha$ are the parameters at the
minimum, which will be different for each hypothesis. In practice
often the variances have to be estimated from the data itself, \eg, $\sigma_i
\approx \sigma'_i\approx \sqrt{x_i}$. In the following we will assume
$\sigma_i = \sigma'_i$. Let us note that generalization to correlated
data is straight forward. The test statistic $T$ from Eq.~\eqref{eq:T}
is then given by $T = X^2(H') - X^2(H)$. In the following we will
derive the distribution of $T$.

\bigskip
{\bf The distributions of $X(H)$ and $X(H')$.}
Let us assume for definiteness that $H$ is true. First we consider
$X^2(H)$, and we derive the well-known result, that $X^2(H)$ is
distributed as $\chi^2$ with $N-P$ d.o.f. Let us define the variables 
\be
y_i(\theta_\alpha) \equiv
\frac{\mu_i(\theta_\alpha) - x_i}{\sigma_i}  \,.
\ee
Under $H$, the $y_i(\theta_\alpha^0) = n_i$ are $N$ standard normal distributed variables
with $\mathcal{N}(0,1)$. Then we have $X^2(H) =
\text{min}_\theta \sum_i [y_i(\theta_\alpha)]^2$. The
minimum condition is 
\be\label{eq:min}
\frac{\partial X^2}{\partial \theta_\alpha} = 2 \sum_i y_i(\hat\theta_\alpha)
\frac{\partial y_i}{\partial \theta_\alpha} = 0 \,.
\ee
Asymptotically the parameter values at the minimum $\hat\theta_\alpha$
will converge to the true values $\theta_\alpha^0$. Therefore 
we assume
\be\label{eq:B}
\left. \frac{\partial y_i}{\partial \theta_\alpha}
\right|_{\hat\theta_\alpha}
\approx
\left. \frac{\partial y_i}{\partial \theta_\alpha} \right|_{\theta_\alpha^0}
\equiv B_{i\alpha} 
\ee
and expand
\be
y_i(\hat\theta_\alpha) = n_i + \sum_\alpha B_{i\alpha} (\hat\theta_\alpha - \theta_\alpha^0) \,.
\ee
Here and in the following sums run over $\alpha,\beta = 1,\ldots, P$
and $i,j, k = 1,\ldots, N$ if not explicitly noted otherwise. Then the
minimum condition Eq.~\eqref{eq:min} becomes
\be\label{eq:min2}
\sum_i
B_{i\alpha} n_i + \sum_{i\beta}
B_{i\alpha} B_{i\beta} (\hat\theta_\beta - \theta_\beta^0)  = 0
\ee
and we obtain
\be
X^2(H) = \sum_i [y_i(\hat\theta_\alpha)]^2 = 
\sum_i n_i^2 - \sum_{i\alpha\beta}
(\hat\theta_\alpha - \theta_\alpha^0)
B_{i\alpha} B_{i\beta} (\hat\theta_\beta - \theta_\beta^0) \,.
\ee
Now we diagonalize the symmetric $P\times P$ matrix $B^TB = (\sum_i
B_{i\alpha} B_{i\beta})$ with the orthogonal matrix $R$ as $B^TB = R^T b^2
R$ with $b = \text{diag}(b_\alpha)$. Then Eq.~\eqref{eq:min2} can be written
as
\be\label{eq:min3}
\sum_\beta
b_\alpha R_{\alpha\beta} (\hat\theta_\beta - \theta_\beta^0) = 
- \sum_i V_{i \alpha} n_i 
\quad\text{with}\quad
V_{i\alpha} \equiv b_\alpha^{-1} \sum_\beta  R_{\alpha \beta} B_{i\beta} 
\ee
and 
\be
X^2(H) = \sum_{ij} n_i
\left(\delta_{ij} - \sum_\alpha
V_{i\alpha}V_{j\alpha}\right) n_j \,.
\ee
The matrix $(V_{i\alpha})$ defined in Eq.~\eqref{eq:min3} is a rectangular
$N\times P$ matrix which per construction obeys the orthogonality condition 
$\sum_i V_{i\alpha} V_{i\beta} = \delta_{\alpha\beta}$. Hence, we can always
complete it by $N-P$ columns to a full orthogonal $N\times N$ matrix such
that $\sum_k V_{ik} V_{jk} = \delta_{ij}$ and $\sum_k V_{ki} V_{kj} =
\delta_{ij}$. Then we have
\be\label{eq:XH}
X^2(H) = \sum_{ij} n_i \left(\sum_{r=P+1}^N V_{ir}V_{jr}\right) n_j
= 
\sum_{r=P+1}^N w_r^2 \,,
\ee
where $w_r \equiv \sum_i V_{ri} n_i$ are $N-P$ independent variables
distributed as $\mathcal{N}(0,1)$. This shows explicitly that if $H$ is true, 
$X^2(H)$ is distributed as a $\chi^2$ with $N-P$ d.o.f.~\cite{Wilks}.

\bigskip

Let us now derive the distribution of $X^2(H')$ under the assumption that
$H$ is true. Again we define
\be
y'_i(\theta'_\alpha) \equiv
\frac{\mu'_i(\theta'_\alpha) - x_i}{\sigma_i}  \,,
\ee
however, now $y'_i$ will not be standard normal
distributed as $\mathcal{N}(0,1)$, since per assumption $x_i$ have
mean $\mu_i(\theta_\alpha^0)$ (and not $\mu'_i$). Nevertheless we can
assume that $y'_i(\theta'_\alpha)$ can be expanded around a fixed
reference point $\theta_\alpha^*$, such that the minimum in the wrong
hypothesis, $\hat\theta'_\alpha$, converges asymptotically towards
it. We write
\begin{align}
\left. \frac{\partial y'_i}{\partial \theta_\alpha} \right|_{\hat\theta'_\alpha}
\approx
\left. \frac{\partial y'_i}{\partial \theta_\alpha} \right|_{\theta_\alpha^*}
\equiv B'_{i\alpha}  \,,\qquad
y'_i(\hat\theta'_\alpha) = y'_i(\theta_\alpha^*)
+ \sum_\alpha B'_{i\alpha} (\hat\theta'_\alpha - \theta_\alpha^*) \,,
\end{align}
and
\be\label{eq:m}
y'_i(\theta_\alpha^*) = \frac{\mu'_i(\theta^*_\alpha) - x_i}{\sigma_i}
= m_i + n_i = n'_i
\quad\text{with}\quad 
m_i \equiv \frac{\mu'_i(\theta^*_\alpha) - \mu_i(\theta^0_\alpha)}{\sigma_i}
\,.
\ee
Here $n_i$ are $\mathcal{N}(0,1)$ as before, but $n'_i$ are $\mathcal{N}(m_i,1)$.
Now the calculation proceeds as before and we arrive at
\be\label{eq:XHp}
X^2(H') = \sum_{r=P'+1}^N (w'_r)^2 \,,
\ee
where $w'_r \equiv \sum_i V'_{ri} n'_i$ are now $N-P'$ independent normal variables
with mean $\langle w'_r\rangle = \sum_i V'_{ri} m_i$. Then $X^2(H')$ has a
so-called non-central $\chi^2$ distribution with $N-P'$ d.o.f.\ and a
non-centrality parameter $\Delta = \sum_{r=P'+1}^N \langle w'_r\rangle^2$.

\bigskip
{\bf The distribution of the test statistic $T$}.
Let us now consider the test statistic $T = X^2(H') - X^2(H)$. 
Using Eqs.~\eqref{eq:XH} and \eqref{eq:XHp} we find:
\begin{align}
T =& 
\sum_{ij} (m_i + n_i)\left(\sum_{r=P'+1}^N V'_{ir}V'_{jr}\right) (m_j + n_j) 
- \sum_{ij} n_i
\left(\sum_{r=P+1}^N V_{ir}V_{jr}\right) n_j \\
=& \sum_{ij} m_i \left(\sum_{r=P'+1}^N V'_{ir}V'_{jr}\right) m_j 
+ 2 \sum_{ij} m_i \left(\sum_{r=P'+1}^N V'_{ir}V'_{jr}\right) n_j \label{eq:T2}\\
&+ \sum_{ij} n_i
 \left( \sum_{\alpha=1}^P V_{i\alpha}V_{j\alpha} - \sum_{\alpha=1}^{P'} V'_{i\alpha}V'_{j\alpha}\right) n_j
\label{eq:T3} 
\end{align}
The first term in Eq.~\eqref{eq:T2} is just a constant, independent of
the data.  Using the definition of $m_i$ in Eq.~\eqref{eq:m} and
comparing with Eq.~\eqref{eq:XH} one can see that this term is identical
to $X^2(H')$ but replacing the data $x_i$ by the prediction for $H$ at
the true values:
\be
\sum_{ij} m_i \left(\sum_{r=P'+1}^N V'_{ir}V'_{jr}\right) m_j 
= \min_{\theta'_\alpha} \sum_i 
\left(\frac{\mu_i'(\theta'_\alpha) - \mu_i(\theta_\alpha^0) }{\sigma_i}\right)^2 
\equiv T_0 \,.
\ee
This is nothing else than the usual ``$\Delta\chi^2$'' between the
two hypotheses without statistical fluctuations, compare Eq.~\eqref{eq:T0}.

The second term in Eq.~\eqref{eq:T2} is a sum of $N$ standard normal
variables, $\sum_i a_i n_i$. This gives a normal variable with variance 
$\sum_i  a_i^2$. It is easy to show from Eq.~\eqref{eq:T2} that the variance is 
$4T_0$ and Eq.~\eqref{eq:T2} can be written as
\be\label{eq:Gauss}
T_0 + 2\sqrt{T_0} n \,,
\ee
with $n$ standard normal. Hence, we find that if the term in
Eq.~\eqref{eq:T3} can be neglected, $T$ is gaussian distributed with
mean $T_0$ and standard deviation $2\sqrt{T_0}$ \cite{Qian:2012zn}.

Consider now the term in Eq.~\eqref{eq:T3}.  Using $\langle n_i
n_j \rangle = \delta_{ij}$ and the orthonormality of $V$ and $V'$ we obtain
that the mean value of this term is $P - P'$. Hence, if the number of
parameters in the two hypotheses is equal (as it is the case for the
neutrino mass ordering) the mean value of $T$ remains $T_0$ as in the
Gaussian approximation, Eq.~\eqref{eq:Gauss}. For testing hypotheses
with different numbers of parameters the mean value will be shifted
from $T_0$. However, even if the mean value remains unaffected, the
higher moments of the distribution can still be modified. Under which
conditions can the term in Eq.~\eqref{eq:T3} be neglected?
\begin{itemize}
\item 
Obviously this term is absent if no parameters are estimated from the
data, $P=P'=0$, \ie, for simple hypotheses. This applies in particular,
if we compare the two hypotheses for fixed parameters.

\item
The term in Eq.~\eqref{eq:T3} will also vanish for
\begin{equation}\label{eq:condV}
\sum_{\alpha = 1}^P V_{i\alpha} V_{j\alpha} = 
\sum_{\alpha = 1}^{P'} V'_{i\alpha} V'_{j\alpha}  
\quad\text{or}\quad
VV^T=V'{V'}^T \,. 
\end{equation}
This condition has a geometrical interpretation. Consider the $N$
dimensional space of data. Varying $P$ parameters $\theta_\alpha$ the
predictions $\mu_i(\theta_\alpha)$ describe a $P$ dimensional subspace
in the $N$ dimensional space. The operator $VV^T$ is a projection
operator into the tangential hyperplane to this subspace at the $X^2$
minimum.  This can be seen by considering the definition of $V$ in
Eq.~\eqref{eq:min3} and of $B$ in Eq.~\eqref{eq:B}, which show that
$V$ is determined by the derivatives
$\partial \mu_i/\partial \theta_\alpha$ at the minimum.  Similar,
$V'{V'}^T$ projects into the $P'$ dimensional tangential hyperplane at
the minimum corresponding to $H'$.  Hence, the
condition \eqref{eq:condV} means that the hyperplanes of the two
hypotheses have to be parallel at the minima. Obviously this condition
can be satisfied only if the dimensions of the hyperplanes are the
same, \ie, $P=P'$.

We note also that the condition \eqref{eq:condV} is invariant under a
change of parameterization, which amounts to $B \to B S$ with
$S_{\alpha\beta} \equiv \partial \theta_\alpha
/ \partial \tilde\theta_\beta$ being a $P\times P$ orthogonal matrix
describing the variable transformation
$\theta_\alpha \to \tilde\theta_\alpha$. Such a transformation would
just change the orthogonal matrix $R$, but leave the operator $VV^T$
invariant. Roughly speaking we can say that sufficiently close to the
respective minima $\theta^0_\alpha$ and $\theta^*_\alpha$, the two
hypotheses should depend on the parameters ``in the same way'', where
the precise meaning is given by Eq.~\eqref{eq:condV}.

\item
Irrespective of the above conditions, we can neglect Eq.~\eqref{eq:T3}
if its variance is much smaller than the variance of the term in
Eq.~\eqref{eq:T2}, which is given by $4T_0$.  Eq.~\eqref{eq:T3} is the
difference of two $\chi^2$ distributions with $P$ and $P'$ d.o.f.,
respectively. The $\chi^2_n$ distribution has a mean and variance of
$n$ and $2n$, respectively. Hence, we should be able to neglect this
term if $T_0 \gg P,P'$, \ie, for high sensitivity experiments.
\end{itemize}

{\bf Example with one parameter.}
To simplify the situation let us consider the case where just one
parameter $\theta$ is estimated from the data, for both $H$ and $H'$. The matrix
$V_{i\alpha}$ defined in Eq.~\eqref{eq:min3} becomes now just a normalized column vector
\begin{equation}\label{eq:V1}
  V_i = \frac{1}{\mathcal{N}}
    \frac{1}{\sigma_i} \frac{\partial \mu_i}{\partial \theta}
\quad\text{with}\quad
 \mathcal{N} = \sqrt{\sum_j \frac{1}{\sigma_j^2}\left(\frac{\partial \mu_j}{\partial \theta}\right)^2}
\end{equation}
and similar for $V'_i$. The term in Eq.~\eqref{eq:T3} is now just the
difference of the square of two standard normal variables: $n^2 -
{n'}^2$, with $n = \sum_i V_i n_i$ and $n' = \sum_i V'_i n_i$. As
mentioned above for the general case, we see that $\langle n^2 -
{n'}^2 \rangle = 0$. The variance of this term is obtained as
\begin{equation}\label{eq:var}
  \langle (n^2 - {n'}^2)^2 \rangle = \sum_{ijkl} 
   \langle n_i n_j n_k n_l \rangle (V_i V_j - V'_i V'_j)(V_k V_l - V'_k V'_l)
   = 4 \left[ 1 - \left(\sum_i V_i V'_i \right)^2 \right]
\end{equation}
where we have used that $\langle n_i^4 \rangle = 3$. We can write
$(V^T V')^2 = \mathrm{Tr}[VV^T V' {V'}^T] = \cos^2\varphi$, where
$\varphi$ is the angle between the two hyperplanes (\ie, lines, in this
case) for $H$ and $H'$. Hence we find that the variance is zero if
$|V^T V'| = 1$, \ie, the lines are parallel. And we have a measure to
estimate when Eq.~\eqref{eq:Gauss} is valid, namely when the variance
of Eq.~\eqref{eq:T3} is small compared to the variance of the second
term in Eq.~\eqref{eq:T2}. In the example of one parameter this means
\begin{equation}\label{eq:cond-1param}
1 - (V^T V')^2 = \sin^2\varphi \ll T_0 \,.  
\end{equation}
Since $\sin^2\varphi \leq 1$ we find that for $T_0 \gg 1$ the gaussian
approximation is expected to be valid if only one parameter is
estimated from data.

\section{Simulation details}
\label{app:simulation}

In the following, we describe the main details that have been used to
simulate the experimental setups considered in this work. Unless
stated otherwise the true values for the oscillation parameters have
been set to the following values~\cite{GonzalezGarcia:2012sz}, and
the $\chi^2$ (or the test statistic $T$) has been minimized with
respect to them by adding Gaussian penalty terms to the
$\chi^2$ with the following $1\sigma$ errors:
\begin{align}
&\theta_{12} = 33.36^\circ \pm 3\% \,,\quad
\sin^22\theta_{13} = 0.089 \pm 0.005 \,,\quad
\sin^22\theta_{23} = 0.97 \pm 0.05 \,,\nonumber\\
&\Delta m^2_{21} = 7.5\times 10^{-5} \,{\rm eV}^2 \pm 2.5\% \,,\quad
\Delta m^2_{31} = \left\{
\begin{array}{rc}
2.47\times 10^{-3} \,{\rm eV}^2 &\,(\NO) \\
-2.43\times 10^{-3} \,{\rm eV}^2& \,(\IO) 
\end{array}
\right\}
 \pm 10\% \,.
\label{eq:osc-params}
\end{align}

Unless otherwise stated, we assume the true value of $\theta_{23}$ to be in the first octant. Nevertheless, the region around $\pi/2-\theta_{23}$ would not be disfavored by the penalty term since it is added in terms of $\sin^22\theta_{23}$ instead of $\theta_{23}$. Therefore, we also look for compatible solutions around $\sim \pi/2 - \theta_{23}$ (the so-called octant degeneracy~\cite{Fogli:1996pv}) and keep the minimum of the $\chi^2$ between the two. 

\subsection{Medium baseline reactor experiment: JUNO} 
\label{app:juno}

We adopt an experimental configuration for the JUNO experiment based
on Refs.~\cite{JUNO, DB2-WWang, Li:2013zyd}, following the analysis
described in Ref.~\cite{Blennow:2013vta}. We normalize the number of events
such that for the default exposure of $\rm 20\, \kt\, \times
36\,GW \times 6\, yr = 4320 \, \kt\, GW \, yr$ we obtain $10^5$
events \cite{JUNO, DB2-WWang}. The energy resolution is assumed to be
$3\%\sqrt{1\,{\rm MeV}/E}$. We perform a $\chi^2$ analysis using 350
bins for the energy spectrum. This number is chosen sufficiently large
such that bins are smaller (or of the order of) the energy
resolution. We take into account an overall normalization uncertainty
of 5\% and a linear energy scale uncertainty of 3\%.  Uncertainties in
the oscillation parameters $\sin^2\theta_{13}$ and $\sin^2\theta_{12}$
are included as pull parameters in the $\chi^2$ using true values and
uncertainties according to Eq.~\eqref{eq:osc-params}, while $|\Delta
m^2_{31}|$ is left free when fitting the data. For this parameter a
dense grid is computed and the minimum is manually searched for. We
have updated the analysis from Ref.~\cite{Blennow:2013vta} by taking into
account the precise baseline distribution of 12 reactor cores as given
in Tab.~1 of Ref.~\cite{Li:2013zyd} (including also the Daya Bay reactors
at 215 and 265~km). This reduces $T_0$ by about 5 units compared to
the idealized situation of a point-like source at 52.47~km (the latter
being the power averaged distance of the 10 reactors not including the
Daya Bay reactors). Adopting the same assumptions as
in Ref.~\cite{Li:2013zyd} we find for a $4320 \rm \, \kt\, GW \, yr$
exposure $T_0 \approx 11.8$, which is in excellent agreement with
their results, see red-dashed curve in Fig.~2 (right)
of Ref.~\cite{Li:2013zyd}.

Our analysis ignores some possible challenges of the experiment, in
particular the effect of a non-linearity in the energy scale
uncertainty \cite{Qian:2012xh}, see also Ref.~\cite{Li:2013zyd,
Capozzi:2013psa}.  While such issues have to be addressed in the
actual analysis of the experiment, our analysis suffices to discuss
the behavior of the relevant test statistic and sensitivity measures.

\subsection{Atmospheric neutrino experiments: PINGU and INO}
\label{app:atm}

For the simulation of the ICal@INO experiment we use the same code as
in Ref.~\cite{Blennow:2012gj}, where further technical details and
references are given. Here we summarize our main assumptions. We
assume a muon threshold of 2~GeV and assume that muon charge
identification is perfect with an efficiency of 85\% above that
threshold. As stressed in Refs.~\cite{Indumathi:2004kd, Petcov:2005rv} the
energy and direction reconstruction resolutions are crucial parameters
for the sensitivity to the mass ordering. We assume here
the ``high'' resolution scenario from Ref.~\cite{Blennow:2012gj}, which
corresponds to a neutrino energy resolution of $\sigma_{E_\nu} = 0.1 E_\nu$
and neutrino angular resolution of $\sigma_{\theta_\nu} = 10^\circ$,
independent of neutrino energy and zenith angle. More realistic
resolutions have been published in Ref.~\cite{Ghosh:2012px}. While those
results are still preliminary, we take our choice to be representative
(maybe slightly optimistic), justified by the fact that we obtain
sensitivities to the mass ordering in good agreement
with Ref.~\cite{Ghosh:2012px}. With our
assumptions we find 242 $\mu$-like events per 50~\kt\,yr exposure
assuming no oscillations (sum of neutrino and anti-neutrino events) in
the zenith angle range $-1 < \cos\theta < -0.1$.  We divide the
simulated data into 20 bins in reconstructed neutrino energy from
2~GeV to 10~GeV, as well as 20 bins in reconstructed zenith angle from
$\cos\theta = -1$ to $\cos\theta = -0.1$. We then fit the
two-dimensional event distribution in the $20\times 20$ bins by using
the appropriate $\chi^2$-definition for Poisson distributed data.  Our
default exposure for INO is a 50~\kt\ detector operated for 10~yr.

For the PINGU simulation we use the same code as
in Ref.~\cite{Blennow:2013vta}, where technical details can be found. In
particular, we adopt the same effective detector mass as a function of
neutrino energy, with the threshold around 3~GeV, and the effective
mass rises to about 4~Mt at 10~GeV and 7~Mt at 35~GeV. For the
reconstruction abilities we assume that neutrino parameters are
reconstructed with a resolution of $\sigma_{E_\nu} = 0.2E_\nu$ and
$\sigma_{\theta_\nu} = 0.5 / \sqrt{E_\nu / \rm GeV}$.  This
corresponds to about $13^\circ \, (9^\circ)$ angular resolution at
$E_\nu = 5$~GeV (10 GeV). We stress that those resolutions (as well as
other experimental parameters) are far from settled. With our choice
we obtain mass ordering sensitivities in good agreement with
Ref.~\cite{Winter:2013ema}, which are somewhat more conservative than
the official PINGU sensitivities from Ref.~\cite{Aartsen:2013aaa}. For a
3~yr exposure and $\theta_{23} = 45^\circ$ we obtain $T_0 \approx
7.5$.

For both, INO and PINGU, we include the following systematic
uncertainties: a 20\% uncertainty on the over-all normalization of
events, and 5\% on each of the neutrino/anti-neutrino event ratio, the
$\nu_\mu$ to $\nu_e$ flux ratio, the zenith-angle dependence, and on
the energy dependence of the fluxes. Moreover, in order to make the
Monte Carlo simulation feasible we set $\Delta m^2_{21} = 0$, which
implies that also $\theta_{12}$ and the CP phase $\delta$ disappear
from the problem. The validity of this approximation and/or the
expected size of $\delta$-induced effects has been studied for
instance in Refs.~\cite{Ghosh:2012px, Winter:2013ema, Blennow:2012gj,
Blennow:2013vta}. Typically $T_0$ varies by roughly 1--2 units as a
function of $\delta$, which is small compared to uncertainties related
to experimental parameters such as reconstruction abilities.  We do
not expect that $\delta$ and $\Delta m^2_{21}$ related effects will
change the statistical behavior of the test statistic $T$
significantly, as also the results of Ref.~\cite{Franco:2013in} seem
to indicate.

\subsection{Long baseline beam experiments: \NOvAt, \LBNE{10}, \LBNE{34}}
\label{app:lbl}

The sensitivity of this type of experiments is largely dependent on
the baseline and neutrino energies considered, which may vary widely
from one setup to another. In this work we have studied three
different setups, \NOvA, \LBNE{10}, \LBNE{34}.

The first setup considered, \NOvA~\cite{Ayres:2004js,Patterson:2012zs}, has a moderate sensitivity to the mass ordering, estimated to reach at most $3\sigma$ (see for instance Refs.~\cite{Patterson:2012zs,Bian:2013saa}). The setup consists of a narrow band beam with neutrino energies around 2 GeV, aiming to a 13~\kt\ Totally Active Scintillator Detector (TASD) placed at a baseline of $L=810$~km. \NOvA\ has recently started taking data. The beam is expected to reach $700$ kW by mid-2014~\cite{Bian:2013saa}, and by the end of its scheduled running time it will have accumulated a total of $3.6\times 10^{21}$~PoT, equally split between $\pi^+$ and $\pi^-$ focusing modes. The detector performance has been simulated following Refs.~\cite{Patterson:2012zs, patterson-priv}. Systematic errors are implemented as bin-to-bin correlated normalization uncertainties over the signal and background rates. These have been set to 5\% and 10\% for the signal and background rates, respectively, for both appearance and disappearance channels.

The second setup considered in this work is the LBNE proposal~\cite{Adams:2013qkq,CDR}. LBNE would use a wide band beam with an energy around 2--3 GeV and a baseline of $L=1300$ km. The first phase of the project (dubbed in this work as \LBNE{10}) consists of a 10~\kt\ Liquid Argon (LAr) detector placed on surface. In a second stage, dubbed in this work as \LBNE{34}, the detector mass would be upgraded to 34~\kt\ and placed underground. The longer baseline and higher neutrino energies make this setup more sensitive to the mass ordering: in its first stage is already expected to reach at least a significance between $~2.5-7\sigma$, depending on the value of $\delta$. The results also depend significantly on the assumptions on systematics and the beam design, see for instance Ref.~\cite{Adams:2013qkq}. In this work, the detector performance has been simulated according to Ref.~\cite{CDR}. Systematic uncertainties have been set at the 5\% level for both signal and background rates in the appearance channels, and at the 5\% (10\%) for the signal (background) rates in the disappearance channels. Tab.~\ref{tab:LBLevents} shows the expected total event rates in the appearance channels for each of the long baseline setups considered in this work. It should be noted the difference in statistics between the \LBNE{10} and \LBNE{34}, which is not only due to the larger detector mass but also to a different neutrino beam design. The first stage of the project, \LBNE{10}, is simulated using the fluxes from the October 2012 Conceptual Design Report, Ref.~\cite{CDR}, while for the upgraded version, \LBNE{34}, we consider the fluxes from Ref.~\cite{Akiri:2011dv}. In both cases the beam power is set to 700~kW.

\begin{table}
\begin{tabular}{l@{\quad}c@{\quad}c}
\hline \hline
   & $\nu_\mu \rightarrow \nu_e $ & $\bar\nu_\mu \rightarrow \bar\nu_e $ \\ \hline 
\NOvA & 61  & 18  \\
\LBNE{10} &  146 & 47  \\ 
\LBNE{34} &  885 & 240  \\ \hline\hline
\end{tabular}
\caption{ Expected total event rates in the appearance channels for the long baseline setups considered in this work. Efficiencies are already accounted for, and the values of the oscillation parameters are set to the central values in Eq.~\eqref{eq:osc-params} and $\delta=0$.  
\label{tab:LBLevents}}
\end{table}

The simulations for the long baseline beam experiments have been
performed using GLoBES~\cite{Huber:2004ka,Huber:2007ji}. In order to
generate random fluctuations in the number of events,
version 1.3 of the MonteCUBES~\cite{Blennow:2009pk} software was used. In
addition to the true values and prior uncertainties for the
oscillation parameters given in Eq.~\eqref{eq:osc-params}, a 2\% uncertainty on the matter density is also considered.


\end{document}